\documentclass[prl, 9pt,twocolumn,superscriptaddress, groupedaddress]{revtex4-2}
\usepackage{bm}
\usepackage{amsmath, amsfonts}
\usepackage{amssymb}
\usepackage{graphicx}
\usepackage{xcolor}
\usepackage[colorlinks=true,allcolors=blue]{hyperref}
\usepackage{cleveref}
\usepackage{mathtools}
\usepackage{newtxtext,newtxmath}

\usepackage[normalem]{ulem}
\begin{document}
\title{High-Order Qubit Dephasing at Sweet Spots by Non-Gaussian Fluctuators:\\ Symmetry Breaking and Floquet Protection}
\author{Ziwen Huang}
\email{zhuang@fnal.gov}
\address{Superconducting Quantum Materials and Systems Center,
Fermi National Accelerator Laboratory (FNAL), Batavia, IL 60510, USA}
\author{Xinyuan You}
\address{Superconducting Quantum Materials and Systems Center,
Fermi National Accelerator Laboratory (FNAL), Batavia, IL 60510, USA}
\author{Ugur Alyanak}
\address{Superconducting Quantum Materials and Systems Center,
Fermi National Accelerator Laboratory (FNAL), Batavia, IL 60510, USA}
\author{Alexander Romanenko}
\address{Superconducting Quantum Materials and Systems Center,
Fermi National Accelerator Laboratory (FNAL), Batavia, IL 60510, USA}
\author{Anna Grassellino}
\address{Superconducting Quantum Materials and Systems Center,
Fermi National Accelerator Laboratory (FNAL), Batavia, IL 60510, USA}
\author{Shaojiang Zhu}
\email{szhu26@fnal.gov}
\address{Superconducting Quantum Materials and Systems Center,
Fermi National Accelerator Laboratory (FNAL), Batavia, IL 60510, USA}

\begin{abstract}
 Although the Gaussian-noise assumption is widely adopted in the study of qubit decoherence, non-Gaussian noise sources, especially the strong discrete fluctuators, have been detected in many qubits. It remains an important task to further understand and mitigate the distinctive decoherence effect of the non-Gaussian noise. Here, we study the qubit dephasing caused by the non-Gaussian fluctuators, and predict a symmetry-breaking effect that is unique to the non-Gaussian noise. This broken symmetry  results in an experimentally measurable mismatch between the extremum points of the dephasing rate and qubit frequency, which demands extra carefulness in characterizing the noise and locating the optimal working point. To further enhance the coherence time at the sweet spot, we propose to suppress the second-order derivative of the qubit frequency by the Floquet engineering. Our simulation with a heavy fluxonium shows  an order of magnitude improvement  of the dephasing time, even after including the noise introduced by the drive.
\end{abstract}
\maketitle

\textit{Introduction.--} Efficient quantum computing relies on high-coherence qubits and high-fidelity quantum operations \cite{DiVincenzoCriteria,Preskill_NISQ_review}. However, further enhancement of the coherence times and gate fidelities toward full quantum error correction, one prerequisite for full-fledged quantum computation \cite{DiVincenzoCriteria,Preskill_NISQ_review,Oliver_scqubits_review,Bruzewicz_trapped_ion_review}, has been set back by the  decoherence caused by the environmental noise \cite{Bruzewicz_trapped_ion_review,Muller_tls_review,Murray_material_review,Gyenis_protected_review}. Superconducting qubits, one of the leading qubit platforms, are especially subjected to the uncontrolled low-frequency environmental fluctuations \cite{Dutta_1/f_review,Muller_tls_review,Weissman_1/f_review,Murray_material_review,McDermott_spin_1/f_noise,Devoret_quasi_particle_decoherence,Yu_critical_current,Weides_transmon_single_fluctuator,Pop_fluxonium_beating,Palmer_CPB_TLS,Ustinov_TLS_TLF,Mariantoni_TLS_TLF,Petukhov_nonGaussianDD,Martinis_flux_noise,Gyenis_protected_review,Devoret_quantronium,Nakamura_charge_qubit,Nakamura_charge_1/f,Fluxonium_hc,Manucharyan_heavy_fluxonium,Schuster_heavy_fluxonium,Schuster_heavy_fluxonium_control,Paladino_TLF_review,Mooij_flux_qubit_2003,Martinis_1/f_spectrum,Oliver_SQUID_1/f,Oliver_flux_qubit_revisited}. For example, the notorious $1/f$ noise limits the coherence times of many superconducting qubits \cite{Martinis_flux_noise,Devoret_quantronium,Nakamura_charge_qubit,Nakamura_charge_1/f,Mooij_flux_qubit_2003,Fluxonium_hc,Martinis_1/f_spectrum,Manucharyan_heavy_fluxonium,Schuster_heavy_fluxonium,Schuster_heavy_fluxonium_control,Paladino_TLF_review,Petukhov_nonGaussianDD,Oliver_SQUID_1/f,Oliver_flux_qubit_revisited}.

This challenge has motivated research efforts on identifying, understanding, and mitigating the contributing noise channels \cite{Pop_fluxonium_beating,Palmer_CPB_TLS,Ustinov_TLS_TLF,Mariantoni_TLS_TLF,Dutta_1/f_review,Muller_tls_review,Weissman_1/f_review,Murray_material_review,McDermott_spin_1/f_noise,Devoret_quasi_particle_decoherence,Yu_critical_current,Weides_transmon_single_fluctuator,Martinis_flux_noise,Gyenis_protected_review,Devoret_quantronium,Nakamura_charge_qubit,Nakamura_charge_1/f,Fluxonium_hc,Manucharyan_heavy_fluxonium,Schuster_heavy_fluxonium,Schuster_heavy_fluxonium_control,Paladino_TLF_review,Mooij_flux_qubit_2003,Martinis_1/f_spectrum,Glazman_quasi_particle_tunneling,Koch_transmon_theory,Groszkowski_Zero_pi_theory,Schon_TLS_ensemble,Paladino_initial_decoherence,Paladino_decoherence_saturation,Shnirman_Keldysh_dephasing,Tokura_dephasing,Altshuler_telegraph_TLF,Matteo_Gaussian_nonGaussian,Bergli_tlf_arbitrary,Galperin_Echo_TLFs,Ithier_decoherence_analysis,Koch_TLS_FD_thoerem,Dynamical_sweet_spot,Dynamical_sweet_spot_exp,Koch_current_mirror,McDermott_spin_1/f_noise,Didier_dynamical_sweet_spot,Didier_dynamical_sweet_spot_exp,Petukhov_nonGaussianDD,Oliver_SQUID_1/f,Oliver_non_Gaussian_spec,Oliver_flux_qubit_revisited,Oliver_flux_qubit_dd,Houck_zero_pi_experiment,Houck_tantalum_qubit,Yu_tantalum_qubit,Fluxonium_Devoret,Clerk_non_Gaussian,Shnirman_Keldysh_dephasing_1,Didier_dynamical_sweet_spot_exp,Didier_dynamical_sweet_spot,Rigetti_ac_sweet_spot_exp,Manucharyan_1ms_fluxonium,Bell_bifluxon,Makhlin_TLS_1/f,Rigetti_ac_sweet_spot,Nori_C_shunt_flux}. Theoretically, the noise is usually assumed Gaussian (sometimes Markovian), while the realistic noise is more complicated. For example, if strong discrete fluctuators are present \cite{Petukhov_nonGaussianDD,Weides_transmon_single_fluctuator,Pop_fluxonium_beating,Palmer_CPB_TLS,Ustinov_TLS_TLF,Mariantoni_TLS_TLF}, the decoherence process of the qubits deviates significantly from the prediction by assuming only Gaussian noise \cite{Weides_transmon_single_fluctuator,Weissman_1/f_review,Weissman_1/f_review,Muller_tls_review,Dutta_1/f_review,Paladino_TLF_review,Galperin_Echo_TLFs,Bergli_tlf_arbitrary,Paladino_initial_decoherence,Paladino_decoherence_saturation}. Since the optimization of the circuit design and control protocols of the superconducting qubits crucially relies on the appropriate estimation of the decoherence rate, it is important to carefully include the realistic environmental noise in the analysis. 

In this Letter, we focus on the qubit dephasing at the sweet spots, the working points where the qubits reach their maximal dephasing time \cite{Manucharyan_1ms_fluxonium,Schuster_heavy_fluxonium_control,Fluxonium_hc,Manucharyan_fluxonium_two_qubit_gate,Houck_zero_pi_experiment,Didier_dynamical_sweet_spot_exp,Oliver_flux_qubit_revisited,Bell_bifluxon}. We report that the non-Gaussian noise introduces a previously unrevealed dephasing feature, the $\mathbb{Z}_2$ symmetry breaking at the qubit sweet spot. The breaking is predicted to cause an otherwise unexpected mismatch between the extremum points of the qubit dephasing rate and its oscillating frequency. Using a realistic non-Gaussian noise model, the strong two-level fluctuators (TLFs), we theoretically demonstrate such mismatch in a concrete qubit model, the heavy fluxonium qubit \cite{Schuster_heavy_fluxonium_control}. This finding provides a simple yet decisive tool to identify the non-Gaussian component in the noise background \cite{Oliver_non_Gaussian_spec}. To further enhance the sweet-spot coherence time limited by such fluctuators (also Gaussian noise), we propose a triple-protection scheme via the Floquet engineering \cite{Dynamical_sweet_spot,Dynamical_sweet_spot_exp,Sillanpaa_hybrid_circuit,Blais_Floquet,Oliver_extensible_circuit}, where the qubit is not only protected from the dc fluctuation to the second order, but also from ac fluctuation to the first order.

\textit{Model.--} The system we study is a tunable qubit subjected to the low-frequency fluctuation. The full Hamiltonian in consideration is given by $\hat{H} = \hat{H}_q(\lambda) + \delta\xi(t) \hat{x}$, where $\hat{H}_q(\lambda)$ denotes the bare qubit Hamiltonian, $\delta\xi(t)$ is the fluctuations due to the environmental noise and $\hat{x}$ is the operator that the fluctuators are coupled to. The qubit is tuned by an external control parameter $\lambda$ according to $\hat{H}_q(\lambda) = \hat{H}_q(0)+\lambda\hat{x}$, which for example corresponds to the  flux (charge) control of the fluxonium qubit (Cooper-pair box) \cite{Koch_transmon_theory,Fluxonium_Devoret}. We assume that the first-order derivative of this qubit vanishes at $\lambda = 0$, which is a result of a $\mathbb{Z}_2$ symmetry. Specifically, the Hamiltonian satisfies $\hat{R}\hat{H}_{q}(0)\hat{R}^\dagger = \hat{H}_{q}(0)$, where $\hat{R}$ is a reflection operation defined by the relation $\hat{R}\hat{x}\hat{R}^\dagger = -\hat{x}$. This symmetry ensures $(\partial \omega^{\mathrm{bare}}_j/\partial \lambda)|_{\lambda = 0} = 0$, where $\omega^{\mathrm{bare}}_j$ is the eigenenergy of the $j$th eigenstate of the bare Hamiltonian $\hat{H}_{q}(\lambda)$. The external control is assumed to only 
modulate the Hamiltonian of the qubit while negligibly affecting the properties of the fluctuation $\delta\xi(t)$ over the small range of $\lambda$ under inspection. 


\textit{$\mathbb{Z}_2$ symmetry and its breaking.--} The symmetry we address is the invariance of the ensemble-averaged qubit density matrix elements, ${\rho}_{ij}(t)|_\lambda$, evaluated in their respective eigenbasis, under the reflection $\lambda\rightarrow -\lambda$. Such a symmetry is strictly preserved if the fluctuation is Gaussian, and is generally broken for non-Gaussian $\delta\xi(t)$.

\begin{figure}
    \centering
    \includegraphics[width = 7cm]{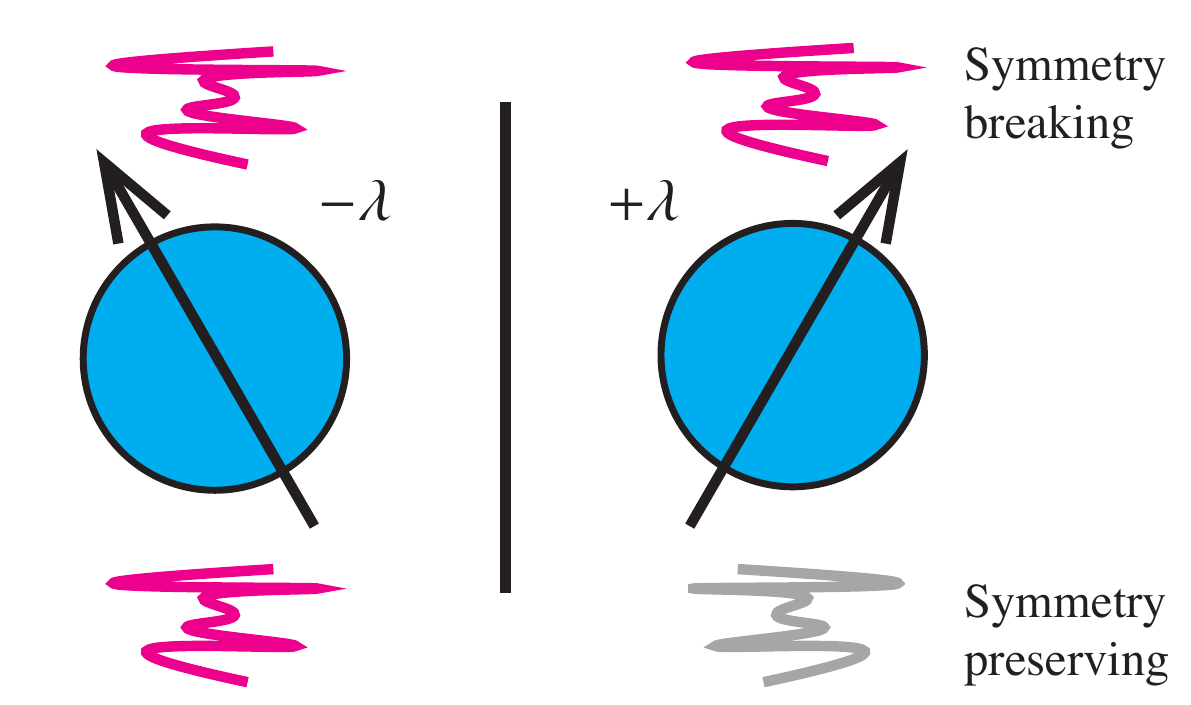}
    \caption{A cartoon illustrating how noise breaks the reflection symmetry of the qubit. Since the fluctuation $\delta\xi(t)$ we consider is independent of the control parameters, the noise term $\hat{H}_I(t)$ will not undergo a similar reflection (bottom gray curve) as the bare Hamiltonian by the external operation $-\lambda\rightarrow\lambda$, which in general breaks the reflection symmetry. Exceptionally, the Gaussian noise still preserves this symmetry because the statistical effect of the Gaussian fluctuator $\delta\xi_G(t)$ on the qubit is identical to that of $-\delta\xi_G(t)$.}
    \label{fig:SymmetryBreaking}
\end{figure}

To rigorously derive the conclusion above, we first inspect the relation between the qubit Hamiltonians and eigenstates for $\pm\lambda$. The Hamiltonians are related by $\hat{H}_q(-\lambda) = \hat{R}\hat{H}_q(\lambda)\hat{R}^\dagger$, and the eigenstates can also be chosen to transform by such reflection,  $\vert j(-\lambda)\rangle = \hat{R}\vert j(\lambda)\rangle$, to avoid unnecessary complication. However, the interaction Hamiltonian $\hat{H}_I(t) \equiv \delta\xi(t)\hat{x}$ does not transform similarly, because $\hat{H}_I(t)$ is independent of the choices of $\lambda$, while a reflection operation on this interaction term flips its sign, i.e., $\hat{R}\hat{H}_I(t)\hat{R}^\dagger = -\hat{H}_I(t)$. As a result of this sign flip, we find the following relation between the propagators:
\begin{align}
    \hat{R}\tilde{U}^{\nu}_I(t)|_{\lambda}\hat{R}^\dagger = (-1)^\nu\tilde{U}_I^\nu(t)|_{-\lambda}.
    \label{eq:propagator_reflection}
\end{align}
Above, $\tilde{U}^{\nu}_{I}(t)$ is the $\nu$th-order term in the Dyson expansion of the interaction propagator $\tilde{U}_I(t) = \mathcal{T}\!\exp[-i\int_0^tdt'\tilde{H}_I(t')]$, where $\tilde{H}_I(t) = \hat{U}_0^\dagger(t)\hat{H}_I(t)\hat{U}_0(t)$ and $\hat{U}_0(t) \equiv \exp[-i\hat{H}_q(\lambda)t]$. The propagator $\tilde{U}_I(t)$ is expanded as $\tilde{U}_I(t)=\sum_{\nu}\tilde{U}^{\nu}_{I}(t)$ according to the Dyson series, where $\tilde{U}^\nu_{I}(t)$ contains a product of $\nu$ times of $\tilde{H}_I(t)$.

The sign flipping is further carried to the density matrix elements, which are expanded as
\begin{align}
{\rho}_{I,jk}(t)|_{\lambda}\! \equiv &\,\overline{\langle j\vert  \tilde{\rho}_{I}(t)\vert k\rangle}|_{\lambda}\! =\! \overline{\langle j\vert  \tilde{U}_I(t)\tilde{\rho}_{I}(0)\tilde{U}^\dagger_I(t)\vert k\rangle}\Big|_{\lambda}\nonumber\\
= &\!\sum_{j'k'\nu}\!{\rho}_{I,j'k'}(0)\,\Pi^{\nu}_{jk\leftarrow j'k'}(t)|_{\lambda}.
\end{align}
Here, we define the interaction-picture density matrix $\tilde{\rho}_{I}(t) = \hat{U}^\dagger_0(t)\hat{\rho}(t)\hat{U}_0(t)$ and the Keldysh projector $\Pi^{\nu}_{jk\leftarrow j'k'}(t)|_{\lambda} \equiv \sum_{\nu'+\nu'' = \nu}\overline{\langle j\vert \tilde{U}^{\nu'}_I(t) \vert j'\rangle \langle k'\vert \, \tilde{U}^{\nu''\dagger}_I(t)\vert k\rangle}|_{\lambda}$. Using Eq.~\eqref{eq:propagator_reflection}, one finds $\Pi^{\nu}_{jk\leftarrow j'k'}(t)|_{\lambda}=(-1)^\nu \Pi^{\nu}_{jk\leftarrow j'k'}(t)|_{-\lambda}$, which implies that ${\rho}_{I,jk}(t)|_{\lambda}$ and ${\rho}_{I,jk}(t)|_{-\lambda}$ are unequal in general. One exception is when all odd projectors are not relevant. The Gaussian noise ensures this condition due to its vanishing odd correlation functions, such as ${\delta\xi_G(t)\delta\xi_G(t_1)\delta\xi_G(0)}=0$. This special property allows us to ignore the complication of flipping signs, and finally find the invariance relation, ${\rho}_{I,jk}(t)|_{-\lambda} = {\rho}_{I,jk}(t)|_{\lambda}$ (the same conclusion in the lab frame), given the identical initial matrix elements $\rho_{I,j'k'}(0)$.  On the other hand, non-Gaussian noise does not in general preserve this equality. We summarize the discussion of the symmetry breaking and preserving  in FIG.~\ref{fig:SymmetryBreaking}.

To theoretically confirm the predicted symmetry breaking, we adopt a simple yet realistic model, i.e., the strong TLFs \cite{Paladino_decoherence_saturation,Paladino_initial_decoherence,Paladino_TLF_review}. This model is motivated by both the microscopic understanding of the materials used in solid-state qubits \cite{Muller_tls_review} and abundant experimental evidence \cite{Petukhov_nonGaussianDD,Weides_transmon_single_fluctuator,Pop_fluxonium_beating,Palmer_CPB_TLS,Ustinov_TLS_TLF,Mariantoni_TLS_TLF}. Therefore, studying and understanding the dephasing rates by the TLFs also have realistic values for research on solid-state qubits. To make our results more broadly applicable, the full fluctuation $\delta\xi(t)$ here consists of both the Gaussian noise and $N_T$ strong TLFs, which specifies $\delta\xi(t) = \sum_{\mu=1}^{N_T} \delta\xi_{T\!\mu}(t)+\delta\xi_G(t)$. The TLFs can only take two values, i.e., $\pm |\xi_{T\!\mu}| - \overline{\xi}_{T\!\mu}$, where $|\xi_{T\!\mu}|$ describes the noise magnitude, and $\overline{\xi}_{T\!\mu}$ is used to cancel the time average of the fluctuator. 

The dephasing of a qubit at its sweet spot by the strong low-frequency noise reaches beyond the description of the master-equation formalism, which is based on the second-order approximation \cite{Breuer_open_quantum_system}. To study the high-order dephasing effect, we use the Keldysh diagrammatic technique \cite{Shnirman_Keldysh_dephasing,Shnirman_Keldysh_dephasing_1,Muller_Keldysh,Clerk_non_Gaussian} to perturbatively calculate the evolution of qubit density matrix \cite{Supplementary}. Our results provide an expression of the free-induced (Ramsey) evolution of the off-diagonal matrix element $\rho_{eg}(t)\approx\rho_{eg}(0)\exp[-i\omega_q't-\Phi(t)]$. Here, the Lamb-shifted qubit oscillation frequency is approximated by $\omega_q' \approx \Delta + D_{2,\lambda=0}\left[\lambda^2 + \int\!d\omega S(\omega)/2\pi\right]/2$, where the coefficient $D_{2,\lambda=0}\equiv( \partial^2\omega_{ge}^{\mathrm{bare}}/\partial \lambda^2)|_{\lambda=0}$ denotes the second-order derivative of the bare qubit frequency and $\Delta$ is the extremum qubit bare frequency $\Delta\equiv (\omega^{\mathrm{bare}}_e-\omega^{\mathrm{bare}}_g)|_{\lambda = 0}$. (Here and below, we omit the integration limits $\pm\infty$ in $\int_{-\infty}^{\infty}$ to save space.) The dephasing profile $\Phi(t)$ can be divided into multiple contributions $\Phi(t) \approx \,\Phi_G(t) +\! \sum_{\mu=1}^{N_T}\Phi_{T\!\mu}(t)  + \sum_{\upsilon\neq\upsilon'}\Phi_{\upsilon\! \upsilon'}(t)$, where $\Phi_G(t)$, $\Phi_{T\!\mu}(t)$ and $\Phi_{\upsilon\!\upsilon'}(t)$ are contributed by the Gaussian noise, the $\mu$th TLF and mutual effect by two different components. Specifically, they are given by
\begin{align}
    \Phi_G(t) = &\, D_{2,\lambda=0}^{2}\,\lambda^2\int\frac{d\omega}{2\pi}S_G(\omega)K^R(\omega,t)\label{eq:dephasing}\\
    +&\frac{D_{2,\lambda=0}^{2}}{2}\!\iint\frac{d\omega}{2\pi}\frac{d\omega'}{2\pi}S_{G}(\omega)S_{G}(\omega')K^R(\omega+\omega',t),\nonumber\\
    \Phi_{T\!\mu}(t) =&\, D_{2,\lambda=0}^{2}\left(\lambda-\overline{\xi}_{T\!\mu}\right)^2\!\int\frac{d\omega}{2\pi}S_{T\!\mu}(\omega)K^R(\omega,t),\nonumber\\
    \Phi_{\upsilon\!\upsilon'}(t) =&\, \frac{D_{2,\lambda=0}^{2}}{2}\!\iint\!\frac{d\omega}{2\pi}\frac{d\omega}{2\pi}'S_{\upsilon}(\omega)S_{\upsilon'}(\omega')K^R(\omega+\omega',t),\nonumber
\end{align}
where $S_{\!\upsilon}(\omega)$ is the Fourier transformation of the two-point correlation functions,  $S_{\!\upsilon}(\omega) = \int\!\! dt\, \overline{\xi_\upsilon(t)\xi_{\upsilon}(0)}e^{i\omega t}$ ($\upsilon = G, T\!\mu$ and $\mu=1,2,\cdots, N_T$), and the sum spectrum is denoted by $S(\omega) = \sum_{\upsilon}S_{\!\upsilon}(\omega)$. These spectra are sampled time-dependently according to the filter function $K^R(\omega,t) \equiv t^2 \mathrm{sinc}^2\left(\omega t/2\right)/2$. The symmetry breaking is introduced by $\Phi_{T\!\mu}(t)$, since  nonzero $\overline{\xi}_{T\!\mu}$ leads to $\Phi_{T\!\mu}(t)|_{\lambda}\neq \Phi_{T\!\mu}(t)|_{-\lambda}$. Meanwhile, up to the leading order, the qubit frequency is still symmetric after including the Lamb shift, implying an unexpected mismatch between the minima of the dephasing rate and frequency. 

In addition to the analytical calculation based on the  perturbation theory, we further confirm our prediction independently with numerical simulation. The method we use to emulate the qubit evolution subjected to low-frequency fluctuation is the stochastic Schrödinger equation (SSE) \cite{Paladino_initial_decoherence,Paladino_decoherence_saturation}. This method averages the ensemble of traces of the qubit evolution according to many realizations of the random fluctuation. We choose  the heavy fluxonium qubit as a concrete model \cite{Fluxonium_Devoret}, and generate the fluctuation traces according to $\delta\xi(t) = \sum_{\mu=1}^{N_T} \delta\xi_{T\!\mu}(t)+\delta\xi_G(t)$.

\begin{figure}
    \centering
    \includegraphics[width = 7.5cm]{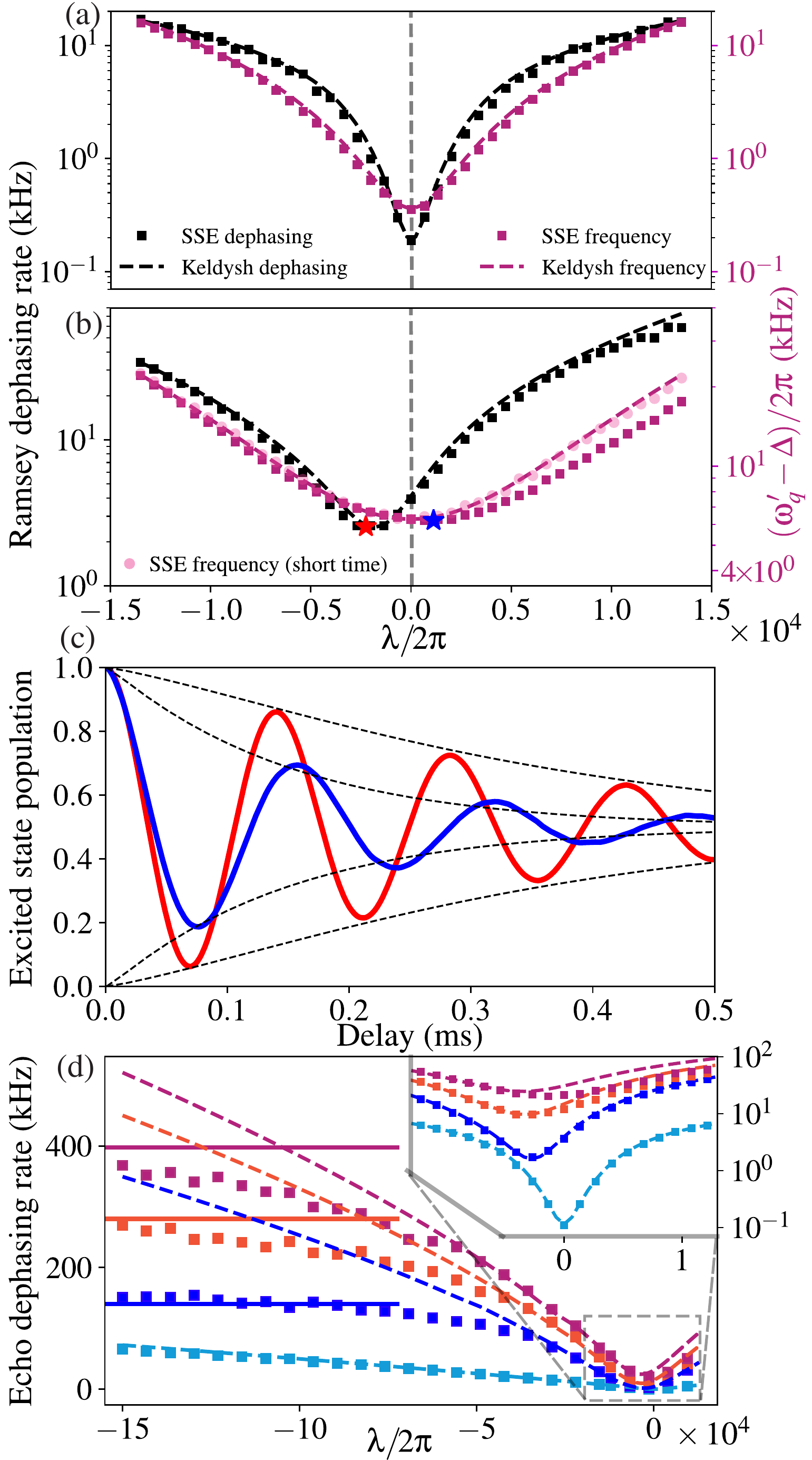}
    \caption{Noise-mediated evolutions of a fluxonium qubit. (a) and (b) show the calculated qubit frequency (purple) and dephasing rates (black) by the Keldysh perturbation theory (dashed lines) and SSE (squares and dots) by simulating the Ramsey measurements. In (c), the red and blue curves are the simulated Ramsey signals at the points marked by red and blue stars (b), respectively. The black dashed curves describe the dephasing profile predicted using Eq.~\eqref{eq:dephasing}. We choose $N_T = 0$ for (a) and $N_T = 1$ for (b) and (c). (d) plots the fitted Echo dephasing rates over a wider range of control parameters, where the dashed curves are from analytical calculation and the squares from SSE. The number of TLFs $N_T$ used in this simulation is varied from 0 to 3 (cyan, blue, red, purple). The solid lines mark the projected saturation rates by the strong TLFs. The details of the qubit and noise are given as follows. The qubit Hamiltonian is given by $\hat{H} \equiv 4E_C\hat{n}^2 + E_L[\hat{\varphi}+\phi_\mathrm{ext} + \delta\xi(t)]^2/2 -E_J\cos\hat{\varphi}$, where $\hat{\varphi}$ and $\hat{n}$ are the conjugate phase and charge operators. The circuit parameters are chosen as $E_C/2\pi = 0.479$ GHz, $E_L/2\pi = 0.132$ GHz and $E_J/2\pi = 3.395$ GHz according to Ref.~\cite{Schuster_heavy_fluxonium_control}. The sweet spot we study is located at $\phi_{\mathrm{ext}} = \pi$, which gives the choices $\lambda = \phi_{\mathrm{ext}}-\pi$ and $\hat{x} \equiv E_L \hat{\varphi}$. The strong TLFs have a magnitude $|\xi_T|/2\pi = 9\times 10^{-5}$ and uneven probability distribution $P_{-(+)}=0.7 (0.3)$. The Gaussian noise bath is approximated by 2001 much weaker and independent TLFs with even probability distributions, which have a $1/\kappa$ distribution \cite{Makhlin_TLS_1/f,Koch_TLS_FD_thoerem} for $\kappa\in[1\, \mathrm{kHz}, 1\, \mathrm{MHz}]$ to mimic the classical $1/f$ noise, and yield a noise magnitude $|{\delta\xi_G(t)}|/2\pi = 2\times 10^{-5}$. }
    \label{fig:doublewell}
\end{figure}

In FIG.~\ref{fig:doublewell} (a) and (b), we plot the calculated qubit frequencies and dephasing rates as functions of the control parameter $\lambda$ for $N_T =0$ (purely Gaussian) and 1 (non-Gaussian), respectively. Both the dephasing rates $\gamma_2$ and the qubit frequency $\omega_q'$ are extracted by fitting the simulated Ramsey measurement using a simple exponential function over the same time range. (Note that the dephasing profile is not exactly exponential given the structured noise we use, but we choose this simplest protocol for consistency in comparing the results from different methods and parameters.) For both cases, we find impressive agreement between our analytical prediction (dashed lines) and numerical simulation (squares). In (a), the Gaussian 
fluctuator $\delta\xi_G(t)$ ensures the $\mathbb{Z}_2$ symmetry under the reflection $-\lambda\rightarrow \lambda$. In (b), we observe a measurable mismatch between the minimum of the dephasing rate and the qubit frequency. For a more intuitive demonstration of such mismatch, we showcase two simulated Ramsey signals in FIG.~\ref{fig:doublewell} (c), taken at the numerically simulated minima of the dephasing rate (red star)  and qubit frequency (blue star) in (b). Although the red signal decays slower, it oscillates faster compared to the blue one. This mismatch demands more carefulness in locating the optimal working point experimentally: the extremum point of the qubit frequency is not necessarily the qubit dephasing sweet spot. Based on our simulation, the Ramsey dephasing rates at the two mismatched minima differ by a factor of 2. Note that different from the analytical prediction, we also find a shift of the minimum of the qubit frequency from $\lambda = 0$ for $N_T=1$. This is attributed to the less accuracy of the perturbation theory at long times ($t\sim 0.5$ ms). [We find much better agreement if we fit for the rate over $0<t<20\,\mu$s, as plotted by light pink circles in (b).]

Farther away from the sweet spot, the dephasing rates are constantly measured to extract the amplitude of the low-frequency noise \cite{Schuster_heavy_fluxonium_control,Fluxonium_hc,Houck_zero_pi_experiment,Oliver_SQUID_1/f}, due to its usually dominant decoherence contribution. In this operating region, the non-Gaussian fluctuators, for example, the two-level fluctuators, can also dephase the qubit distinctively from its Gaussian counterpart \cite{Altshuler_telegraph_TLF,Paladino_decoherence_saturation,Paladino_initial_decoherence,Galperin_Echo_TLFs,Bergli_tlf_arbitrary}. Using the theory developed above, we extend our calculation to this parameter region to reveal such distinction in detail, especially the different dependence of the dephasing rate on $\lambda$. Because the strong discrete noise can introduce beatings in the Ramsey signals \cite{Altshuler_telegraph_TLF,Paladino_decoherence_saturation,Paladino_initial_decoherence}, we switch to the Echo protocol to extract the dephasing rates. We plot the calculated results using the analytical (dashed lines) and numerical methods (squares) in FIG.~\ref{fig:doublewell} (d). From the bottom to top, the number of TLFs is varied from 0 to 3 (cyan, blue, red, purple). For the analytical method, the only difference from the Ramsey protocol is the replacement of $K^R\!(\omega,t)$ by $K^E\!(\omega,t)\! \equiv\!t^2\mathrm{sinc}^2(\omega t/4)\sin^2(\omega t/4)/2$ in Eq.~\eqref{eq:dephasing} \cite{Supplementary}. We find great agreement in the rates obtained by the analytical (dashed curves) and numerical (squares) methods for $N_T = 0$  and in the vicinity of $\lambda\approx 0$ for nonzero $N_T$'s,  but the deviation is prominent for larger $|\lambda|$ in the non-Gaussian ($N_T>0$) cases. In that regime, the dephasing rate contributed by the TLFs saturates \cite{Altshuler_telegraph_TLF,Paladino_decoherence_saturation,Paladino_initial_decoherence,Galperin_Echo_TLFs} and does not further grow with larger qubit linear dispersion.  Using a method based on the rotating-wave approximation \cite{Supplementary}, we derive the saturated dephasing rates as $\overline{\kappa} = \sum_{\vec{\eta}}\overline{\kappa}_{\vec{\eta}}P_{\vec{\eta}}$, where $\vec{\eta}$ is one configuration of the group of strong TLFs, and $\kappa_{\vec{\eta}}$ and $P_{\vec{\eta}}$ are the flipping rate corresponding to that configuration and the probability of finding such configuration, respectively. These predicted saturation rates (solid lines) well capture the  behaviors of the dephasing rates at the far left part of the FIG.~\ref{fig:doublewell} (d). Therefore, if non-Gaussian fluctuators are present, the estimated noise amplitude will be inconsistent at different $\lambda$ if one still assumes the noise as Gaussian.

\begin{figure}[b!t!]
    \centering
    \includegraphics[width = 8.1cm]{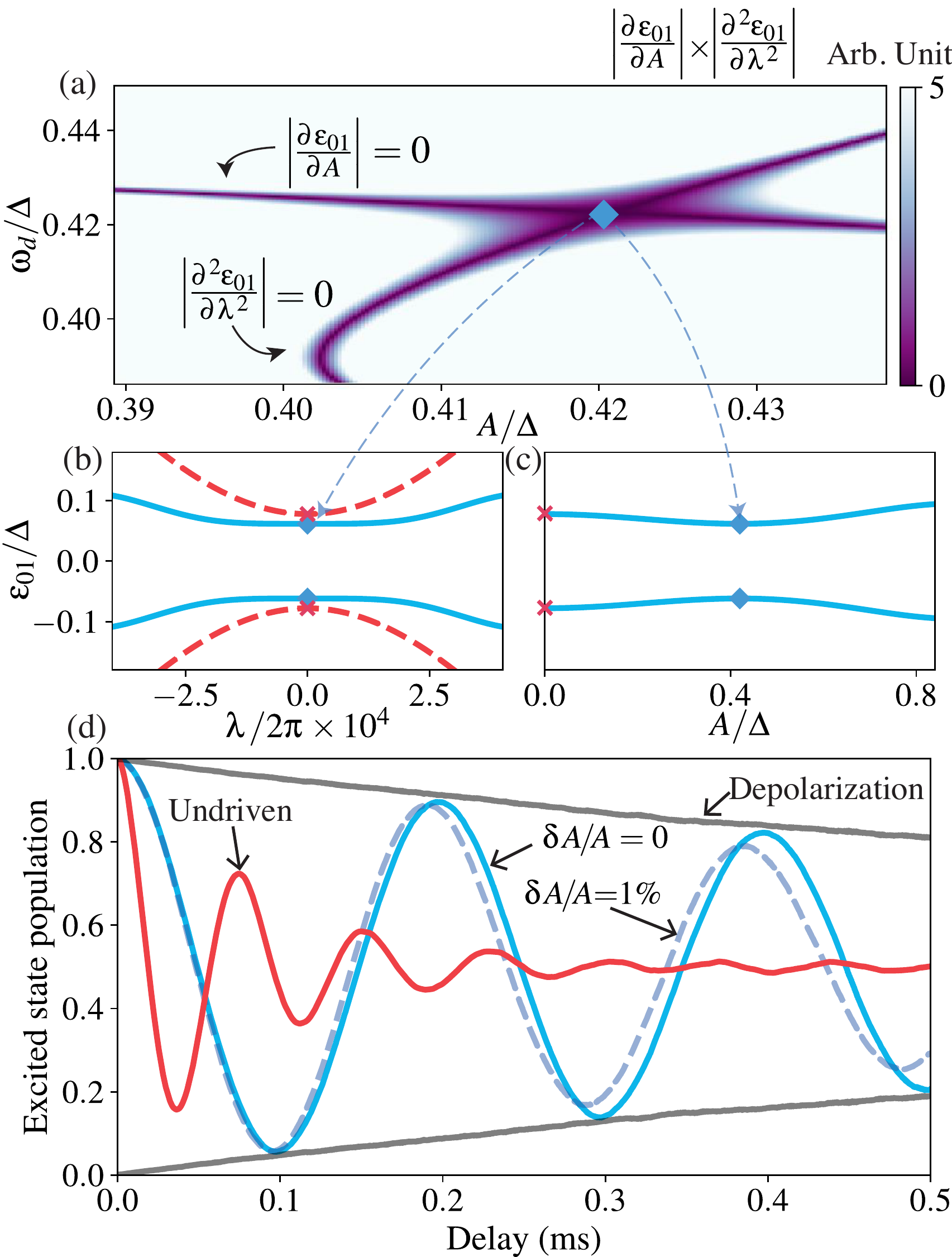}
    \caption{Floquet engineering of the triple sweet spots. (a) plots the product  of the derivatives, i.e., $|\partial\varepsilon^2_{01}/\partial \lambda^2|\times|\partial\varepsilon_{01}/\partial A|$, as a function of the drive amplitude $A$ and frequency $\omega_d$. The two dark curves  correspond to the vanishing of the two derivatives, respectively, and  the crossing hosts the triple sweet spot (cyan diamond). (b) and (c) show the quasi-energy spectra as functions of the dc control parameter $\lambda$ and the drive amplitude $A$, respectively, to show the flattening of the spectra visually. The red dashed lines describe the quasi-energies of the undriven bare fluxonium and the crosses mark the location of static working points ($A=0$). (d) compares the simulated Ramsey measurements between the Floquet engineered triple sweet spot (cyan) and the static one (red). The dashed cyan curve corresponds to the scenario where the drive amplitude has a $1\%$ random fluctuation. The gray solid curves are the simulated Floquet excited state population for $1\!\!\rightarrow\!\! 0$ decay (top) and $0\!\!\rightarrow\!\!1$ excitation (bottom), which we use to confirm its dominant contribution to the Floquet dephasing.}
    \label{fig:Floquet}
\end{figure}

\textit{Triple protection by Floquet engineering.--} Our calculations suggest the need to explore beyond the first-order protection from the low-frequency noise at traditional sweet spots. Specifically, Eq.~\eqref{eq:dephasing} points out the crucial role of the second-order derivative $D_{2,\lambda = 0}$ in limiting the optimal coherence times, for the presence of both the Gaussian noise and non-Gaussian TLFs. From the hardware level, it is ideal to design a qubit with flatter spectrum \cite{Koch_transmon_theory,Nori_C_shunt_flux,Groszkowski_Zero_pi_theory,Houck_zero_pi_experiment,Koch_current_mirror}.  However, the experimental implementation of some of these qubits, especially the protected qubits which are predicted to possess both long $T_1$ and $T_2$, encounters challenges from the experimentally achievable range of circuit parameters for full protection \cite{Houck_zero_pi_experiment,Gyenis_protected_review}. In the following, we present a drive-based protocol that can suppress the second-order derivative of the qubit, which can circumvent the limitations from the hardware level.

Different from the dynamical sweet spots studied in \cite{Dynamical_sweet_spot,Dynamical_sweet_spot_exp}, here we fix the dc control parameter at its static sweet spot ($\lambda = 0$), and search for operating points where both $\partial\varepsilon_{01}/\partial\lambda$ and $\partial^2\varepsilon_{01}/\partial^2\lambda$ vanish. (Here, $\varepsilon_{01}$ denotes the \textit{bare} quasi-energy difference.)  In fact, these operating points are not difficult to find. For example, this derivative vanishes if the qubit is driven resonantly by  $\hat{H}_d(t) = A\cos(\omega_d t)\, \hat{x}$  (choosing $\omega_d\approx\Delta$) at its sweet spot \cite{Dynamical_sweet_spot}, where $A$ and $\omega_d$ denote the drive amplitude and frequency, respectively. However, as pointed out in Refs.~\cite{Dynamical_sweet_spot_exp,Rigetti_ac_sweet_spot_exp,Didier_dynamical_sweet_spot_exp,Didier_dynamical_sweet_spot,Rigetti_ac_sweet_spot,Cohen_drive_noise}, the fluctuation of the drive amplitude also causes qubit dephasing. It is therefore useful to enable protection from the ac noise as well. 

The need to protect qubits from both dc and ac noises motivates us to target operating points with $\partial\varepsilon_{01}/\partial\lambda = 0$, $\partial^2\varepsilon_{01}/\partial^2\lambda = 0$ and $\partial\varepsilon_{01}/\partial A = 0$, which we call the triple sweet spots. These operating points are more difficult to find. For the fluxonium qubit we work with, we do not find such triple protection with the simple sinusoidal drive. (We have confirmed this absence numerically over a wide range of the drive parameters.) However, a slightly more complex periodic drive solves this problem. For example, we find the vanishing of the three derivatives using $\hat{H}_d(t) = A[\cos\omega_d t + \alpha \cos(2n+1)\omega_d t]\hat{x}$ $(n\in \mathbb{N}^+)$. (We choose an odd integer $2n+1$ to ensure the vanishing of $\partial\epsilon_{01}/\partial\lambda$ \cite{Supplementary}.) This driving protocol is similar to those chosen in Refs.~\cite{Didier_dynamical_sweet_spot,Didier_dynamical_sweet_spot_exp}, although we are targeting working points with an extra second-order protection.  We show one example of such sweet spots in FIG.~\ref{fig:Floquet} (a), where we choose $\alpha = 1$ and $n=1$. Beside the position of this sweet spot, we also visually demonstrate the flattening of the spectrum.  In FIG.~\ref{fig:Floquet} (b), we plot the quasi-energies of the qubit as a function of the dc control parameter $\lambda$, with the ac drive amplitude $A$ fixed at the sweet spot, and the other way around in (c). Both plots show suppressed variation of the qubit frequency around the sweet spot. In (b), the variation of the frequency of the undriven qubit (subtracted by integer multiples of the drive frequency for comparison) is  plotted (red-dashed curve) to contrast the second-order suppression. Similar sweet spots for other choices for $\alpha$ and $n$ have also been found. According to Eq.~\eqref{eq:dephasing}, this protection protocol should mitigate dephasing caused by both the Gaussian noise and the non-Gaussian TLFs. 

To numerically confirm the improvement of coherence times, we use the SSE method to simulate the Floquet Ramsey evolution \cite{Dynamical_sweet_spot_exp}, and compare the dephasing time with that in the undriven case in FIG.~\ref{fig:Floquet} (d). Still using the noise model $\delta\xi(t)$ ($N_T =2$ for this simulation), the dephasing time at this triple protection point (solid cyan curve) is improved by 10 times compared to that at the static sweet spot  (solid red curve). This is directly related to the suppression of the second-order derivative [FIG.~\ref{fig:Floquet} (b)]. The limiting factor for the coherence time at this protection point is the depolarization (solid gray curves) by the noise at the smaller qubit frequency, which can potentially be further mitigated by optimizing the qubit parameters and drive shapes. To emulate the low-frequency drive noise,  we further include random fluctuation of the drive amplitude $A$ before each evolution trace in the SSE simulation. Remarkably, we still find 8 times improvement (cyan dashed curve) for up to $1\%$ fluctuation of the drive amplitude. This insensitivity is owing to the first-order insensitivity of the quasi-energy difference to the drive amplitude [FIG.~\ref{fig:Floquet} (c)].

\textit{Conclusion.--} We study the high-order dephasing effect in a qubit by non-Gaussian fluctuators. Our calculation predicts a symmetry breaking that is unique to the non-Gaussian noise. Concretely, we show that the strong TLFs, a popular non-Gaussian noise model, dramatically change the behavior of the dephasing rates and cause an unexpected mismatch between the minima of the dephasing rate and the qubit frequency. These findings challenge the usually assumed equivalence between the two minima, call for extra carefulness in locating the optimal working point, and caution the use of the Gaussian model for noise characterization. Finally, we propose a triple-protection scheme to suppress both the first-order and the second-order sensitivity of the qubit energy to these fluctuators, where the qubit is also first-order protected  from the low-frequency drive noise. Our simulation demonstrates an order of magnitude improvement of the dephasing time based on the parameters of an experimentally realized heavy fluxonium qubit.

\textit{Acknowledgements.--} This material is based upon work supported by the U.S. Department of Energy, Office of Science, National Quantum Information Science Research Centers, Superconducting Quantum Materials and Systems Center (SQMS) under contract number DE-AC02-07CH11359. We thank Jens Koch, David I. Schuster, Andr\'as Gyenis, Peter Groszkowski, Wei-Ting Lin, Pranav S. Mundada and Helin Zhang for constructive discussions.

\bibliography{mybib}

\begin{thebibliography}{72}%
\makeatletter
\providecommand \@ifxundefined [1]{%
 \@ifx{#1\undefined}
}%
\providecommand \@ifnum [1]{%
 \ifnum #1\expandafter \@firstoftwo
 \else \expandafter \@secondoftwo
 \fi
}%
\providecommand \@ifx [1]{%
 \ifx #1\expandafter \@firstoftwo
 \else \expandafter \@secondoftwo
 \fi
}%
\providecommand \natexlab [1]{#1}%
\providecommand \enquote  [1]{``#1''}%
\providecommand \bibnamefont  [1]{#1}%
\providecommand \bibfnamefont [1]{#1}%
\providecommand \citenamefont [1]{#1}%
\providecommand \href@noop [0]{\@secondoftwo}%
\providecommand \href [0]{\begingroup \@sanitize@url \@href}%
\providecommand \@href[1]{\@@startlink{#1}\@@href}%
\providecommand \@@href[1]{\endgroup#1\@@endlink}%
\providecommand \@sanitize@url [0]{\catcode `\\12\catcode `\$12\catcode
  `\&12\catcode `\#12\catcode `\^12\catcode `\_12\catcode `\%12\relax}%
\providecommand \@@startlink[1]{}%
\providecommand \@@endlink[0]{}%
\providecommand \url  [0]{\begingroup\@sanitize@url \@url }%
\providecommand \@url [1]{\endgroup\@href {#1}{\urlprefix }}%
\providecommand \urlprefix  [0]{URL }%
\providecommand \Eprint [0]{\href }%
\providecommand \doibase [0]{https://doi.org/}%
\providecommand \selectlanguage [0]{\@gobble}%
\providecommand \bibinfo  [0]{\@secondoftwo}%
\providecommand \bibfield  [0]{\@secondoftwo}%
\providecommand \translation [1]{[#1]}%
\providecommand \BibitemOpen [0]{}%
\providecommand \bibitemStop [0]{}%
\providecommand \bibitemNoStop [0]{.\EOS\space}%
\providecommand \EOS [0]{\spacefactor3000\relax}%
\providecommand \BibitemShut  [1]{\csname bibitem#1\endcsname}%
\let\auto@bib@innerbib\@empty
\bibitem [{\citenamefont {DiVincenzo}(2000)}]{DiVincenzoCriteria}%
  \BibitemOpen
  \bibfield  {author} {\bibinfo {author} {\bibfnamefont {D.~P.}\ \bibnamefont
  {DiVincenzo}},\ }\bibfield  {title} {\bibinfo {title} {\textit{The Physical
  Implementation of Quantum Computation}},\ }\href
  {https://doi.org/https://doi.org/10.1002/1521-3978(200009)48:9/11<771::AID-PROP771>3.0.CO;2-E}
  {\bibfield  {journal} {\bibinfo  {journal} {Fortschr. Phys.}\ }\textbf
  {\bibinfo {volume} {48}},\ \bibinfo {pages} {771} (\bibinfo {year}
  {2000})}\BibitemShut {NoStop}%
\bibitem [{\citenamefont {Preskill}(2018)}]{Preskill_NISQ_review}%
  \BibitemOpen
  \bibfield  {author} {\bibinfo {author} {\bibfnamefont {J.}~\bibnamefont
  {Preskill}},\ }\bibfield  {title} {\bibinfo {title} {\textit{Quantum
  Computing in the NISQ Era and Beyond}},\ }\href
  {https://doi.org/10.22331/q-2018-08-06-79} {\bibfield  {journal} {\bibinfo
  {journal} {Quantum}\ }\textbf {\bibinfo {volume} {2}},\ \bibinfo {pages} {79}
  (\bibinfo {year} {2018})}\BibitemShut {NoStop}%
\bibitem [{\citenamefont {Kjaergaard}\ \emph {et~al.}(2020)\citenamefont
  {Kjaergaard}, \citenamefont {Schwartz}, \citenamefont {Braumüller},
  \citenamefont {Krantz}, \citenamefont {Wang}, \citenamefont {Gustavsson},\
  and\ \citenamefont {Oliver}}]{Oliver_scqubits_review}%
  \BibitemOpen
  \bibfield  {author} {\bibinfo {author} {\bibfnamefont {M.}~\bibnamefont
  {Kjaergaard}}, \bibinfo {author} {\bibfnamefont {M.~E.}\ \bibnamefont
  {Schwartz}}, \bibinfo {author} {\bibfnamefont {J.}~\bibnamefont
  {Braumüller}}, \bibinfo {author} {\bibfnamefont {P.}~\bibnamefont {Krantz}},
  \bibinfo {author} {\bibfnamefont {J.~I.-J.}\ \bibnamefont {Wang}}, \bibinfo
  {author} {\bibfnamefont {S.}~\bibnamefont {Gustavsson}},\ and\ \bibinfo
  {author} {\bibfnamefont {W.~D.}\ \bibnamefont {Oliver}},\ }\bibfield  {title}
  {\bibinfo {title} {\textit{Superconducting Qubits: Current State of Play}},\
  }\href {https://doi.org/10.1146/annurev-conmatphys-031119-050605} {\bibfield
  {journal} {\bibinfo  {journal} {Annu. Rev. Condens. Matter Phys.}\ }\textbf
  {\bibinfo {volume} {11}},\ \bibinfo {pages} {369} (\bibinfo {year}
  {2020})}\BibitemShut {NoStop}%
\bibitem [{\citenamefont {Bruzewicz}\ \emph {et~al.}(2019)\citenamefont
  {Bruzewicz}, \citenamefont {Chiaverini}, \citenamefont {McConnell},\ and\
  \citenamefont {Sage}}]{Bruzewicz_trapped_ion_review}%
  \BibitemOpen
  \bibfield  {author} {\bibinfo {author} {\bibfnamefont {C.~D.}\ \bibnamefont
  {Bruzewicz}}, \bibinfo {author} {\bibfnamefont {J.}~\bibnamefont
  {Chiaverini}}, \bibinfo {author} {\bibfnamefont {R.}~\bibnamefont
  {McConnell}},\ and\ \bibinfo {author} {\bibfnamefont {J.~M.}\ \bibnamefont
  {Sage}},\ }\bibfield  {title} {\bibinfo {title} {\textit{Trapped-Ion Quantum
  Computing: Progress and Challenges}},\ }\href
  {https://doi.org/10.1063/1.5088164} {\bibfield  {journal} {\bibinfo
  {journal} {Appl. Phys. Rev.}\ }\textbf {\bibinfo {volume} {6}},\ \bibinfo
  {pages} {021314} (\bibinfo {year} {2019})}\BibitemShut {NoStop}%
\bibitem [{\citenamefont {Müller}\ \emph {et~al.}(2019)\citenamefont
  {Müller}, \citenamefont {Cole},\ and\ \citenamefont
  {Lisenfeld}}]{Muller_tls_review}%
  \BibitemOpen
  \bibfield  {author} {\bibinfo {author} {\bibfnamefont {C.}~\bibnamefont
  {Müller}}, \bibinfo {author} {\bibfnamefont {J.~H.}\ \bibnamefont {Cole}},\
  and\ \bibinfo {author} {\bibfnamefont {J.}~\bibnamefont {Lisenfeld}},\
  }\bibfield  {title} {\bibinfo {title} {\textit{Towards Understanding
  Two-Level-Systems in Amorphous Solids: Insights from Quantum Circuits}},\
  }\href {https://doi.org/10.1088/1361-6633/ab3a7e} {\bibfield  {journal}
  {\bibinfo  {journal} {Rep. Prog. Phys.}\ }\textbf {\bibinfo {volume} {82}},\
  \bibinfo {pages} {124501} (\bibinfo {year} {2019})}\BibitemShut {NoStop}%
\bibitem [{\citenamefont {Murray}(2021)}]{Murray_material_review}%
  \BibitemOpen
  \bibfield  {author} {\bibinfo {author} {\bibfnamefont {C.~E.}\ \bibnamefont
  {Murray}},\ }\bibfield  {title} {\bibinfo {title} {\textit{Material Matters
  in Superconducting Qubits}},\ }\href
  {https://doi.org/https://doi.org/10.1016/j.mser.2021.100646} {\bibfield
  {journal} {\bibinfo  {journal} {Mater. Sci. Eng. R Rep.}\ }\textbf {\bibinfo
  {volume} {146}},\ \bibinfo {pages} {100646} (\bibinfo {year}
  {2021})}\BibitemShut {NoStop}%
\bibitem [{\citenamefont {Gyenis}\ \emph
  {et~al.}(2021{\natexlab{a}})\citenamefont {Gyenis}, \citenamefont {Di~Paolo},
  \citenamefont {Koch}, \citenamefont {Blais}, \citenamefont {Houck},\ and\
  \citenamefont {Schuster}}]{Gyenis_protected_review}%
  \BibitemOpen
  \bibfield  {author} {\bibinfo {author} {\bibfnamefont {A.}~\bibnamefont
  {Gyenis}}, \bibinfo {author} {\bibfnamefont {A.}~\bibnamefont {Di~Paolo}},
  \bibinfo {author} {\bibfnamefont {J.}~\bibnamefont {Koch}}, \bibinfo {author}
  {\bibfnamefont {A.}~\bibnamefont {Blais}}, \bibinfo {author} {\bibfnamefont
  {A.~A.}\ \bibnamefont {Houck}},\ and\ \bibinfo {author} {\bibfnamefont
  {D.~I.}\ \bibnamefont {Schuster}},\ }\bibfield  {title} {\bibinfo {title}
  {\textit{Moving beyond the Transmon: Noise-Protected Superconducting Quantum
  Circuits }},\ }\href {https://doi.org/10.1103/PRXQuantum.2.030101} {\bibfield
   {journal} {\bibinfo  {journal} {PRX Quantum}\ }\textbf {\bibinfo {volume}
  {2}},\ \bibinfo {pages} {030101} (\bibinfo {year}
  {2021}{\natexlab{a}})}\BibitemShut {NoStop}%
\bibitem [{\citenamefont {Dutta}\ and\ \citenamefont
  {Horn}(1981)}]{Dutta_1/f_review}%
  \BibitemOpen
  \bibfield  {author} {\bibinfo {author} {\bibfnamefont {P.}~\bibnamefont
  {Dutta}}\ and\ \bibinfo {author} {\bibfnamefont {P.~M.}\ \bibnamefont
  {Horn}},\ }\bibfield  {title} {\bibinfo {title} {\textit{Low-Frequency
  Fluctuations in Solids: $\frac{1}{f}$ Noise}},\ }\href
  {https://doi.org/10.1103/RevModPhys.53.497} {\bibfield  {journal} {\bibinfo
  {journal} {Rev. Mod. Phys.}\ }\textbf {\bibinfo {volume} {53}},\ \bibinfo
  {pages} {497} (\bibinfo {year} {1981})}\BibitemShut {NoStop}%
\bibitem [{\citenamefont {Weissman}(1988)}]{Weissman_1/f_review}%
  \BibitemOpen
  \bibfield  {author} {\bibinfo {author} {\bibfnamefont {M.~B.}\ \bibnamefont
  {Weissman}},\ }\bibfield  {title} {\bibinfo {title} {\textit{$\frac{1}{f}$
  Noise and Other Slow, Nonexponential Kinetics in Condensed Matter}},\ }\href
  {https://doi.org/10.1103/RevModPhys.60.537} {\bibfield  {journal} {\bibinfo
  {journal} {Rev. Mod. Phys.}\ }\textbf {\bibinfo {volume} {60}},\ \bibinfo
  {pages} {537} (\bibinfo {year} {1988})}\BibitemShut {NoStop}%
\bibitem [{\citenamefont {Kumar}\ \emph {et~al.}(2016)\citenamefont {Kumar},
  \citenamefont {Sendelbach}, \citenamefont {Beck}, \citenamefont {Freeland},
  \citenamefont {Wang}, \citenamefont {Wang}, \citenamefont {Yu}, \citenamefont
  {Wu}, \citenamefont {Pappas},\ and\ \citenamefont
  {McDermott}}]{McDermott_spin_1/f_noise}%
  \BibitemOpen
  \bibfield  {author} {\bibinfo {author} {\bibfnamefont {P.}~\bibnamefont
  {Kumar}}, \bibinfo {author} {\bibfnamefont {S.}~\bibnamefont {Sendelbach}},
  \bibinfo {author} {\bibfnamefont {M.~A.}\ \bibnamefont {Beck}}, \bibinfo
  {author} {\bibfnamefont {J.~W.}\ \bibnamefont {Freeland}}, \bibinfo {author}
  {\bibfnamefont {Z.}~\bibnamefont {Wang}}, \bibinfo {author} {\bibfnamefont
  {H.}~\bibnamefont {Wang}}, \bibinfo {author} {\bibfnamefont {C.~C.}\
  \bibnamefont {Yu}}, \bibinfo {author} {\bibfnamefont {R.~Q.}\ \bibnamefont
  {Wu}}, \bibinfo {author} {\bibfnamefont {D.~P.}\ \bibnamefont {Pappas}},\
  and\ \bibinfo {author} {\bibfnamefont {R.}~\bibnamefont {McDermott}},\
  }\bibfield  {title} {\bibinfo {title} {\textit{Origin and Reduction of $1/f$
  Magnetic Flux Noise in Superconducting Devices}},\ }\href
  {https://doi.org/10.1103/PhysRevApplied.6.041001} {\bibfield  {journal}
  {\bibinfo  {journal} {Phys. Rev. Applied}\ }\textbf {\bibinfo {volume} {6}},\
  \bibinfo {pages} {041001} (\bibinfo {year} {2016})}\BibitemShut {NoStop}%
\bibitem [{\citenamefont {Serniak}\ \emph {et~al.}(2018)\citenamefont
  {Serniak}, \citenamefont {Hays}, \citenamefont {de~Lange}, \citenamefont
  {Diamond}, \citenamefont {Shankar}, \citenamefont {Burkhart}, \citenamefont
  {Frunzio}, \citenamefont {Houzet},\ and\ \citenamefont
  {Devoret}}]{Devoret_quasi_particle_decoherence}%
  \BibitemOpen
  \bibfield  {author} {\bibinfo {author} {\bibfnamefont {K.}~\bibnamefont
  {Serniak}}, \bibinfo {author} {\bibfnamefont {M.}~\bibnamefont {Hays}},
  \bibinfo {author} {\bibfnamefont {G.}~\bibnamefont {de~Lange}}, \bibinfo
  {author} {\bibfnamefont {S.}~\bibnamefont {Diamond}}, \bibinfo {author}
  {\bibfnamefont {S.}~\bibnamefont {Shankar}}, \bibinfo {author} {\bibfnamefont
  {L.~D.}\ \bibnamefont {Burkhart}}, \bibinfo {author} {\bibfnamefont
  {L.}~\bibnamefont {Frunzio}}, \bibinfo {author} {\bibfnamefont
  {M.}~\bibnamefont {Houzet}},\ and\ \bibinfo {author} {\bibfnamefont {M.~H.}\
  \bibnamefont {Devoret}},\ }\bibfield  {title} {\bibinfo {title} {\textit{Hot
  Nonequilibrium Quasiparticles in Transmon Qubits}},\ }\href
  {https://doi.org/10.1103/PhysRevLett.121.157701} {\bibfield  {journal}
  {\bibinfo  {journal} {Phys. Rev. Lett.}\ }\textbf {\bibinfo {volume} {121}},\
  \bibinfo {pages} {157701} (\bibinfo {year} {2018})}\BibitemShut {NoStop}%
\bibitem [{\citenamefont {Constantin}\ and\ \citenamefont
  {Yu}(2007)}]{Yu_critical_current}%
  \BibitemOpen
  \bibfield  {author} {\bibinfo {author} {\bibfnamefont {M.}~\bibnamefont
  {Constantin}}\ and\ \bibinfo {author} {\bibfnamefont {C.~C.}\ \bibnamefont
  {Yu}},\ }\bibfield  {title} {\bibinfo {title} {\textit{Microscopic Model of
  Critical Current Noise in Josephson Junctions}},\ }\href
  {https://doi.org/10.1103/PhysRevLett.99.207001} {\bibfield  {journal}
  {\bibinfo  {journal} {Phys. Rev. Lett.}\ }\textbf {\bibinfo {volume} {99}},\
  \bibinfo {pages} {207001} (\bibinfo {year} {2007})}\BibitemShut {NoStop}%
\bibitem [{\citenamefont {Schl\"or}\ \emph {et~al.}(2019)\citenamefont
  {Schl\"or}, \citenamefont {Lisenfeld}, \citenamefont {M\"uller},
  \citenamefont {Bilmes}, \citenamefont {Schneider}, \citenamefont {Pappas},
  \citenamefont {Ustinov},\ and\ \citenamefont
  {Weides}}]{Weides_transmon_single_fluctuator}%
  \BibitemOpen
  \bibfield  {author} {\bibinfo {author} {\bibfnamefont {S.}~\bibnamefont
  {Schl\"or}}, \bibinfo {author} {\bibfnamefont {J.}~\bibnamefont {Lisenfeld}},
  \bibinfo {author} {\bibfnamefont {C.}~\bibnamefont {M\"uller}}, \bibinfo
  {author} {\bibfnamefont {A.}~\bibnamefont {Bilmes}}, \bibinfo {author}
  {\bibfnamefont {A.}~\bibnamefont {Schneider}}, \bibinfo {author}
  {\bibfnamefont {D.~P.}\ \bibnamefont {Pappas}}, \bibinfo {author}
  {\bibfnamefont {A.~V.}\ \bibnamefont {Ustinov}},\ and\ \bibinfo {author}
  {\bibfnamefont {M.}~\bibnamefont {Weides}},\ }\bibfield  {title} {\bibinfo
  {title} {\textit{Correlating Decoherence in Transmon Qubits: Low Frequency
  Noise by Single Fluctuators}},\ }\href
  {https://doi.org/10.1103/PhysRevLett.123.190502} {\bibfield  {journal}
  {\bibinfo  {journal} {Phys. Rev. Lett.}\ }\textbf {\bibinfo {volume} {123}},\
  \bibinfo {pages} {190502} (\bibinfo {year} {2019})}\BibitemShut {NoStop}%
\bibitem [{\citenamefont {Rieger}\ \emph {et~al.}()\citenamefont {Rieger},
  \citenamefont {Günzler}, \citenamefont {Spiecker}, \citenamefont {Paluch},
  \citenamefont {Winkel}, \citenamefont {Hahn}, \citenamefont {Hohmann},
  \citenamefont {Bacher}, \citenamefont {Wernsdorfer},\ and\ \citenamefont
  {Pop}}]{Pop_fluxonium_beating}%
  \BibitemOpen
  \bibfield  {author} {\bibinfo {author} {\bibfnamefont {D.}~\bibnamefont
  {Rieger}}, \bibinfo {author} {\bibfnamefont {S.}~\bibnamefont {Günzler}},
  \bibinfo {author} {\bibfnamefont {M.}~\bibnamefont {Spiecker}}, \bibinfo
  {author} {\bibfnamefont {P.}~\bibnamefont {Paluch}}, \bibinfo {author}
  {\bibfnamefont {P.}~\bibnamefont {Winkel}}, \bibinfo {author} {\bibfnamefont
  {L.}~\bibnamefont {Hahn}}, \bibinfo {author} {\bibfnamefont {J.~K.}\
  \bibnamefont {Hohmann}}, \bibinfo {author} {\bibfnamefont {A.}~\bibnamefont
  {Bacher}}, \bibinfo {author} {\bibfnamefont {W.}~\bibnamefont
  {Wernsdorfer}},\ and\ \bibinfo {author} {\bibfnamefont {I.~M.}\ \bibnamefont
  {Pop}},\ }\bibfield  {title} {\bibinfo {title} {\textit{Gralmonium: Granular
  Aluminum Nano-Junction Fluxonium Qubit} (2022)},\ }\Eprint
  {https://arxiv.org/abs/arXiv:2202.01776} {arXiv:2202.01776} \BibitemShut
  {NoStop}%
\bibitem [{\citenamefont {Zaretskey}\ \emph {et~al.}(2013)\citenamefont
  {Zaretskey}, \citenamefont {Suri}, \citenamefont {Novikov}, \citenamefont
  {Wellstood},\ and\ \citenamefont {Palmer}}]{Palmer_CPB_TLS}%
  \BibitemOpen
  \bibfield  {author} {\bibinfo {author} {\bibfnamefont {V.}~\bibnamefont
  {Zaretskey}}, \bibinfo {author} {\bibfnamefont {B.}~\bibnamefont {Suri}},
  \bibinfo {author} {\bibfnamefont {S.}~\bibnamefont {Novikov}}, \bibinfo
  {author} {\bibfnamefont {F.~C.}\ \bibnamefont {Wellstood}},\ and\ \bibinfo
  {author} {\bibfnamefont {B.~S.}\ \bibnamefont {Palmer}},\ }\bibfield  {title}
  {\bibinfo {title} {\textit{Spectroscopy of a Cooper-Pair Box Coupled to a
  Two-Level System via Charge and Critical Current}},\ }\href
  {https://doi.org/10.1103/PhysRevB.87.174522} {\bibfield  {journal} {\bibinfo
  {journal} {Phys. Rev. B}\ }\textbf {\bibinfo {volume} {87}},\ \bibinfo
  {pages} {174522} (\bibinfo {year} {2013})}\BibitemShut {NoStop}%
\bibitem [{\citenamefont {Lisenfeld}\ \emph {et~al.}(2015)\citenamefont
  {Lisenfeld}, \citenamefont {Grabovskij}, \citenamefont {M{\"u}ller},
  \citenamefont {Cole}, \citenamefont {Weiss},\ and\ \citenamefont
  {Ustinov}}]{Ustinov_TLS_TLF}%
  \BibitemOpen
  \bibfield  {author} {\bibinfo {author} {\bibfnamefont {J.}~\bibnamefont
  {Lisenfeld}}, \bibinfo {author} {\bibfnamefont {G.~J.}\ \bibnamefont
  {Grabovskij}}, \bibinfo {author} {\bibfnamefont {C.}~\bibnamefont
  {M{\"u}ller}}, \bibinfo {author} {\bibfnamefont {J.~H.}\ \bibnamefont
  {Cole}}, \bibinfo {author} {\bibfnamefont {G.}~\bibnamefont {Weiss}},\ and\
  \bibinfo {author} {\bibfnamefont {A.~V.}\ \bibnamefont {Ustinov}},\
  }\bibfield  {title} {\bibinfo {title} {\textit{Observation of Directly
  Interacting Coherent Two-Level Systems in an Amorphous Material}},\ }\href
  {https://doi.org/10.1038/ncomms7182} {\bibfield  {journal} {\bibinfo
  {journal} {Nat. Commun.}\ }\textbf {\bibinfo {volume} {6}},\ \bibinfo {pages}
  {6182} (\bibinfo {year} {2015})}\BibitemShut {NoStop}%
\bibitem [{\citenamefont {B\'ejanin}\ \emph {et~al.}(2021)\citenamefont
  {B\'ejanin}, \citenamefont {Earnest}, \citenamefont {Sharafeldin},\ and\
  \citenamefont {Mariantoni}}]{Mariantoni_TLS_TLF}%
  \BibitemOpen
  \bibfield  {author} {\bibinfo {author} {\bibfnamefont {J.~H.}\ \bibnamefont
  {B\'ejanin}}, \bibinfo {author} {\bibfnamefont {C.~T.}\ \bibnamefont
  {Earnest}}, \bibinfo {author} {\bibfnamefont {A.~S.}\ \bibnamefont
  {Sharafeldin}},\ and\ \bibinfo {author} {\bibfnamefont {M.}~\bibnamefont
  {Mariantoni}},\ }\bibfield  {title} {\bibinfo {title} {\textit{Interacting
  Defects Generate Stochastic Fluctuations in Superconducting Qubits}},\ }\href
  {https://doi.org/10.1103/PhysRevB.104.094106} {\bibfield  {journal} {\bibinfo
   {journal} {Phys. Rev. B}\ }\textbf {\bibinfo {volume} {104}},\ \bibinfo
  {pages} {094106} (\bibinfo {year} {2021})}\BibitemShut {NoStop}%
\bibitem [{\citenamefont {McCourt}\ \emph {et~al.}()\citenamefont {McCourt},
  \citenamefont {Neill}, \citenamefont {Lee}, \citenamefont {Quintana},
  \citenamefont {Chen}, \citenamefont {Kellyand}, \citenamefont {Smelyanskiy},
  \citenamefont {Dykman}, \citenamefont {Korotkov}, \citenamefont {Chuang},\
  and\ \citenamefont {Petukhov}}]{Petukhov_nonGaussianDD}%
  \BibitemOpen
  \bibfield  {author} {\bibinfo {author} {\bibfnamefont {T.}~\bibnamefont
  {McCourt}}, \bibinfo {author} {\bibfnamefont {C.}~\bibnamefont {Neill}},
  \bibinfo {author} {\bibfnamefont {K.}~\bibnamefont {Lee}}, \bibinfo {author}
  {\bibfnamefont {C.}~\bibnamefont {Quintana}}, \bibinfo {author}
  {\bibfnamefont {Y.}~\bibnamefont {Chen}}, \bibinfo {author} {\bibfnamefont
  {J.}~\bibnamefont {Kellyand}}, \bibinfo {author} {\bibfnamefont {V.~N.}\
  \bibnamefont {Smelyanskiy}}, \bibinfo {author} {\bibfnamefont {M.~I.}\
  \bibnamefont {Dykman}}, \bibinfo {author} {\bibfnamefont {A.}~\bibnamefont
  {Korotkov}}, \bibinfo {author} {\bibfnamefont {I.~L.}\ \bibnamefont
  {Chuang}},\ and\ \bibinfo {author} {\bibfnamefont {A.~G.}\ \bibnamefont
  {Petukhov}},\ }\bibfield  {title} {\bibinfo {title} {\textit{Learning Noise
  via Dynamical Decoupling of Entangled Qubits } (2022)},\ }\Eprint
  {https://arxiv.org/abs/arXiv:2201.11173} {arXiv:2201.11173} \BibitemShut
  {NoStop}%
\bibitem [{\citenamefont {Bialczak}\ \emph {et~al.}(2007)\citenamefont
  {Bialczak}, \citenamefont {McDermott}, \citenamefont {Ansmann}, \citenamefont
  {Hofheinz}, \citenamefont {Katz}, \citenamefont {Lucero}, \citenamefont
  {Neeley}, \citenamefont {O'Connell}, \citenamefont {Wang}, \citenamefont
  {Cleland},\ and\ \citenamefont {Martinis}}]{Martinis_flux_noise}%
  \BibitemOpen
  \bibfield  {author} {\bibinfo {author} {\bibfnamefont {R.~C.}\ \bibnamefont
  {Bialczak}}, \bibinfo {author} {\bibfnamefont {R.}~\bibnamefont {McDermott}},
  \bibinfo {author} {\bibfnamefont {M.}~\bibnamefont {Ansmann}}, \bibinfo
  {author} {\bibfnamefont {M.}~\bibnamefont {Hofheinz}}, \bibinfo {author}
  {\bibfnamefont {N.}~\bibnamefont {Katz}}, \bibinfo {author} {\bibfnamefont
  {E.}~\bibnamefont {Lucero}}, \bibinfo {author} {\bibfnamefont
  {M.}~\bibnamefont {Neeley}}, \bibinfo {author} {\bibfnamefont {A.~D.}\
  \bibnamefont {O'Connell}}, \bibinfo {author} {\bibfnamefont {H.}~\bibnamefont
  {Wang}}, \bibinfo {author} {\bibfnamefont {A.~N.}\ \bibnamefont {Cleland}},\
  and\ \bibinfo {author} {\bibfnamefont {J.~M.}\ \bibnamefont {Martinis}},\
  }\bibfield  {title} {\bibinfo {title} {\textit{$1/f$ Flux Noise in Josephson
  Phase Qubits}},\ }\href {https://doi.org/10.1103/PhysRevLett.99.187006}
  {\bibfield  {journal} {\bibinfo  {journal} {Phys. Rev. Lett.}\ }\textbf
  {\bibinfo {volume} {99}},\ \bibinfo {pages} {187006} (\bibinfo {year}
  {2007})}\BibitemShut {NoStop}%
\bibitem [{\citenamefont {Vion}\ \emph {et~al.}(2002)\citenamefont {Vion},
  \citenamefont {Aassime}, \citenamefont {Cottet}, \citenamefont {Joyez},
  \citenamefont {Pothier}, \citenamefont {Urbina}, \citenamefont {Esteve},\
  and\ \citenamefont {Devoret}}]{Devoret_quantronium}%
  \BibitemOpen
  \bibfield  {author} {\bibinfo {author} {\bibfnamefont {D.}~\bibnamefont
  {Vion}}, \bibinfo {author} {\bibfnamefont {A.}~\bibnamefont {Aassime}},
  \bibinfo {author} {\bibfnamefont {A.}~\bibnamefont {Cottet}}, \bibinfo
  {author} {\bibfnamefont {P.}~\bibnamefont {Joyez}}, \bibinfo {author}
  {\bibfnamefont {H.}~\bibnamefont {Pothier}}, \bibinfo {author} {\bibfnamefont
  {C.}~\bibnamefont {Urbina}}, \bibinfo {author} {\bibfnamefont
  {D.}~\bibnamefont {Esteve}},\ and\ \bibinfo {author} {\bibfnamefont {M.~H.}\
  \bibnamefont {Devoret}},\ }\bibfield  {title} {\bibinfo {title}
  {\textit{Manipulating the Quantum State of an Electrical Circuit}},\ }\href
  {https://doi.org/10.1126/science.1069372} {\bibfield  {journal} {\bibinfo
  {journal} {Science}\ }\textbf {\bibinfo {volume} {296}},\ \bibinfo {pages}
  {886} (\bibinfo {year} {2002})}\BibitemShut {NoStop}%
\bibitem [{\citenamefont {Nakamura}\ \emph {et~al.}(1999)\citenamefont
  {Nakamura}, \citenamefont {Pashkin},\ and\ \citenamefont
  {Tsai}}]{Nakamura_charge_qubit}%
  \BibitemOpen
  \bibfield  {author} {\bibinfo {author} {\bibfnamefont {Y.}~\bibnamefont
  {Nakamura}}, \bibinfo {author} {\bibfnamefont {Y.~A.}\ \bibnamefont
  {Pashkin}},\ and\ \bibinfo {author} {\bibfnamefont {J.~S.}\ \bibnamefont
  {Tsai}},\ }\bibfield  {title} {\bibinfo {title} {\textit{Coherent Control of
  Macroscopic Quantum States in a Single-Cooper-Pair Box}},\ }\href
  {https://doi.org/https://doi.org/10.1038/19718} {\bibfield  {journal}
  {\bibinfo  {journal} {Nature}\ }\textbf {\bibinfo {volume} {398}},\ \bibinfo
  {pages} {786} (\bibinfo {year} {1999})}\BibitemShut {NoStop}%
\bibitem [{\citenamefont {Nakamura}\ \emph {et~al.}(2002)\citenamefont
  {Nakamura}, \citenamefont {Pashkin}, \citenamefont {Yamamoto},\ and\
  \citenamefont {Tsai}}]{Nakamura_charge_1/f}%
  \BibitemOpen
  \bibfield  {author} {\bibinfo {author} {\bibfnamefont {Y.}~\bibnamefont
  {Nakamura}}, \bibinfo {author} {\bibfnamefont {Y.~A.}\ \bibnamefont
  {Pashkin}}, \bibinfo {author} {\bibfnamefont {T.}~\bibnamefont {Yamamoto}},\
  and\ \bibinfo {author} {\bibfnamefont {J.~S.}\ \bibnamefont {Tsai}},\
  }\bibfield  {title} {\bibinfo {title} {\textit{Charge Echo in a Cooper-Pair
  Box}},\ }\href {https://doi.org/10.1103/PhysRevLett.88.047901} {\bibfield
  {journal} {\bibinfo  {journal} {Phys. Rev. Lett.}\ }\textbf {\bibinfo
  {volume} {88}},\ \bibinfo {pages} {047901} (\bibinfo {year}
  {2002})}\BibitemShut {NoStop}%
\bibitem [{\citenamefont {Nguyen}\ \emph {et~al.}(2019)\citenamefont {Nguyen},
  \citenamefont {Lin}, \citenamefont {Somoroff}, \citenamefont {Mencia},
  \citenamefont {Grabon},\ and\ \citenamefont {Manucharyan}}]{Fluxonium_hc}%
  \BibitemOpen
  \bibfield  {author} {\bibinfo {author} {\bibfnamefont {L.~B.}\ \bibnamefont
  {Nguyen}}, \bibinfo {author} {\bibfnamefont {Y.-H.}\ \bibnamefont {Lin}},
  \bibinfo {author} {\bibfnamefont {A.}~\bibnamefont {Somoroff}}, \bibinfo
  {author} {\bibfnamefont {R.}~\bibnamefont {Mencia}}, \bibinfo {author}
  {\bibfnamefont {N.}~\bibnamefont {Grabon}},\ and\ \bibinfo {author}
  {\bibfnamefont {V.~E.}\ \bibnamefont {Manucharyan}},\ }\bibfield  {title}
  {\bibinfo {title} {\textit{High-Coherence Fluxonium Qubit}},\ }\href
  {https://doi.org/10.1103/PhysRevX.9.041041} {\bibfield  {journal} {\bibinfo
  {journal} {Phys. Rev. X}\ }\textbf {\bibinfo {volume} {9}},\ \bibinfo {pages}
  {041041} (\bibinfo {year} {2019})}\BibitemShut {NoStop}%
\bibitem [{\citenamefont {Lin}\ \emph {et~al.}(2018)\citenamefont {Lin},
  \citenamefont {Nguyen}, \citenamefont {Grabon}, \citenamefont {San~Miguel},
  \citenamefont {Pankratova},\ and\ \citenamefont
  {Manucharyan}}]{Manucharyan_heavy_fluxonium}%
  \BibitemOpen
  \bibfield  {author} {\bibinfo {author} {\bibfnamefont {Y.-H.}\ \bibnamefont
  {Lin}}, \bibinfo {author} {\bibfnamefont {L.~B.}\ \bibnamefont {Nguyen}},
  \bibinfo {author} {\bibfnamefont {N.}~\bibnamefont {Grabon}}, \bibinfo
  {author} {\bibfnamefont {J.}~\bibnamefont {San~Miguel}}, \bibinfo {author}
  {\bibfnamefont {N.}~\bibnamefont {Pankratova}},\ and\ \bibinfo {author}
  {\bibfnamefont {V.~E.}\ \bibnamefont {Manucharyan}},\ }\bibfield  {title}
  {\bibinfo {title} {\textit{Demonstration of Protection of a Superconducting
  Qubit from Energy Decay}},\ }\href
  {https://doi.org/10.1103/PhysRevLett.120.150503} {\bibfield  {journal}
  {\bibinfo  {journal} {Phys. Rev. Lett.}\ }\textbf {\bibinfo {volume} {120}},\
  \bibinfo {pages} {150503} (\bibinfo {year} {2018})}\BibitemShut {NoStop}%
\bibitem [{\citenamefont {Earnest}\ \emph {et~al.}(2018)\citenamefont
  {Earnest}, \citenamefont {Chakram}, \citenamefont {Lu}, \citenamefont
  {Irons}, \citenamefont {Naik}, \citenamefont {Leung}, \citenamefont {Ocola},
  \citenamefont {Czaplewski}, \citenamefont {Baker}, \citenamefont {Lawrence},
  \citenamefont {Koch},\ and\ \citenamefont
  {Schuster}}]{Schuster_heavy_fluxonium}%
  \BibitemOpen
  \bibfield  {author} {\bibinfo {author} {\bibfnamefont {N.}~\bibnamefont
  {Earnest}}, \bibinfo {author} {\bibfnamefont {S.}~\bibnamefont {Chakram}},
  \bibinfo {author} {\bibfnamefont {Y.}~\bibnamefont {Lu}}, \bibinfo {author}
  {\bibfnamefont {N.}~\bibnamefont {Irons}}, \bibinfo {author} {\bibfnamefont
  {R.~K.}\ \bibnamefont {Naik}}, \bibinfo {author} {\bibfnamefont
  {N.}~\bibnamefont {Leung}}, \bibinfo {author} {\bibfnamefont
  {L.}~\bibnamefont {Ocola}}, \bibinfo {author} {\bibfnamefont {D.~A.}\
  \bibnamefont {Czaplewski}}, \bibinfo {author} {\bibfnamefont
  {B.}~\bibnamefont {Baker}}, \bibinfo {author} {\bibfnamefont
  {J.}~\bibnamefont {Lawrence}}, \bibinfo {author} {\bibfnamefont
  {J.}~\bibnamefont {Koch}},\ and\ \bibinfo {author} {\bibfnamefont {D.~I.}\
  \bibnamefont {Schuster}},\ }\bibfield  {title} {\bibinfo {title}
  {\textit{Realization of a $\mathrm{\ensuremath{\Lambda}}$ System with
  Metastable States of a Capacitively Shunted Fluxonium}},\ }\href
  {https://doi.org/10.1103/PhysRevLett.120.150504} {\bibfield  {journal}
  {\bibinfo  {journal} {Phys. Rev. Lett.}\ }\textbf {\bibinfo {volume} {120}},\
  \bibinfo {pages} {150504} (\bibinfo {year} {2018})}\BibitemShut {NoStop}%
\bibitem [{\citenamefont {Zhang}\ \emph {et~al.}(2021)\citenamefont {Zhang},
  \citenamefont {Chakram}, \citenamefont {Roy}, \citenamefont {Earnest},
  \citenamefont {Lu}, \citenamefont {Huang}, \citenamefont {Weiss},
  \citenamefont {Koch},\ and\ \citenamefont
  {Schuster}}]{Schuster_heavy_fluxonium_control}%
  \BibitemOpen
  \bibfield  {author} {\bibinfo {author} {\bibfnamefont {H.}~\bibnamefont
  {Zhang}}, \bibinfo {author} {\bibfnamefont {S.}~\bibnamefont {Chakram}},
  \bibinfo {author} {\bibfnamefont {T.}~\bibnamefont {Roy}}, \bibinfo {author}
  {\bibfnamefont {N.}~\bibnamefont {Earnest}}, \bibinfo {author} {\bibfnamefont
  {Y.}~\bibnamefont {Lu}}, \bibinfo {author} {\bibfnamefont {Z.}~\bibnamefont
  {Huang}}, \bibinfo {author} {\bibfnamefont {D.~K.}\ \bibnamefont {Weiss}},
  \bibinfo {author} {\bibfnamefont {J.}~\bibnamefont {Koch}},\ and\ \bibinfo
  {author} {\bibfnamefont {D.~I.}\ \bibnamefont {Schuster}},\ }\bibfield
  {title} {\bibinfo {title} {\textit{Universal Fast-Flux Control of a Coherent,
  Low-Frequency Qubit}},\ }\href {https://doi.org/10.1103/PhysRevX.11.011010}
  {\bibfield  {journal} {\bibinfo  {journal} {Phys. Rev. X}\ }\textbf {\bibinfo
  {volume} {11}},\ \bibinfo {pages} {011010} (\bibinfo {year}
  {2021})}\BibitemShut {NoStop}%
\bibitem [{\citenamefont {Paladino}\ \emph {et~al.}(2014)\citenamefont
  {Paladino}, \citenamefont {Galperin}, \citenamefont {Falci},\ and\
  \citenamefont {Altshuler}}]{Paladino_TLF_review}%
  \BibitemOpen
  \bibfield  {author} {\bibinfo {author} {\bibfnamefont {E.}~\bibnamefont
  {Paladino}}, \bibinfo {author} {\bibfnamefont {Y.~M.}\ \bibnamefont
  {Galperin}}, \bibinfo {author} {\bibfnamefont {G.}~\bibnamefont {Falci}},\
  and\ \bibinfo {author} {\bibfnamefont {B.~L.}\ \bibnamefont {Altshuler}},\
  }\bibfield  {title} {\bibinfo {title} {$1/f$ \textit{Noise: Implications for
  Solid-State Quantum Information}},\ }\href
  {https://doi.org/10.1103/RevModPhys.86.361} {\bibfield  {journal} {\bibinfo
  {journal} {Rev. Mod. Phys.}\ }\textbf {\bibinfo {volume} {86}},\ \bibinfo
  {pages} {361} (\bibinfo {year} {2014})}\BibitemShut {NoStop}%
\bibitem [{\citenamefont {Chiorescu}\ \emph {et~al.}(2003)\citenamefont
  {Chiorescu}, \citenamefont {Nakamura}, \citenamefont {Harmans},\ and\
  \citenamefont {Mooij}}]{Mooij_flux_qubit_2003}%
  \BibitemOpen
  \bibfield  {author} {\bibinfo {author} {\bibfnamefont {I.}~\bibnamefont
  {Chiorescu}}, \bibinfo {author} {\bibfnamefont {Y.}~\bibnamefont {Nakamura}},
  \bibinfo {author} {\bibfnamefont {C.~J. P.~M.}\ \bibnamefont {Harmans}},\
  and\ \bibinfo {author} {\bibfnamefont {J.~E.}\ \bibnamefont {Mooij}},\
  }\bibfield  {title} {\bibinfo {title} {\textit{Coherent Quantum Dynamics of a
  Superconducting Flux Qubit}},\ }\href
  {https://doi.org/10.1126/science.1081045} {\bibfield  {journal} {\bibinfo
  {journal} {Science}\ }\textbf {\bibinfo {volume} {299}},\ \bibinfo {pages}
  {1869} (\bibinfo {year} {2003})}\BibitemShut {NoStop}%
\bibitem [{\citenamefont {Quintana}\ \emph {et~al.}(2017)\citenamefont
  {Quintana}, \citenamefont {Chen}, \citenamefont {Sank}, \citenamefont
  {Petukhov}, \citenamefont {White}, \citenamefont {Kafri}, \citenamefont
  {Chiaro}, \citenamefont {Megrant}, \citenamefont {Barends}, \citenamefont
  {Campbell}, \citenamefont {Chen}, \citenamefont {Dunsworth}, \citenamefont
  {Fowler}, \citenamefont {Graff}, \citenamefont {Jeffrey}, \citenamefont
  {Kelly}, \citenamefont {Lucero}, \citenamefont {Mutus}, \citenamefont
  {Neeley}, \citenamefont {Neill}, \citenamefont {O'Malley}, \citenamefont
  {Roushan}, \citenamefont {Shabani}, \citenamefont {Smelyanskiy},
  \citenamefont {Vainsencher}, \citenamefont {Wenner}, \citenamefont {Neven},\
  and\ \citenamefont {Martinis}}]{Martinis_1/f_spectrum}%
  \BibitemOpen
  \bibfield  {author} {\bibinfo {author} {\bibfnamefont {C.~M.}\ \bibnamefont
  {Quintana}}, \bibinfo {author} {\bibfnamefont {Y.}~\bibnamefont {Chen}},
  \bibinfo {author} {\bibfnamefont {D.}~\bibnamefont {Sank}}, \bibinfo {author}
  {\bibfnamefont {A.~G.}\ \bibnamefont {Petukhov}}, \bibinfo {author}
  {\bibfnamefont {T.~C.}\ \bibnamefont {White}}, \bibinfo {author}
  {\bibfnamefont {D.}~\bibnamefont {Kafri}}, \bibinfo {author} {\bibfnamefont
  {B.}~\bibnamefont {Chiaro}}, \bibinfo {author} {\bibfnamefont
  {A.}~\bibnamefont {Megrant}}, \bibinfo {author} {\bibfnamefont
  {R.}~\bibnamefont {Barends}}, \bibinfo {author} {\bibfnamefont
  {B.}~\bibnamefont {Campbell}}, \bibinfo {author} {\bibfnamefont
  {Z.}~\bibnamefont {Chen}}, \bibinfo {author} {\bibfnamefont {A.}~\bibnamefont
  {Dunsworth}}, \bibinfo {author} {\bibfnamefont {A.~G.}\ \bibnamefont
  {Fowler}}, \bibinfo {author} {\bibfnamefont {R.}~\bibnamefont {Graff}},
  \bibinfo {author} {\bibfnamefont {E.}~\bibnamefont {Jeffrey}}, \bibinfo
  {author} {\bibfnamefont {J.}~\bibnamefont {Kelly}}, \bibinfo {author}
  {\bibfnamefont {E.}~\bibnamefont {Lucero}}, \bibinfo {author} {\bibfnamefont
  {J.~Y.}\ \bibnamefont {Mutus}}, \bibinfo {author} {\bibfnamefont
  {M.}~\bibnamefont {Neeley}}, \bibinfo {author} {\bibfnamefont
  {C.}~\bibnamefont {Neill}}, \bibinfo {author} {\bibfnamefont {P.~J.~J.}\
  \bibnamefont {O'Malley}}, \bibinfo {author} {\bibfnamefont {P.}~\bibnamefont
  {Roushan}}, \bibinfo {author} {\bibfnamefont {A.}~\bibnamefont {Shabani}},
  \bibinfo {author} {\bibfnamefont {V.~N.}\ \bibnamefont {Smelyanskiy}},
  \bibinfo {author} {\bibfnamefont {A.}~\bibnamefont {Vainsencher}}, \bibinfo
  {author} {\bibfnamefont {J.}~\bibnamefont {Wenner}}, \bibinfo {author}
  {\bibfnamefont {H.}~\bibnamefont {Neven}},\ and\ \bibinfo {author}
  {\bibfnamefont {J.~M.}\ \bibnamefont {Martinis}},\ }\bibfield  {title}
  {\bibinfo {title} {\textit{Observation of Classical-Quantum Crossover of
  $1/f$ Flux Noise and Its Paramagnetic Temperature Dependence}},\ }\href
  {https://doi.org/10.1103/PhysRevLett.118.057702} {\bibfield  {journal}
  {\bibinfo  {journal} {Phys. Rev. Lett.}\ }\textbf {\bibinfo {volume} {118}},\
  \bibinfo {pages} {057702} (\bibinfo {year} {2017})}\BibitemShut {NoStop}%
\bibitem [{\citenamefont {Braum\"uller}\ \emph {et~al.}(2020)\citenamefont
  {Braum\"uller}, \citenamefont {Ding}, \citenamefont {Veps\"al\"ainen},
  \citenamefont {Sung}, \citenamefont {Kjaergaard}, \citenamefont {Menke},
  \citenamefont {Winik}, \citenamefont {Kim}, \citenamefont {Niedzielski},
  \citenamefont {Melville}, \citenamefont {Yoder}, \citenamefont
  {Hirjibehedin}, \citenamefont {Orlando}, \citenamefont {Gustavsson},\ and\
  \citenamefont {Oliver}}]{Oliver_SQUID_1/f}%
  \BibitemOpen
  \bibfield  {author} {\bibinfo {author} {\bibfnamefont {J.}~\bibnamefont
  {Braum\"uller}}, \bibinfo {author} {\bibfnamefont {L.}~\bibnamefont {Ding}},
  \bibinfo {author} {\bibfnamefont {A.~P.}\ \bibnamefont {Veps\"al\"ainen}},
  \bibinfo {author} {\bibfnamefont {Y.}~\bibnamefont {Sung}}, \bibinfo {author}
  {\bibfnamefont {M.}~\bibnamefont {Kjaergaard}}, \bibinfo {author}
  {\bibfnamefont {T.}~\bibnamefont {Menke}}, \bibinfo {author} {\bibfnamefont
  {R.}~\bibnamefont {Winik}}, \bibinfo {author} {\bibfnamefont
  {D.}~\bibnamefont {Kim}}, \bibinfo {author} {\bibfnamefont {B.~M.}\
  \bibnamefont {Niedzielski}}, \bibinfo {author} {\bibfnamefont
  {A.}~\bibnamefont {Melville}}, \bibinfo {author} {\bibfnamefont {J.~L.}\
  \bibnamefont {Yoder}}, \bibinfo {author} {\bibfnamefont {C.~F.}\ \bibnamefont
  {Hirjibehedin}}, \bibinfo {author} {\bibfnamefont {T.~P.}\ \bibnamefont
  {Orlando}}, \bibinfo {author} {\bibfnamefont {S.}~\bibnamefont
  {Gustavsson}},\ and\ \bibinfo {author} {\bibfnamefont {W.~D.}\ \bibnamefont
  {Oliver}},\ }\bibfield  {title} {\bibinfo {title} {\textit{Characterizing and
  Optimizing Qubit Coherence Based on SQUID Geometry}},\ }\href
  {https://doi.org/10.1103/PhysRevApplied.13.054079} {\bibfield  {journal}
  {\bibinfo  {journal} {Phys. Rev. Applied}\ }\textbf {\bibinfo {volume}
  {13}},\ \bibinfo {pages} {054079} (\bibinfo {year} {2020})}\BibitemShut
  {NoStop}%
\bibitem [{\citenamefont {Yan}\ \emph {et~al.}(2016)\citenamefont {Yan},
  \citenamefont {Gustavsson}, \citenamefont {Kamal}, \citenamefont {Birenbaum},
  \citenamefont {Sears}, \citenamefont {Hover}, \citenamefont {Gudmundsen},
  \citenamefont {Rosenberg}, \citenamefont {Samach}, \citenamefont {Weber},
  \citenamefont {Yoder}, \citenamefont {Orlando}, \citenamefont {Clarke},
  \citenamefont {Kerman},\ and\ \citenamefont
  {Oliver}}]{Oliver_flux_qubit_revisited}%
  \BibitemOpen
  \bibfield  {author} {\bibinfo {author} {\bibfnamefont {F.}~\bibnamefont
  {Yan}}, \bibinfo {author} {\bibfnamefont {S.}~\bibnamefont {Gustavsson}},
  \bibinfo {author} {\bibfnamefont {A.}~\bibnamefont {Kamal}}, \bibinfo
  {author} {\bibfnamefont {J.}~\bibnamefont {Birenbaum}}, \bibinfo {author}
  {\bibfnamefont {A.~P.}\ \bibnamefont {Sears}}, \bibinfo {author}
  {\bibfnamefont {D.}~\bibnamefont {Hover}}, \bibinfo {author} {\bibfnamefont
  {T.~J.}\ \bibnamefont {Gudmundsen}}, \bibinfo {author} {\bibfnamefont
  {D.}~\bibnamefont {Rosenberg}}, \bibinfo {author} {\bibfnamefont
  {G.}~\bibnamefont {Samach}}, \bibinfo {author} {\bibfnamefont
  {S.}~\bibnamefont {Weber}}, \bibinfo {author} {\bibfnamefont {J.~L.}\
  \bibnamefont {Yoder}}, \bibinfo {author} {\bibfnamefont {T.~P.}\ \bibnamefont
  {Orlando}}, \bibinfo {author} {\bibfnamefont {J.}~\bibnamefont {Clarke}},
  \bibinfo {author} {\bibfnamefont {A.~J.}\ \bibnamefont {Kerman}},\ and\
  \bibinfo {author} {\bibfnamefont {W.~D.}\ \bibnamefont {Oliver}},\ }\bibfield
   {title} {\bibinfo {title} {\textit{The Flux Qubit Revisited to Enhance
  Coherence and Reproducibility}},\ }\href
  {https://doi.org/10.1038/ncomms12964} {\bibfield  {journal} {\bibinfo
  {journal} {Nat. Commun.}\ }\textbf {\bibinfo {volume} {7}},\ \bibinfo {pages}
  {12964} (\bibinfo {year} {2016})}\BibitemShut {NoStop}%
\bibitem [{\citenamefont {Catelani}\ \emph {et~al.}(2012)\citenamefont
  {Catelani}, \citenamefont {Nigg}, \citenamefont {Girvin}, \citenamefont
  {Schoelkopf},\ and\ \citenamefont
  {Glazman}}]{Glazman_quasi_particle_tunneling}%
  \BibitemOpen
  \bibfield  {author} {\bibinfo {author} {\bibfnamefont {G.}~\bibnamefont
  {Catelani}}, \bibinfo {author} {\bibfnamefont {S.~E.}\ \bibnamefont {Nigg}},
  \bibinfo {author} {\bibfnamefont {S.~M.}\ \bibnamefont {Girvin}}, \bibinfo
  {author} {\bibfnamefont {R.~J.}\ \bibnamefont {Schoelkopf}},\ and\ \bibinfo
  {author} {\bibfnamefont {L.~I.}\ \bibnamefont {Glazman}},\ }\bibfield
  {title} {\bibinfo {title} {\textit{Decoherence of Superconducting Qubits
  Caused by Quasiparticle Tunneling}},\ }\href
  {https://doi.org/10.1103/PhysRevB.86.184514} {\bibfield  {journal} {\bibinfo
  {journal} {Phys. Rev. B}\ }\textbf {\bibinfo {volume} {86}},\ \bibinfo
  {pages} {184514} (\bibinfo {year} {2012})}\BibitemShut {NoStop}%
\bibitem [{\citenamefont {Koch}\ \emph {et~al.}(2007)\citenamefont {Koch},
  \citenamefont {Yu}, \citenamefont {Gambetta}, \citenamefont {Houck},
  \citenamefont {Schuster}, \citenamefont {Majer}, \citenamefont {Blais},
  \citenamefont {Devoret}, \citenamefont {Girvin},\ and\ \citenamefont
  {Schoelkopf}}]{Koch_transmon_theory}%
  \BibitemOpen
  \bibfield  {author} {\bibinfo {author} {\bibfnamefont {J.}~\bibnamefont
  {Koch}}, \bibinfo {author} {\bibfnamefont {T.~M.}\ \bibnamefont {Yu}},
  \bibinfo {author} {\bibfnamefont {J.}~\bibnamefont {Gambetta}}, \bibinfo
  {author} {\bibfnamefont {A.~A.}\ \bibnamefont {Houck}}, \bibinfo {author}
  {\bibfnamefont {D.~I.}\ \bibnamefont {Schuster}}, \bibinfo {author}
  {\bibfnamefont {J.}~\bibnamefont {Majer}}, \bibinfo {author} {\bibfnamefont
  {A.}~\bibnamefont {Blais}}, \bibinfo {author} {\bibfnamefont {M.~H.}\
  \bibnamefont {Devoret}}, \bibinfo {author} {\bibfnamefont {S.~M.}\
  \bibnamefont {Girvin}},\ and\ \bibinfo {author} {\bibfnamefont {R.~J.}\
  \bibnamefont {Schoelkopf}},\ }\bibfield  {title} {\bibinfo {title}
  {\textit{Charge-Insensitive Qubit Design Derived from the Cooper Pair Box}},\
  }\href {https://doi.org/10.1103/PhysRevA.76.042319} {\bibfield  {journal}
  {\bibinfo  {journal} {Phys. Rev. A}\ }\textbf {\bibinfo {volume} {76}},\
  \bibinfo {pages} {042319} (\bibinfo {year} {2007})}\BibitemShut {NoStop}%
\bibitem [{\citenamefont {Groszkowski}\ \emph {et~al.}(2018)\citenamefont
  {Groszkowski}, \citenamefont {Di~Paolo}, \citenamefont {Grimsmo},
  \citenamefont {Blais}, \citenamefont {Schuster}, \citenamefont {Houck},\ and\
  \citenamefont {Koch}}]{Groszkowski_Zero_pi_theory}%
  \BibitemOpen
  \bibfield  {author} {\bibinfo {author} {\bibfnamefont {P.}~\bibnamefont
  {Groszkowski}}, \bibinfo {author} {\bibfnamefont {A.}~\bibnamefont
  {Di~Paolo}}, \bibinfo {author} {\bibfnamefont {A.~L.}\ \bibnamefont
  {Grimsmo}}, \bibinfo {author} {\bibfnamefont {A.}~\bibnamefont {Blais}},
  \bibinfo {author} {\bibfnamefont {D.~I.}\ \bibnamefont {Schuster}}, \bibinfo
  {author} {\bibfnamefont {A.~A.}\ \bibnamefont {Houck}},\ and\ \bibinfo
  {author} {\bibfnamefont {J.}~\bibnamefont {Koch}},\ }\bibfield  {title}
  {\bibinfo {title} {\textit{Coherence Properties of the 0-$\pi$ Qubit}},\
  }\href {https://doi.org/10.1088/1367-2630/aab7cd} {\bibfield  {journal}
  {\bibinfo  {journal} {New J. Phys.}\ }\textbf {\bibinfo {volume} {20}},\
  \bibinfo {pages} {043053} (\bibinfo {year} {2018})}\BibitemShut {NoStop}%
\bibitem [{\citenamefont {Schriefl}\ \emph {et~al.}(2006)\citenamefont
  {Schriefl}, \citenamefont {Makhlin}, \citenamefont {Shnirman},\ and\
  \citenamefont {Schön}}]{Schon_TLS_ensemble}%
  \BibitemOpen
  \bibfield  {author} {\bibinfo {author} {\bibfnamefont {J.}~\bibnamefont
  {Schriefl}}, \bibinfo {author} {\bibfnamefont {Y.}~\bibnamefont {Makhlin}},
  \bibinfo {author} {\bibfnamefont {A.}~\bibnamefont {Shnirman}},\ and\
  \bibinfo {author} {\bibfnamefont {G.}~\bibnamefont {Schön}},\ }\bibfield
  {title} {\bibinfo {title} {\textit{Decoherence from Ensembles of Two-Level
  Fluctuators}},\ }\href {https://doi.org/10.1088/1367-2630/8/1/001} {\bibfield
   {journal} {\bibinfo  {journal} {New J. Phys.}\ }\textbf {\bibinfo {volume}
  {8}},\ \bibinfo {pages} {1} (\bibinfo {year} {2006})}\BibitemShut {NoStop}%
\bibitem [{\citenamefont {Falci}\ \emph {et~al.}(2005)\citenamefont {Falci},
  \citenamefont {D'Arrigo}, \citenamefont {Mastellone},\ and\ \citenamefont
  {Paladino}}]{Paladino_initial_decoherence}%
  \BibitemOpen
  \bibfield  {author} {\bibinfo {author} {\bibfnamefont {G.}~\bibnamefont
  {Falci}}, \bibinfo {author} {\bibfnamefont {A.}~\bibnamefont {D'Arrigo}},
  \bibinfo {author} {\bibfnamefont {A.}~\bibnamefont {Mastellone}},\ and\
  \bibinfo {author} {\bibfnamefont {E.}~\bibnamefont {Paladino}},\ }\bibfield
  {title} {\bibinfo {title} {\textit{Initial Decoherence in Solid State
  Qubits}},\ }\href {https://doi.org/10.1103/PhysRevLett.94.167002} {\bibfield
  {journal} {\bibinfo  {journal} {Phys. Rev. Lett.}\ }\textbf {\bibinfo
  {volume} {94}},\ \bibinfo {pages} {167002} (\bibinfo {year}
  {2005})}\BibitemShut {NoStop}%
\bibitem [{\citenamefont {Paladino}\ \emph {et~al.}(2002)\citenamefont
  {Paladino}, \citenamefont {Faoro}, \citenamefont {Falci},\ and\ \citenamefont
  {Fazio}}]{Paladino_decoherence_saturation}%
  \BibitemOpen
  \bibfield  {author} {\bibinfo {author} {\bibfnamefont {E.}~\bibnamefont
  {Paladino}}, \bibinfo {author} {\bibfnamefont {L.}~\bibnamefont {Faoro}},
  \bibinfo {author} {\bibfnamefont {G.}~\bibnamefont {Falci}},\ and\ \bibinfo
  {author} {\bibfnamefont {R.}~\bibnamefont {Fazio}},\ }\bibfield  {title}
  {\bibinfo {title} {\textit{Decoherence and $1/\mathit{f}$ Noise in Josephson
  Qubits}},\ }\href {https://doi.org/10.1103/PhysRevLett.88.228304} {\bibfield
  {journal} {\bibinfo  {journal} {Phys. Rev. Lett.}\ }\textbf {\bibinfo
  {volume} {88}},\ \bibinfo {pages} {228304} (\bibinfo {year}
  {2002})}\BibitemShut {NoStop}%
\bibitem [{\citenamefont {Makhlin}\ and\ \citenamefont
  {Shnirman}(2004)}]{Shnirman_Keldysh_dephasing}%
  \BibitemOpen
  \bibfield  {author} {\bibinfo {author} {\bibfnamefont {Y.}~\bibnamefont
  {Makhlin}}\ and\ \bibinfo {author} {\bibfnamefont {A.}~\bibnamefont
  {Shnirman}},\ }\bibfield  {title} {\bibinfo {title} {\textit{Dephasing of
  Solid-State Qubits at Optimal Points}},\ }\href
  {https://doi.org/10.1103/PhysRevLett.92.178301} {\bibfield  {journal}
  {\bibinfo  {journal} {Phys. Rev. Lett.}\ }\textbf {\bibinfo {volume} {92}},\
  \bibinfo {pages} {178301} (\bibinfo {year} {2004})}\BibitemShut {NoStop}%
\bibitem [{\citenamefont {Itakura}\ and\ \citenamefont
  {Tokura}(2003)}]{Tokura_dephasing}%
  \BibitemOpen
  \bibfield  {author} {\bibinfo {author} {\bibfnamefont {T.}~\bibnamefont
  {Itakura}}\ and\ \bibinfo {author} {\bibfnamefont {Y.}~\bibnamefont
  {Tokura}},\ }\bibfield  {title} {\bibinfo {title} {\textit{Dephasing due to
  Background Charge Fluctuations}},\ }\href
  {https://doi.org/10.1103/PhysRevB.67.195320} {\bibfield  {journal} {\bibinfo
  {journal} {Phys. Rev. B}\ }\textbf {\bibinfo {volume} {67}},\ \bibinfo
  {pages} {195320} (\bibinfo {year} {2003})}\BibitemShut {NoStop}%
\bibitem [{\citenamefont {Bergli}\ \emph {et~al.}(2009)\citenamefont {Bergli},
  \citenamefont {Galperin},\ and\ \citenamefont
  {Altshuler}}]{Altshuler_telegraph_TLF}%
  \BibitemOpen
  \bibfield  {author} {\bibinfo {author} {\bibfnamefont {J.}~\bibnamefont
  {Bergli}}, \bibinfo {author} {\bibfnamefont {Y.~M.}\ \bibnamefont
  {Galperin}},\ and\ \bibinfo {author} {\bibfnamefont {B.~L.}\ \bibnamefont
  {Altshuler}},\ }\bibfield  {title} {\bibinfo {title} {\textit{Decoherence in
  Qubits Due to Low-Frequency Noise}},\ }\href
  {https://doi.org/10.1088/1367-2630/11/2/025002} {\bibfield  {journal}
  {\bibinfo  {journal} {New J. Phys.}\ }\textbf {\bibinfo {volume} {11}},\
  \bibinfo {pages} {025002} (\bibinfo {year} {2009})}\BibitemShut {NoStop}%
\bibitem [{\citenamefont {Rossi}\ and\ \citenamefont
  {Paris}(2016)}]{Matteo_Gaussian_nonGaussian}%
  \BibitemOpen
  \bibfield  {author} {\bibinfo {author} {\bibfnamefont {M.~A.~C.}\
  \bibnamefont {Rossi}}\ and\ \bibinfo {author} {\bibfnamefont {M.~G.~A.}\
  \bibnamefont {Paris}},\ }\bibfield  {title} {\bibinfo {title}
  {\textit{Non-Markovian Dynamics of Single- and Two-Qubit Systems Interacting
  with Gaussian and non-Gaussian Fluctuating Transverse Environments}},\ }\href
  {https://doi.org/10.1063/1.4939733} {\bibfield  {journal} {\bibinfo
  {journal} {J. Chem. Phys.}\ }\textbf {\bibinfo {volume} {144}},\ \bibinfo
  {pages} {024113} (\bibinfo {year} {2016})}\BibitemShut {NoStop}%
\bibitem [{\citenamefont {Bergli}\ \emph {et~al.}(2006)\citenamefont {Bergli},
  \citenamefont {Galperin},\ and\ \citenamefont
  {Altshuler}}]{Bergli_tlf_arbitrary}%
  \BibitemOpen
  \bibfield  {author} {\bibinfo {author} {\bibfnamefont {J.}~\bibnamefont
  {Bergli}}, \bibinfo {author} {\bibfnamefont {Y.~M.}\ \bibnamefont
  {Galperin}},\ and\ \bibinfo {author} {\bibfnamefont {B.~L.}\ \bibnamefont
  {Altshuler}},\ }\bibfield  {title} {\bibinfo {title} {\textit{Decoherence of
  a Qubit by Non-Gaussian Noise at an Arbitrary Working Point}},\ }\href
  {https://doi.org/10.1103/PhysRevB.74.024509} {\bibfield  {journal} {\bibinfo
  {journal} {Phys. Rev. B}\ }\textbf {\bibinfo {volume} {74}},\ \bibinfo
  {pages} {024509} (\bibinfo {year} {2006})}\BibitemShut {NoStop}%
\bibitem [{\citenamefont {Galperin}\ \emph {et~al.}(2006)\citenamefont
  {Galperin}, \citenamefont {Altshuler}, \citenamefont {Bergli},\ and\
  \citenamefont {Shantsev}}]{Galperin_Echo_TLFs}%
  \BibitemOpen
  \bibfield  {author} {\bibinfo {author} {\bibfnamefont {Y.~M.}\ \bibnamefont
  {Galperin}}, \bibinfo {author} {\bibfnamefont {B.~L.}\ \bibnamefont
  {Altshuler}}, \bibinfo {author} {\bibfnamefont {J.}~\bibnamefont {Bergli}},\
  and\ \bibinfo {author} {\bibfnamefont {D.~V.}\ \bibnamefont {Shantsev}},\
  }\bibfield  {title} {\bibinfo {title} {\textit{Non-Gaussian Low-Frequency
  Noise as a Source of Qubit Decoherence}},\ }\href
  {https://doi.org/10.1103/PhysRevLett.96.097009} {\bibfield  {journal}
  {\bibinfo  {journal} {Phys. Rev. Lett.}\ }\textbf {\bibinfo {volume} {96}},\
  \bibinfo {pages} {097009} (\bibinfo {year} {2006})}\BibitemShut {NoStop}%
\bibitem [{\citenamefont {Ithier}\ \emph {et~al.}(2005)\citenamefont {Ithier},
  \citenamefont {Collin}, \citenamefont {Joyez}, \citenamefont {Meeson},
  \citenamefont {Vion}, \citenamefont {Esteve}, \citenamefont {Chiarello},
  \citenamefont {Shnirman}, \citenamefont {Makhlin}, \citenamefont {Schriefl},\
  and\ \citenamefont {Sch\"on}}]{Ithier_decoherence_analysis}%
  \BibitemOpen
  \bibfield  {author} {\bibinfo {author} {\bibfnamefont {G.}~\bibnamefont
  {Ithier}}, \bibinfo {author} {\bibfnamefont {E.}~\bibnamefont {Collin}},
  \bibinfo {author} {\bibfnamefont {P.}~\bibnamefont {Joyez}}, \bibinfo
  {author} {\bibfnamefont {P.~J.}\ \bibnamefont {Meeson}}, \bibinfo {author}
  {\bibfnamefont {D.}~\bibnamefont {Vion}}, \bibinfo {author} {\bibfnamefont
  {D.}~\bibnamefont {Esteve}}, \bibinfo {author} {\bibfnamefont
  {F.}~\bibnamefont {Chiarello}}, \bibinfo {author} {\bibfnamefont
  {A.}~\bibnamefont {Shnirman}}, \bibinfo {author} {\bibfnamefont
  {Y.}~\bibnamefont {Makhlin}}, \bibinfo {author} {\bibfnamefont
  {J.}~\bibnamefont {Schriefl}},\ and\ \bibinfo {author} {\bibfnamefont
  {G.}~\bibnamefont {Sch\"on}},\ }\bibfield  {title} {\bibinfo {title}
  {\textit{Decoherence in a Superconducting Quantum Bit Circuit}},\ }\href
  {https://doi.org/10.1103/PhysRevB.72.134519} {\bibfield  {journal} {\bibinfo
  {journal} {Phys. Rev. B}\ }\textbf {\bibinfo {volume} {72}},\ \bibinfo
  {pages} {134519} (\bibinfo {year} {2005})}\BibitemShut {NoStop}%
\bibitem [{\citenamefont {You}\ \emph {et~al.}(2021)\citenamefont {You},
  \citenamefont {Clerk},\ and\ \citenamefont {Koch}}]{Koch_TLS_FD_thoerem}%
  \BibitemOpen
  \bibfield  {author} {\bibinfo {author} {\bibfnamefont {X.}~\bibnamefont
  {You}}, \bibinfo {author} {\bibfnamefont {A.~A.}\ \bibnamefont {Clerk}},\
  and\ \bibinfo {author} {\bibfnamefont {J.}~\bibnamefont {Koch}},\ }\bibfield
  {title} {\bibinfo {title} {\textit{Positive- and Negative-Frequency Noise
  from an Ensemble of Two-Level Fluctuators}},\ }\href
  {https://doi.org/10.1103/PhysRevResearch.3.013045} {\bibfield  {journal}
  {\bibinfo  {journal} {Phys. Rev. Research}\ }\textbf {\bibinfo {volume}
  {3}},\ \bibinfo {pages} {013045} (\bibinfo {year} {2021})}\BibitemShut
  {NoStop}%
\bibitem [{\citenamefont {Huang}\ \emph {et~al.}(2021)\citenamefont {Huang},
  \citenamefont {Mundada}, \citenamefont {Gyenis}, \citenamefont {Schuster},
  \citenamefont {Houck},\ and\ \citenamefont {Koch}}]{Dynamical_sweet_spot}%
  \BibitemOpen
  \bibfield  {author} {\bibinfo {author} {\bibfnamefont {Z.}~\bibnamefont
  {Huang}}, \bibinfo {author} {\bibfnamefont {P.~S.}\ \bibnamefont {Mundada}},
  \bibinfo {author} {\bibfnamefont {A.}~\bibnamefont {Gyenis}}, \bibinfo
  {author} {\bibfnamefont {D.~I.}\ \bibnamefont {Schuster}}, \bibinfo {author}
  {\bibfnamefont {A.~A.}\ \bibnamefont {Houck}},\ and\ \bibinfo {author}
  {\bibfnamefont {J.}~\bibnamefont {Koch}},\ }\bibfield  {title} {\bibinfo
  {title} {\textit{Engineering Dynamical Sweet Spots to Protect Qubits from
  $1/f$ Noise}},\ }\href {https://doi.org/10.1103/PhysRevApplied.15.034065}
  {\bibfield  {journal} {\bibinfo  {journal} {Phys. Rev. Applied}\ }\textbf
  {\bibinfo {volume} {15}},\ \bibinfo {pages} {034065} (\bibinfo {year}
  {2021})}\BibitemShut {NoStop}%
\bibitem [{\citenamefont {Mundada}\ \emph {et~al.}(2020)\citenamefont
  {Mundada}, \citenamefont {Gyenis}, \citenamefont {Huang}, \citenamefont
  {Koch},\ and\ \citenamefont {Houck}}]{Dynamical_sweet_spot_exp}%
  \BibitemOpen
  \bibfield  {author} {\bibinfo {author} {\bibfnamefont {P.~S.}\ \bibnamefont
  {Mundada}}, \bibinfo {author} {\bibfnamefont {A.}~\bibnamefont {Gyenis}},
  \bibinfo {author} {\bibfnamefont {Z.}~\bibnamefont {Huang}}, \bibinfo
  {author} {\bibfnamefont {J.}~\bibnamefont {Koch}},\ and\ \bibinfo {author}
  {\bibfnamefont {A.~A.}\ \bibnamefont {Houck}},\ }\bibfield  {title} {\bibinfo
  {title} {\textit{Floquet-Engineered Enhancement of Coherence Times in a
  Driven Fluxonium Qubit}},\ }\href
  {https://doi.org/10.1103/PhysRevApplied.14.054033} {\bibfield  {journal}
  {\bibinfo  {journal} {Phys. Rev. Applied}\ }\textbf {\bibinfo {volume}
  {14}},\ \bibinfo {pages} {054033} (\bibinfo {year} {2020})}\BibitemShut
  {NoStop}%
\bibitem [{\citenamefont {Weiss}\ \emph {et~al.}(2019)\citenamefont {Weiss},
  \citenamefont {Li}, \citenamefont {Ferguson},\ and\ \citenamefont
  {Koch}}]{Koch_current_mirror}%
  \BibitemOpen
  \bibfield  {author} {\bibinfo {author} {\bibfnamefont {D.~K.}\ \bibnamefont
  {Weiss}}, \bibinfo {author} {\bibfnamefont {A.~C.~Y.}\ \bibnamefont {Li}},
  \bibinfo {author} {\bibfnamefont {D.~G.}\ \bibnamefont {Ferguson}},\ and\
  \bibinfo {author} {\bibfnamefont {J.}~\bibnamefont {Koch}},\ }\bibfield
  {title} {\bibinfo {title} {\textit{Spectrum and Coherence Properties of the
  Current-Mirror Qubit}},\ }\href {https://doi.org/10.1103/PhysRevB.100.224507}
  {\bibfield  {journal} {\bibinfo  {journal} {Phys. Rev. B}\ }\textbf {\bibinfo
  {volume} {100}},\ \bibinfo {pages} {224507} (\bibinfo {year}
  {2019})}\BibitemShut {NoStop}%
\bibitem [{\citenamefont {Didier}()}]{Didier_dynamical_sweet_spot}%
  \BibitemOpen
  \bibfield  {author} {\bibinfo {author} {\bibfnamefont {N.}~\bibnamefont
  {Didier}},\ }\bibfield  {title} {\bibinfo {title} {\textit{Flux Control of
  Superconducting Qubits at Dynamical Sweet Spot} (2019)},\ }\Eprint
  {https://arxiv.org/abs/arXiv:1912.09416} {arXiv:1912.09416} \BibitemShut
  {NoStop}%
\bibitem [{\citenamefont {Valery}\ \emph {et~al.}(2022)\citenamefont {Valery},
  \citenamefont {Chowdhury}, \citenamefont {Jones},\ and\ \citenamefont
  {Didier}}]{Didier_dynamical_sweet_spot_exp}%
  \BibitemOpen
  \bibfield  {author} {\bibinfo {author} {\bibfnamefont {J.~A.}\ \bibnamefont
  {Valery}}, \bibinfo {author} {\bibfnamefont {S.}~\bibnamefont {Chowdhury}},
  \bibinfo {author} {\bibfnamefont {G.}~\bibnamefont {Jones}},\ and\ \bibinfo
  {author} {\bibfnamefont {N.}~\bibnamefont {Didier}},\ }\bibfield  {title}
  {\bibinfo {title} {Dynamical sweet spot engineering via two-tone flux
  modulation of superconducting qubits},\ }\href
  {https://doi.org/10.1103/PRXQuantum.3.020337} {\bibfield  {journal} {\bibinfo
   {journal} {PRX Quantum}\ }\textbf {\bibinfo {volume} {3}},\ \bibinfo {pages}
  {020337} (\bibinfo {year} {2022})}\BibitemShut {NoStop}%
\bibitem [{\citenamefont {Sung}\ \emph {et~al.}(2019)\citenamefont {Sung},
  \citenamefont {Beaudoin}, \citenamefont {Norris}, \citenamefont {Yan},
  \citenamefont {Kim}, \citenamefont {Qiu}, \citenamefont {von L{\"u}pke},
  \citenamefont {Yoder}, \citenamefont {Orlando}, \citenamefont {Gustavsson},
  \citenamefont {Viola},\ and\ \citenamefont
  {Oliver}}]{Oliver_non_Gaussian_spec}%
  \BibitemOpen
  \bibfield  {author} {\bibinfo {author} {\bibfnamefont {Y.}~\bibnamefont
  {Sung}}, \bibinfo {author} {\bibfnamefont {F.}~\bibnamefont {Beaudoin}},
  \bibinfo {author} {\bibfnamefont {L.~M.}\ \bibnamefont {Norris}}, \bibinfo
  {author} {\bibfnamefont {F.}~\bibnamefont {Yan}}, \bibinfo {author}
  {\bibfnamefont {D.~K.}\ \bibnamefont {Kim}}, \bibinfo {author} {\bibfnamefont
  {J.~Y.}\ \bibnamefont {Qiu}}, \bibinfo {author} {\bibfnamefont
  {U.}~\bibnamefont {von L{\"u}pke}}, \bibinfo {author} {\bibfnamefont {J.~L.}\
  \bibnamefont {Yoder}}, \bibinfo {author} {\bibfnamefont {T.~P.}\ \bibnamefont
  {Orlando}}, \bibinfo {author} {\bibfnamefont {S.}~\bibnamefont {Gustavsson}},
  \bibinfo {author} {\bibfnamefont {L.}~\bibnamefont {Viola}},\ and\ \bibinfo
  {author} {\bibfnamefont {W.~D.}\ \bibnamefont {Oliver}},\ }\bibfield  {title}
  {\bibinfo {title} {\textit{Non-Gaussian Noise Spectroscopy with a
  Superconducting Qubit Sensor}},\ }\href
  {https://doi.org/10.1038/s41467-019-11699-4} {\bibfield  {journal} {\bibinfo
  {journal} {Nat. Commun.}\ }\textbf {\bibinfo {volume} {10}},\ \bibinfo
  {pages} {3715} (\bibinfo {year} {2019})}\BibitemShut {NoStop}%
\bibitem [{\citenamefont {Bylander}\ \emph {et~al.}(2011)\citenamefont
  {Bylander}, \citenamefont {Gustavsson}, \citenamefont {Yan}, \citenamefont
  {Yoshihara}, \citenamefont {Harrabi}, \citenamefont {Fitch}, \citenamefont
  {Cory}, \citenamefont {Nakamura}, \citenamefont {Tsai},\ and\ \citenamefont
  {Oliver}}]{Oliver_flux_qubit_dd}%
  \BibitemOpen
  \bibfield  {author} {\bibinfo {author} {\bibfnamefont {J.}~\bibnamefont
  {Bylander}}, \bibinfo {author} {\bibfnamefont {S.}~\bibnamefont
  {Gustavsson}}, \bibinfo {author} {\bibfnamefont {F.}~\bibnamefont {Yan}},
  \bibinfo {author} {\bibfnamefont {F.}~\bibnamefont {Yoshihara}}, \bibinfo
  {author} {\bibfnamefont {K.}~\bibnamefont {Harrabi}}, \bibinfo {author}
  {\bibfnamefont {G.}~\bibnamefont {Fitch}}, \bibinfo {author} {\bibfnamefont
  {D.~G.}\ \bibnamefont {Cory}}, \bibinfo {author} {\bibfnamefont
  {Y.}~\bibnamefont {Nakamura}}, \bibinfo {author} {\bibfnamefont {J.-S.}\
  \bibnamefont {Tsai}},\ and\ \bibinfo {author} {\bibfnamefont {W.~D.}\
  \bibnamefont {Oliver}},\ }\bibfield  {title} {\bibinfo {title} {\textit{Noise
  Spectroscopy Through Dynamical Decoupling with a Superconducting Flux
  Qubit}},\ }\href {https://doi.org/10.1038/nphys1994} {\bibfield  {journal}
  {\bibinfo  {journal} {Nat. Phys.}\ }\textbf {\bibinfo {volume} {7}},\
  \bibinfo {pages} {565} (\bibinfo {year} {2011})}\BibitemShut {NoStop}%
\bibitem [{\citenamefont {Gyenis}\ \emph
  {et~al.}(2021{\natexlab{b}})\citenamefont {Gyenis}, \citenamefont {Mundada},
  \citenamefont {Di~Paolo}, \citenamefont {Hazard}, \citenamefont {You},
  \citenamefont {Schuster}, \citenamefont {Koch}, \citenamefont {Blais},\ and\
  \citenamefont {Houck}}]{Houck_zero_pi_experiment}%
  \BibitemOpen
  \bibfield  {author} {\bibinfo {author} {\bibfnamefont {A.}~\bibnamefont
  {Gyenis}}, \bibinfo {author} {\bibfnamefont {P.~S.}\ \bibnamefont {Mundada}},
  \bibinfo {author} {\bibfnamefont {A.}~\bibnamefont {Di~Paolo}}, \bibinfo
  {author} {\bibfnamefont {T.~M.}\ \bibnamefont {Hazard}}, \bibinfo {author}
  {\bibfnamefont {X.}~\bibnamefont {You}}, \bibinfo {author} {\bibfnamefont
  {D.~I.}\ \bibnamefont {Schuster}}, \bibinfo {author} {\bibfnamefont
  {J.}~\bibnamefont {Koch}}, \bibinfo {author} {\bibfnamefont {A.}~\bibnamefont
  {Blais}},\ and\ \bibinfo {author} {\bibfnamefont {A.~A.}\ \bibnamefont
  {Houck}},\ }\bibfield  {title} {\bibinfo {title} {\textit{Experimental
  Realization of a Protected Superconducting Circuit Derived from the 0-$\pi$
  Qubit}},\ }\href {https://doi.org/10.1103/PRXQuantum.2.010339} {\bibfield
  {journal} {\bibinfo  {journal} {PRX Quantum}\ }\textbf {\bibinfo {volume}
  {2}},\ \bibinfo {pages} {010339} (\bibinfo {year}
  {2021}{\natexlab{b}})}\BibitemShut {NoStop}%
\bibitem [{\citenamefont {Place}\ \emph {et~al.}(2021)\citenamefont {Place},
  \citenamefont {Rodgers}, \citenamefont {Mundada}, \citenamefont {Smitham},
  \citenamefont {Fitzpatrick}, \citenamefont {Leng}, \citenamefont {Premkumar},
  \citenamefont {Bryon}, \citenamefont {Vrajitoarea}, \citenamefont {Sussman},
  \citenamefont {Cheng}, \citenamefont {Madhavan}, \citenamefont {Babla},
  \citenamefont {Le}, \citenamefont {Gang}, \citenamefont {J{\"a}ck},
  \citenamefont {Gyenis}, \citenamefont {Yao}, \citenamefont {Cava},
  \citenamefont {de~Leon},\ and\ \citenamefont {Houck}}]{Houck_tantalum_qubit}%
  \BibitemOpen
  \bibfield  {author} {\bibinfo {author} {\bibfnamefont {A.~P.~M.}\
  \bibnamefont {Place}}, \bibinfo {author} {\bibfnamefont {L.~V.~H.}\
  \bibnamefont {Rodgers}}, \bibinfo {author} {\bibfnamefont {P.}~\bibnamefont
  {Mundada}}, \bibinfo {author} {\bibfnamefont {B.~M.}\ \bibnamefont
  {Smitham}}, \bibinfo {author} {\bibfnamefont {M.}~\bibnamefont
  {Fitzpatrick}}, \bibinfo {author} {\bibfnamefont {Z.}~\bibnamefont {Leng}},
  \bibinfo {author} {\bibfnamefont {A.}~\bibnamefont {Premkumar}}, \bibinfo
  {author} {\bibfnamefont {J.}~\bibnamefont {Bryon}}, \bibinfo {author}
  {\bibfnamefont {A.}~\bibnamefont {Vrajitoarea}}, \bibinfo {author}
  {\bibfnamefont {S.}~\bibnamefont {Sussman}}, \bibinfo {author} {\bibfnamefont
  {G.}~\bibnamefont {Cheng}}, \bibinfo {author} {\bibfnamefont
  {T.}~\bibnamefont {Madhavan}}, \bibinfo {author} {\bibfnamefont {H.~K.}\
  \bibnamefont {Babla}}, \bibinfo {author} {\bibfnamefont {X.~H.}\ \bibnamefont
  {Le}}, \bibinfo {author} {\bibfnamefont {Y.}~\bibnamefont {Gang}}, \bibinfo
  {author} {\bibfnamefont {B.}~\bibnamefont {J{\"a}ck}}, \bibinfo {author}
  {\bibfnamefont {A.}~\bibnamefont {Gyenis}}, \bibinfo {author} {\bibfnamefont
  {N.}~\bibnamefont {Yao}}, \bibinfo {author} {\bibfnamefont {R.~J.}\
  \bibnamefont {Cava}}, \bibinfo {author} {\bibfnamefont {N.~P.}\ \bibnamefont
  {de~Leon}},\ and\ \bibinfo {author} {\bibfnamefont {A.~A.}\ \bibnamefont
  {Houck}},\ }\bibfield  {title} {\bibinfo {title} {\textit{New Material
  Platform for Superconducting Transmon Qubits with Coherence Times Exceeding
  0.3 Milliseconds}},\ }\href {https://doi.org/10.1038/s41467-021-22030-5}
  {\bibfield  {journal} {\bibinfo  {journal} {Nat. Commun.}\ }\textbf {\bibinfo
  {volume} {12}},\ \bibinfo {pages} {1779} (\bibinfo {year}
  {2021})}\BibitemShut {NoStop}%
\bibitem [{\citenamefont {Wang}\ \emph {et~al.}(2022)\citenamefont {Wang},
  \citenamefont {Li}, \citenamefont {Xu}, \citenamefont {Li}, \citenamefont
  {Wang}, \citenamefont {Yang}, \citenamefont {Mi}, \citenamefont {Liang},
  \citenamefont {Su}, \citenamefont {Yang}, \citenamefont {Wang}, \citenamefont
  {Wang}, \citenamefont {Li}, \citenamefont {Chen}, \citenamefont {Li},
  \citenamefont {Linghu}, \citenamefont {Han}, \citenamefont {Zhang},
  \citenamefont {Feng}, \citenamefont {Song}, \citenamefont {Ma}, \citenamefont
  {Zhang}, \citenamefont {Wang}, \citenamefont {Zhao}, \citenamefont {Liu},
  \citenamefont {Xue}, \citenamefont {Jin},\ and\ \citenamefont
  {Yu}}]{Yu_tantalum_qubit}%
  \BibitemOpen
  \bibfield  {author} {\bibinfo {author} {\bibfnamefont {C.}~\bibnamefont
  {Wang}}, \bibinfo {author} {\bibfnamefont {X.}~\bibnamefont {Li}}, \bibinfo
  {author} {\bibfnamefont {H.}~\bibnamefont {Xu}}, \bibinfo {author}
  {\bibfnamefont {Z.}~\bibnamefont {Li}}, \bibinfo {author} {\bibfnamefont
  {J.}~\bibnamefont {Wang}}, \bibinfo {author} {\bibfnamefont {Z.}~\bibnamefont
  {Yang}}, \bibinfo {author} {\bibfnamefont {Z.}~\bibnamefont {Mi}}, \bibinfo
  {author} {\bibfnamefont {X.}~\bibnamefont {Liang}}, \bibinfo {author}
  {\bibfnamefont {T.}~\bibnamefont {Su}}, \bibinfo {author} {\bibfnamefont
  {C.}~\bibnamefont {Yang}}, \bibinfo {author} {\bibfnamefont {G.}~\bibnamefont
  {Wang}}, \bibinfo {author} {\bibfnamefont {W.}~\bibnamefont {Wang}}, \bibinfo
  {author} {\bibfnamefont {Y.}~\bibnamefont {Li}}, \bibinfo {author}
  {\bibfnamefont {M.}~\bibnamefont {Chen}}, \bibinfo {author} {\bibfnamefont
  {C.}~\bibnamefont {Li}}, \bibinfo {author} {\bibfnamefont {K.}~\bibnamefont
  {Linghu}}, \bibinfo {author} {\bibfnamefont {J.}~\bibnamefont {Han}},
  \bibinfo {author} {\bibfnamefont {Y.}~\bibnamefont {Zhang}}, \bibinfo
  {author} {\bibfnamefont {Y.}~\bibnamefont {Feng}}, \bibinfo {author}
  {\bibfnamefont {Y.}~\bibnamefont {Song}}, \bibinfo {author} {\bibfnamefont
  {T.}~\bibnamefont {Ma}}, \bibinfo {author} {\bibfnamefont {J.}~\bibnamefont
  {Zhang}}, \bibinfo {author} {\bibfnamefont {R.}~\bibnamefont {Wang}},
  \bibinfo {author} {\bibfnamefont {P.}~\bibnamefont {Zhao}}, \bibinfo {author}
  {\bibfnamefont {W.}~\bibnamefont {Liu}}, \bibinfo {author} {\bibfnamefont
  {G.}~\bibnamefont {Xue}}, \bibinfo {author} {\bibfnamefont {Y.}~\bibnamefont
  {Jin}},\ and\ \bibinfo {author} {\bibfnamefont {H.}~\bibnamefont {Yu}},\
  }\bibfield  {title} {\bibinfo {title} {\textit{Towards Practical Quantum
  Computers: Transmon Qubit with a Lifetime Approaching 0.5 Milliseconds}},\
  }\href {https://doi.org/10.1038/s41534-021-00510-2} {\bibfield  {journal}
  {\bibinfo  {journal} {npj Quantum Inf.}\ }\textbf {\bibinfo {volume} {8}},\
  \bibinfo {pages} {3} (\bibinfo {year} {2022})}\BibitemShut {NoStop}%
\bibitem [{\citenamefont {Manucharyan}\ \emph {et~al.}(2009)\citenamefont
  {Manucharyan}, \citenamefont {Koch}, \citenamefont {Glazman},\ and\
  \citenamefont {Devoret}}]{Fluxonium_Devoret}%
  \BibitemOpen
  \bibfield  {author} {\bibinfo {author} {\bibfnamefont {V.~E.}\ \bibnamefont
  {Manucharyan}}, \bibinfo {author} {\bibfnamefont {J.}~\bibnamefont {Koch}},
  \bibinfo {author} {\bibfnamefont {L.~I.}\ \bibnamefont {Glazman}},\ and\
  \bibinfo {author} {\bibfnamefont {M.~H.}\ \bibnamefont {Devoret}},\
  }\bibfield  {title} {\bibinfo {title} {\textit{Fluxonium: Single Cooper-Pair
  Circuit Free of Charge Offsets}},\ }\href
  {https://doi.org/10.1126/science.1175552} {\bibfield  {journal} {\bibinfo
  {journal} {Science}\ }\textbf {\bibinfo {volume} {326}},\ \bibinfo {pages}
  {113} (\bibinfo {year} {2009})}\BibitemShut {NoStop}%
\bibitem [{\citenamefont {Wang}\ and\ \citenamefont
  {Clerk}(2020)}]{Clerk_non_Gaussian}%
  \BibitemOpen
  \bibfield  {author} {\bibinfo {author} {\bibfnamefont {Y.-X.}\ \bibnamefont
  {Wang}}\ and\ \bibinfo {author} {\bibfnamefont {A.~A.}\ \bibnamefont
  {Clerk}},\ }\bibfield  {title} {\bibinfo {title} {\textit{Spectral
  Characterization of Non-Gaussian Quantum Noise: Keldysh Approach and
  Application to Photon Shot Noise}},\ }\href
  {https://doi.org/10.1103/PhysRevResearch.2.033196} {\bibfield  {journal}
  {\bibinfo  {journal} {Phys. Rev. Research}\ }\textbf {\bibinfo {volume}
  {2}},\ \bibinfo {pages} {033196} (\bibinfo {year} {2020})}\BibitemShut
  {NoStop}%
\bibitem [{\citenamefont {Makhlin}\ and\ \citenamefont
  {Shnirman}(2003)}]{Shnirman_Keldysh_dephasing_1}%
  \BibitemOpen
  \bibfield  {author} {\bibinfo {author} {\bibfnamefont {Y.}~\bibnamefont
  {Makhlin}}\ and\ \bibinfo {author} {\bibfnamefont {A.}~\bibnamefont
  {Shnirman}},\ }\bibfield  {title} {\bibinfo {title} {\textit{Dephasing of
  Qubits by Transverse Low-Frequency Noise}},\ }\href
  {https://doi.org/10.1134/1.1637702} {\bibfield  {journal} {\bibinfo
  {journal} {J. Exp. Theor. Phys.}\ }\textbf {\bibinfo {volume} {78}},\
  \bibinfo {pages} {497} (\bibinfo {year} {2003})}\BibitemShut {NoStop}%
\bibitem [{\citenamefont {Hong}\ \emph {et~al.}(2020)\citenamefont {Hong},
  \citenamefont {Papageorge}, \citenamefont {Sivarajah}, \citenamefont
  {Crossman}, \citenamefont {Didier}, \citenamefont {Polloreno}, \citenamefont
  {Sete}, \citenamefont {Turkowski}, \citenamefont {da~Silva},\ and\
  \citenamefont {Johnson}}]{Rigetti_ac_sweet_spot_exp}%
  \BibitemOpen
  \bibfield  {author} {\bibinfo {author} {\bibfnamefont {S.~S.}\ \bibnamefont
  {Hong}}, \bibinfo {author} {\bibfnamefont {A.~T.}\ \bibnamefont
  {Papageorge}}, \bibinfo {author} {\bibfnamefont {P.}~\bibnamefont
  {Sivarajah}}, \bibinfo {author} {\bibfnamefont {G.}~\bibnamefont {Crossman}},
  \bibinfo {author} {\bibfnamefont {N.}~\bibnamefont {Didier}}, \bibinfo
  {author} {\bibfnamefont {A.~M.}\ \bibnamefont {Polloreno}}, \bibinfo {author}
  {\bibfnamefont {E.~A.}\ \bibnamefont {Sete}}, \bibinfo {author}
  {\bibfnamefont {S.~W.}\ \bibnamefont {Turkowski}}, \bibinfo {author}
  {\bibfnamefont {M.~P.}\ \bibnamefont {da~Silva}},\ and\ \bibinfo {author}
  {\bibfnamefont {B.~R.}\ \bibnamefont {Johnson}},\ }\bibfield  {title}
  {\bibinfo {title} {\textit{Demonstration of a Parametrically Activated
  Entangling Gate Protected from Flux Noise}},\ }\href
  {https://doi.org/10.1103/PhysRevA.101.012302} {\bibfield  {journal} {\bibinfo
   {journal} {Phys. Rev. A}\ }\textbf {\bibinfo {volume} {101}},\ \bibinfo
  {pages} {012302} (\bibinfo {year} {2020})}\BibitemShut {NoStop}%
\bibitem [{\citenamefont {Somoroff}\ \emph {et~al.}()\citenamefont {Somoroff},
  \citenamefont {Ficheux}, \citenamefont {Mencia}, \citenamefont {Xiong},
  \citenamefont {Kuzmin},\ and\ \citenamefont
  {Manucharyan}}]{Manucharyan_1ms_fluxonium}%
  \BibitemOpen
  \bibfield  {author} {\bibinfo {author} {\bibfnamefont {A.}~\bibnamefont
  {Somoroff}}, \bibinfo {author} {\bibfnamefont {Q.}~\bibnamefont {Ficheux}},
  \bibinfo {author} {\bibfnamefont {R.~A.}\ \bibnamefont {Mencia}}, \bibinfo
  {author} {\bibfnamefont {H.}~\bibnamefont {Xiong}}, \bibinfo {author}
  {\bibfnamefont {R.~V.}\ \bibnamefont {Kuzmin}},\ and\ \bibinfo {author}
  {\bibfnamefont {V.~E.}\ \bibnamefont {Manucharyan}},\ }\bibfield  {title}
  {\bibinfo {title} {\textit{Millisecond Coherence in a Superconducting Qubit}
  (2021)},\ }\Eprint {https://arxiv.org/abs/arXiv:2103.08578}
  {arXiv:2103.08578} \BibitemShut {NoStop}%
\bibitem [{\citenamefont {Kalashnikov}\ \emph {et~al.}(2020)\citenamefont
  {Kalashnikov}, \citenamefont {Hsieh}, \citenamefont {Zhang}, \citenamefont
  {Lu}, \citenamefont {Kamenov}, \citenamefont {Di~Paolo}, \citenamefont
  {Blais}, \citenamefont {Gershenson},\ and\ \citenamefont
  {Bell}}]{Bell_bifluxon}%
  \BibitemOpen
  \bibfield  {author} {\bibinfo {author} {\bibfnamefont {K.}~\bibnamefont
  {Kalashnikov}}, \bibinfo {author} {\bibfnamefont {W.~T.}\ \bibnamefont
  {Hsieh}}, \bibinfo {author} {\bibfnamefont {W.}~\bibnamefont {Zhang}},
  \bibinfo {author} {\bibfnamefont {W.-S.}\ \bibnamefont {Lu}}, \bibinfo
  {author} {\bibfnamefont {P.}~\bibnamefont {Kamenov}}, \bibinfo {author}
  {\bibfnamefont {A.}~\bibnamefont {Di~Paolo}}, \bibinfo {author}
  {\bibfnamefont {A.}~\bibnamefont {Blais}}, \bibinfo {author} {\bibfnamefont
  {M.~E.}\ \bibnamefont {Gershenson}},\ and\ \bibinfo {author} {\bibfnamefont
  {M.}~\bibnamefont {Bell}},\ }\bibfield  {title} {\bibinfo {title} {Bifluxon:
  Fluxon-parity-protected superconducting qubit},\ }\href
  {https://doi.org/10.1103/PRXQuantum.1.010307} {\bibfield  {journal} {\bibinfo
   {journal} {PRX Quantum}\ }\textbf {\bibinfo {volume} {1}},\ \bibinfo {pages}
  {010307} (\bibinfo {year} {2020})}\BibitemShut {NoStop}%
\bibitem [{\citenamefont {Shnirman}\ \emph {et~al.}(2005)\citenamefont
  {Shnirman}, \citenamefont {Sch\"on}, \citenamefont {Martin},\ and\
  \citenamefont {Makhlin}}]{Makhlin_TLS_1/f}%
  \BibitemOpen
  \bibfield  {author} {\bibinfo {author} {\bibfnamefont {A.}~\bibnamefont
  {Shnirman}}, \bibinfo {author} {\bibfnamefont {G.}~\bibnamefont {Sch\"on}},
  \bibinfo {author} {\bibfnamefont {I.}~\bibnamefont {Martin}},\ and\ \bibinfo
  {author} {\bibfnamefont {Y.}~\bibnamefont {Makhlin}},\ }\bibfield  {title}
  {\bibinfo {title} {\textit{Low- and High-Frequency Noise from Coherent
  Two-Level Systems}},\ }\href {https://doi.org/10.1103/PhysRevLett.94.127002}
  {\bibfield  {journal} {\bibinfo  {journal} {Phys. Rev. Lett.}\ }\textbf
  {\bibinfo {volume} {94}},\ \bibinfo {pages} {127002} (\bibinfo {year}
  {2005})}\BibitemShut {NoStop}%
\bibitem [{\citenamefont {Didier}\ \emph {et~al.}(2019)\citenamefont {Didier},
  \citenamefont {Sete}, \citenamefont {Combes},\ and\ \citenamefont
  {da~Silva}}]{Rigetti_ac_sweet_spot}%
  \BibitemOpen
  \bibfield  {author} {\bibinfo {author} {\bibfnamefont {N.}~\bibnamefont
  {Didier}}, \bibinfo {author} {\bibfnamefont {E.~A.}\ \bibnamefont {Sete}},
  \bibinfo {author} {\bibfnamefont {J.}~\bibnamefont {Combes}},\ and\ \bibinfo
  {author} {\bibfnamefont {M.~P.}\ \bibnamefont {da~Silva}},\ }\bibfield
  {title} {\bibinfo {title} {\textit{ac Flux Sweet Spots in Parametrically
  Modulated Superconducting Qubits}},\ }\href
  {https://doi.org/10.1103/PhysRevApplied.12.054015} {\bibfield  {journal}
  {\bibinfo  {journal} {Phys. Rev. Applied}\ }\textbf {\bibinfo {volume}
  {12}},\ \bibinfo {pages} {054015} (\bibinfo {year} {2019})}\BibitemShut
  {NoStop}%
\bibitem [{\citenamefont {You}\ \emph {et~al.}(2007)\citenamefont {You},
  \citenamefont {Hu}, \citenamefont {Ashhab},\ and\ \citenamefont
  {Nori}}]{Nori_C_shunt_flux}%
  \BibitemOpen
  \bibfield  {author} {\bibinfo {author} {\bibfnamefont {J.~Q.}\ \bibnamefont
  {You}}, \bibinfo {author} {\bibfnamefont {X.}~\bibnamefont {Hu}}, \bibinfo
  {author} {\bibfnamefont {S.}~\bibnamefont {Ashhab}},\ and\ \bibinfo {author}
  {\bibfnamefont {F.}~\bibnamefont {Nori}},\ }\bibfield  {title} {\bibinfo
  {title} {Low-decoherence flux qubit},\ }\href
  {https://doi.org/10.1103/PhysRevB.75.140515} {\bibfield  {journal} {\bibinfo
  {journal} {Phys. Rev. B}\ }\textbf {\bibinfo {volume} {75}},\ \bibinfo
  {pages} {140515} (\bibinfo {year} {2007})}\BibitemShut {NoStop}%
\bibitem [{\citenamefont {Ficheux}\ \emph {et~al.}(2021)\citenamefont
  {Ficheux}, \citenamefont {Nguyen}, \citenamefont {Somoroff}, \citenamefont
  {Xiong}, \citenamefont {Nesterov}, \citenamefont {Vavilov},\ and\
  \citenamefont {Manucharyan}}]{Manucharyan_fluxonium_two_qubit_gate}%
  \BibitemOpen
  \bibfield  {author} {\bibinfo {author} {\bibfnamefont {Q.}~\bibnamefont
  {Ficheux}}, \bibinfo {author} {\bibfnamefont {L.~B.}\ \bibnamefont {Nguyen}},
  \bibinfo {author} {\bibfnamefont {A.}~\bibnamefont {Somoroff}}, \bibinfo
  {author} {\bibfnamefont {H.}~\bibnamefont {Xiong}}, \bibinfo {author}
  {\bibfnamefont {K.~N.}\ \bibnamefont {Nesterov}}, \bibinfo {author}
  {\bibfnamefont {M.~G.}\ \bibnamefont {Vavilov}},\ and\ \bibinfo {author}
  {\bibfnamefont {V.~E.}\ \bibnamefont {Manucharyan}},\ }\bibfield  {title}
  {\bibinfo {title} {\textit{Fast Logic with Slow Qubits: Microwave-Activated
  Controlled-Z Gate on Low-Frequency Fluxoniums}},\ }\href
  {https://doi.org/10.1103/PhysRevX.11.021026} {\bibfield  {journal} {\bibinfo
  {journal} {Phys. Rev. X}\ }\textbf {\bibinfo {volume} {11}},\ \bibinfo
  {pages} {021026} (\bibinfo {year} {2021})}\BibitemShut {NoStop}%
\bibitem [{\citenamefont {Pirkkalainen}\ \emph {et~al.}(2013)\citenamefont
  {Pirkkalainen}, \citenamefont {Cho}, \citenamefont {Li}, \citenamefont
  {Paraoanu}, \citenamefont {Hakonen},\ and\ \citenamefont
  {Sillanp{\"a}{\"a}}}]{Sillanpaa_hybrid_circuit}%
  \BibitemOpen
  \bibfield  {author} {\bibinfo {author} {\bibfnamefont {J.~M.}\ \bibnamefont
  {Pirkkalainen}}, \bibinfo {author} {\bibfnamefont {S.~U.}\ \bibnamefont
  {Cho}}, \bibinfo {author} {\bibfnamefont {J.}~\bibnamefont {Li}}, \bibinfo
  {author} {\bibfnamefont {G.~S.}\ \bibnamefont {Paraoanu}}, \bibinfo {author}
  {\bibfnamefont {P.~J.}\ \bibnamefont {Hakonen}},\ and\ \bibinfo {author}
  {\bibfnamefont {M.~A.}\ \bibnamefont {Sillanp{\"a}{\"a}}},\ }\bibfield
  {title} {\bibinfo {title} {\textit{Hybrid Circuit Cavity Quantum
  Electrodynamics with a Micromechanical Resonator}},\ }\href
  {https://doi.org/10.1038/nature11821} {\bibfield  {journal} {\bibinfo
  {journal} {Nature}\ }\textbf {\bibinfo {volume} {494}},\ \bibinfo {pages}
  {211} (\bibinfo {year} {2013})}\BibitemShut {NoStop}%
\bibitem [{\citenamefont {Gandon}\ \emph {et~al.}(2022)\citenamefont {Gandon},
  \citenamefont {Le~Calonnec}, \citenamefont {Shillito}, \citenamefont
  {Petrescu},\ and\ \citenamefont {Blais}}]{Blais_Floquet}%
  \BibitemOpen
  \bibfield  {author} {\bibinfo {author} {\bibfnamefont {A.}~\bibnamefont
  {Gandon}}, \bibinfo {author} {\bibfnamefont {C.}~\bibnamefont {Le~Calonnec}},
  \bibinfo {author} {\bibfnamefont {R.}~\bibnamefont {Shillito}}, \bibinfo
  {author} {\bibfnamefont {A.}~\bibnamefont {Petrescu}},\ and\ \bibinfo
  {author} {\bibfnamefont {A.}~\bibnamefont {Blais}},\ }\bibfield  {title}
  {\bibinfo {title} {\textit{Engineering, Control, and Longitudinal Readout of
  Floquet Qubits}},\ }\href {https://doi.org/10.1103/PhysRevApplied.17.064006}
  {\bibfield  {journal} {\bibinfo  {journal} {Phys. Rev. Applied}\ }\textbf
  {\bibinfo {volume} {17}},\ \bibinfo {pages} {064006} (\bibinfo {year}
  {2022})}\BibitemShut {NoStop}%
\bibitem [{\citenamefont {Di~Paolo}\ \emph {et~al.}()\citenamefont {Di~Paolo},
  \citenamefont {Leroux}, \citenamefont {Hazard}, \citenamefont {Serniak},
  \citenamefont {Gustavsson}, \citenamefont {Blais},\ and\ \citenamefont
  {Oliver}}]{Oliver_extensible_circuit}%
  \BibitemOpen
  \bibfield  {author} {\bibinfo {author} {\bibfnamefont {A.}~\bibnamefont
  {Di~Paolo}}, \bibinfo {author} {\bibfnamefont {C.}~\bibnamefont {Leroux}},
  \bibinfo {author} {\bibfnamefont {T.~M.}\ \bibnamefont {Hazard}}, \bibinfo
  {author} {\bibfnamefont {K.}~\bibnamefont {Serniak}}, \bibinfo {author}
  {\bibfnamefont {S.}~\bibnamefont {Gustavsson}}, \bibinfo {author}
  {\bibfnamefont {A.}~\bibnamefont {Blais}},\ and\ \bibinfo {author}
  {\bibfnamefont {W.~D.}\ \bibnamefont {Oliver}},\ }\bibfield  {title}
  {\bibinfo {title} {\textit{Extensible Aircuit-QED Architecture via Amplitude-
  and Frequency-Variable Microwaves} (2022)},\ }\Eprint
  {https://arxiv.org/abs/arXiv:2204.08098} {arXiv:2204.08098} \BibitemShut
  {NoStop}%
\bibitem [{\citenamefont {Breuer}\ and\ \citenamefont
  {Petruccione}(2007)}]{Breuer_open_quantum_system}%
  \BibitemOpen
  \bibfield  {author} {\bibinfo {author} {\bibfnamefont {H.}~\bibnamefont
  {Breuer}}\ and\ \bibinfo {author} {\bibfnamefont {F.}~\bibnamefont
  {Petruccione}},\ }\href@noop {} {\emph {\bibinfo {title} {``\textit{The
  Theory of Open Quantum Systems}''}}}\ (\bibinfo  {publisher} {Oxford
  University Press, New York},\ \bibinfo {year} {2007})\BibitemShut {NoStop}%
\bibitem [{\citenamefont {M\"uller}\ and\ \citenamefont
  {Stace}(2017)}]{Muller_Keldysh}%
  \BibitemOpen
  \bibfield  {author} {\bibinfo {author} {\bibfnamefont {C.}~\bibnamefont
  {M\"uller}}\ and\ \bibinfo {author} {\bibfnamefont {T.~M.}\ \bibnamefont
  {Stace}},\ }\bibfield  {title} {\bibinfo {title} {\textit{Deriving Lindblad
  Master Equations with Keldysh Diagrams: Correlated Gain and Loss in Higher
  Order Perturbation Theory}},\ }\href
  {https://doi.org/10.1103/PhysRevA.95.013847} {\bibfield  {journal} {\bibinfo
  {journal} {Phys. Rev. A}\ }\textbf {\bibinfo {volume} {95}},\ \bibinfo
  {pages} {013847} (\bibinfo {year} {2017})}\BibitemShut {NoStop}%
\bibitem [{Sup()}]{Supplementary}%
  \BibitemOpen
  \href@noop {} {}\bibinfo {note} {See Supplemental Material at [URL will be
  inserted by publisher] for the analytical derivation of the dephasing profile
  of the qubit around its sweet spot for the static and driven
  cases.}\BibitemShut {Stop}%
\bibitem [{\citenamefont {Cohen}\ \emph {et~al.}(2017)\citenamefont {Cohen},
  \citenamefont {Aharon},\ and\ \citenamefont {Retzker}}]{Cohen_drive_noise}%
  \BibitemOpen
  \bibfield  {author} {\bibinfo {author} {\bibfnamefont {I.}~\bibnamefont
  {Cohen}}, \bibinfo {author} {\bibfnamefont {N.}~\bibnamefont {Aharon}},\ and\
  \bibinfo {author} {\bibfnamefont {A.}~\bibnamefont {Retzker}},\ }\bibfield
  {title} {\bibinfo {title} {\textit{Continuous Dynamical Decoupling Utilizing
  Time-Dependent Detuning}},\ }\href
  {https://doi.org/https://doi.org/10.1002/prop.201600071} {\bibfield
  {journal} {\bibinfo  {journal} {Fortschr. Phys.}\ }\textbf {\bibinfo {volume}
  {65}},\ \bibinfo {pages} {1600071} (\bibinfo {year} {2017})}\BibitemShut
  {NoStop}%
\end{thebibliography}%


\begin{thebibliography}{12}%
\makeatletter
\providecommand \@ifxundefined [1]{%
 \@ifx{#1\undefined}
}%
\providecommand \@ifnum [1]{%
 \ifnum #1\expandafter \@firstoftwo
 \else \expandafter \@secondoftwo
 \fi
}%
\providecommand \@ifx [1]{%
 \ifx #1\expandafter \@firstoftwo
 \else \expandafter \@secondoftwo
 \fi
}%
\providecommand \natexlab [1]{#1}%
\providecommand \enquote  [1]{``#1''}%
\providecommand \bibnamefont  [1]{#1}%
\providecommand \bibfnamefont [1]{#1}%
\providecommand \citenamefont [1]{#1}%
\providecommand \href@noop [0]{\@secondoftwo}%
\providecommand \href [0]{\begingroup \@sanitize@url \@href}%
\providecommand \@href[1]{\@@startlink{#1}\@@href}%
\providecommand \@@href[1]{\endgroup#1\@@endlink}%
\providecommand \@sanitize@url [0]{\catcode `\\12\catcode `\$12\catcode
  `\&12\catcode `\#12\catcode `\^12\catcode `\_12\catcode `\%12\relax}%
\providecommand \@@startlink[1]{}%
\providecommand \@@endlink[0]{}%
\providecommand \url  [0]{\begingroup\@sanitize@url \@url }%
\providecommand \@url [1]{\endgroup\@href {#1}{\urlprefix }}%
\providecommand \urlprefix  [0]{URL }%
\providecommand \Eprint [0]{\href }%
\providecommand \doibase [0]{https://doi.org/}%
\providecommand \selectlanguage [0]{\@gobble}%
\providecommand \bibinfo  [0]{\@secondoftwo}%
\providecommand \bibfield  [0]{\@secondoftwo}%
\providecommand \translation [1]{[#1]}%
\providecommand \BibitemOpen [0]{}%
\providecommand \bibitemStop [0]{}%
\providecommand \bibitemNoStop [0]{.\EOS\space}%
\providecommand \EOS [0]{\spacefactor3000\relax}%
\providecommand \BibitemShut  [1]{\csname bibitem#1\endcsname}%
\let\auto@bib@innerbib\@empty
\bibitem [{\citenamefont {Wang}\ and\ \citenamefont
  {Clerk}(2020)}]{Clerk_non_Gaussian}%
  \BibitemOpen
  \bibfield  {author} {\bibinfo {author} {\bibfnamefont {Y.-X.}\ \bibnamefont
  {Wang}}\ and\ \bibinfo {author} {\bibfnamefont {A.~A.}\ \bibnamefont
  {Clerk}},\ }\bibfield  {title} {\bibinfo {title} {\textit{Spectral
  Characterization of Non-Gaussian Quantum Noise: Keldysh Approach and
  Application to Photon Shot Noise}},\ }\href
  {https://doi.org/10.1103/PhysRevResearch.2.033196} {\bibfield  {journal}
  {\bibinfo  {journal} {Phys. Rev. Research}\ }\textbf {\bibinfo {volume}
  {2}},\ \bibinfo {pages} {033196} (\bibinfo {year} {2020})}\BibitemShut
  {NoStop}%
\bibitem [{\citenamefont {Makhlin}\ and\ \citenamefont
  {Shnirman}(2003)}]{Shnirman_Keldysh_dephasing_1}%
  \BibitemOpen
  \bibfield  {author} {\bibinfo {author} {\bibfnamefont {Y.}~\bibnamefont
  {Makhlin}}\ and\ \bibinfo {author} {\bibfnamefont {A.}~\bibnamefont
  {Shnirman}},\ }\bibfield  {title} {\bibinfo {title} {\textit{Dephasing of
  Qubits by Transverse Low-Frequency Noise}},\ }\href
  {https://doi.org/10.1134/1.1637702} {\bibfield  {journal} {\bibinfo
  {journal} {J. Exp. Theor. Phys.}\ }\textbf {\bibinfo {volume} {78}},\
  \bibinfo {pages} {497} (\bibinfo {year} {2003})}\BibitemShut {NoStop}%
\bibitem [{\citenamefont {Makhlin}\ and\ \citenamefont
  {Shnirman}(2004)}]{Shnirman_Keldysh_dephasing}%
  \BibitemOpen
  \bibfield  {author} {\bibinfo {author} {\bibfnamefont {Y.}~\bibnamefont
  {Makhlin}}\ and\ \bibinfo {author} {\bibfnamefont {A.}~\bibnamefont
  {Shnirman}},\ }\bibfield  {title} {\bibinfo {title} {\textit{Dephasing of
  Solid-State Qubits at Optimal Points}},\ }\href
  {https://doi.org/10.1103/PhysRevLett.92.178301} {\bibfield  {journal}
  {\bibinfo  {journal} {Phys. Rev. Lett.}\ }\textbf {\bibinfo {volume} {92}},\
  \bibinfo {pages} {178301} (\bibinfo {year} {2004})}\BibitemShut {NoStop}%
\bibitem [{\citenamefont {Koch}\ \emph {et~al.}(2007)\citenamefont {Koch},
  \citenamefont {Yu}, \citenamefont {Gambetta}, \citenamefont {Houck},
  \citenamefont {Schuster}, \citenamefont {Majer}, \citenamefont {Blais},
  \citenamefont {Devoret}, \citenamefont {Girvin},\ and\ \citenamefont
  {Schoelkopf}}]{Koch_transmon_theory}%
  \BibitemOpen
  \bibfield  {author} {\bibinfo {author} {\bibfnamefont {J.}~\bibnamefont
  {Koch}}, \bibinfo {author} {\bibfnamefont {T.~M.}\ \bibnamefont {Yu}},
  \bibinfo {author} {\bibfnamefont {J.}~\bibnamefont {Gambetta}}, \bibinfo
  {author} {\bibfnamefont {A.~A.}\ \bibnamefont {Houck}}, \bibinfo {author}
  {\bibfnamefont {D.~I.}\ \bibnamefont {Schuster}}, \bibinfo {author}
  {\bibfnamefont {J.}~\bibnamefont {Majer}}, \bibinfo {author} {\bibfnamefont
  {A.}~\bibnamefont {Blais}}, \bibinfo {author} {\bibfnamefont {M.~H.}\
  \bibnamefont {Devoret}}, \bibinfo {author} {\bibfnamefont {S.~M.}\
  \bibnamefont {Girvin}},\ and\ \bibinfo {author} {\bibfnamefont {R.~J.}\
  \bibnamefont {Schoelkopf}},\ }\bibfield  {title} {\bibinfo {title}
  {\textit{Charge-Insensitive Qubit Design Derived from the Cooper Pair Box}},\
  }\href {https://doi.org/10.1103/PhysRevA.76.042319} {\bibfield  {journal}
  {\bibinfo  {journal} {Phys. Rev. A}\ }\textbf {\bibinfo {volume} {76}},\
  \bibinfo {pages} {042319} (\bibinfo {year} {2007})}\BibitemShut {NoStop}%
\bibitem [{\citenamefont {Ithier}\ \emph {et~al.}(2005)\citenamefont {Ithier},
  \citenamefont {Collin}, \citenamefont {Joyez}, \citenamefont {Meeson},
  \citenamefont {Vion}, \citenamefont {Esteve}, \citenamefont {Chiarello},
  \citenamefont {Shnirman}, \citenamefont {Makhlin}, \citenamefont {Schriefl},\
  and\ \citenamefont {Sch\"on}}]{Ithier_decoherence_analysis}%
  \BibitemOpen
  \bibfield  {author} {\bibinfo {author} {\bibfnamefont {G.}~\bibnamefont
  {Ithier}}, \bibinfo {author} {\bibfnamefont {E.}~\bibnamefont {Collin}},
  \bibinfo {author} {\bibfnamefont {P.}~\bibnamefont {Joyez}}, \bibinfo
  {author} {\bibfnamefont {P.~J.}\ \bibnamefont {Meeson}}, \bibinfo {author}
  {\bibfnamefont {D.}~\bibnamefont {Vion}}, \bibinfo {author} {\bibfnamefont
  {D.}~\bibnamefont {Esteve}}, \bibinfo {author} {\bibfnamefont
  {F.}~\bibnamefont {Chiarello}}, \bibinfo {author} {\bibfnamefont
  {A.}~\bibnamefont {Shnirman}}, \bibinfo {author} {\bibfnamefont
  {Y.}~\bibnamefont {Makhlin}}, \bibinfo {author} {\bibfnamefont
  {J.}~\bibnamefont {Schriefl}},\ and\ \bibinfo {author} {\bibfnamefont
  {G.}~\bibnamefont {Sch\"on}},\ }\bibfield  {title} {\bibinfo {title}
  {\textit{Decoherence in a Superconducting Quantum Bit Circuit}},\ }\href
  {https://doi.org/10.1103/PhysRevB.72.134519} {\bibfield  {journal} {\bibinfo
  {journal} {Phys. Rev. B}\ }\textbf {\bibinfo {volume} {72}},\ \bibinfo
  {pages} {134519} (\bibinfo {year} {2005})}\BibitemShut {NoStop}%
\bibitem [{\citenamefont {M\"uller}\ and\ \citenamefont
  {Stace}(2017)}]{Muller_Keldysh}%
  \BibitemOpen
  \bibfield  {author} {\bibinfo {author} {\bibfnamefont {C.}~\bibnamefont
  {M\"uller}}\ and\ \bibinfo {author} {\bibfnamefont {T.~M.}\ \bibnamefont
  {Stace}},\ }\bibfield  {title} {\bibinfo {title} {\textit{Deriving Lindblad
  Master Equations with Keldysh Diagrams: Correlated Gain and Loss in Higher
  Order Perturbation Theory}},\ }\href
  {https://doi.org/10.1103/PhysRevA.95.013847} {\bibfield  {journal} {\bibinfo
  {journal} {Phys. Rev. A}\ }\textbf {\bibinfo {volume} {95}},\ \bibinfo
  {pages} {013847} (\bibinfo {year} {2017})}\BibitemShut {NoStop}%
\bibitem [{\citenamefont {Huang}\ \emph {et~al.}(2021)\citenamefont {Huang},
  \citenamefont {Mundada}, \citenamefont {Gyenis}, \citenamefont {Schuster},
  \citenamefont {Houck},\ and\ \citenamefont {Koch}}]{Dynamical_sweet_spot}%
  \BibitemOpen
  \bibfield  {author} {\bibinfo {author} {\bibfnamefont {Z.}~\bibnamefont
  {Huang}}, \bibinfo {author} {\bibfnamefont {P.~S.}\ \bibnamefont {Mundada}},
  \bibinfo {author} {\bibfnamefont {A.}~\bibnamefont {Gyenis}}, \bibinfo
  {author} {\bibfnamefont {D.~I.}\ \bibnamefont {Schuster}}, \bibinfo {author}
  {\bibfnamefont {A.~A.}\ \bibnamefont {Houck}},\ and\ \bibinfo {author}
  {\bibfnamefont {J.}~\bibnamefont {Koch}},\ }\bibfield  {title} {\bibinfo
  {title} {\textit{Engineering Dynamical Sweet Spots to Protect Qubits from
  $1/f$ Noise}},\ }\href {https://doi.org/10.1103/PhysRevApplied.15.034065}
  {\bibfield  {journal} {\bibinfo  {journal} {Phys. Rev. Applied}\ }\textbf
  {\bibinfo {volume} {15}},\ \bibinfo {pages} {034065} (\bibinfo {year}
  {2021})}\BibitemShut {NoStop}%
\bibitem [{\citenamefont {Bergli}\ \emph {et~al.}(2009)\citenamefont {Bergli},
  \citenamefont {Galperin},\ and\ \citenamefont
  {Altshuler}}]{Altshuler_telegraph_TLF}%
  \BibitemOpen
  \bibfield  {author} {\bibinfo {author} {\bibfnamefont {J.}~\bibnamefont
  {Bergli}}, \bibinfo {author} {\bibfnamefont {Y.~M.}\ \bibnamefont
  {Galperin}},\ and\ \bibinfo {author} {\bibfnamefont {B.~L.}\ \bibnamefont
  {Altshuler}},\ }\bibfield  {title} {\bibinfo {title} {\textit{Decoherence in
  Qubits Due to Low-Frequency Noise}},\ }\href
  {https://doi.org/10.1088/1367-2630/11/2/025002} {\bibfield  {journal}
  {\bibinfo  {journal} {New J. Phys.}\ }\textbf {\bibinfo {volume} {11}},\
  \bibinfo {pages} {025002} (\bibinfo {year} {2009})}\BibitemShut {NoStop}%
\bibitem [{\citenamefont {You}\ \emph {et~al.}(2021)\citenamefont {You},
  \citenamefont {Clerk},\ and\ \citenamefont {Koch}}]{Koch_TLS_FD_thoerem}%
  \BibitemOpen
  \bibfield  {author} {\bibinfo {author} {\bibfnamefont {X.}~\bibnamefont
  {You}}, \bibinfo {author} {\bibfnamefont {A.~A.}\ \bibnamefont {Clerk}},\
  and\ \bibinfo {author} {\bibfnamefont {J.}~\bibnamefont {Koch}},\ }\bibfield
  {title} {\bibinfo {title} {\textit{Positive- and Negative-Frequency Noise
  from an Ensemble of Two-Level Fluctuators}},\ }\href
  {https://doi.org/10.1103/PhysRevResearch.3.013045} {\bibfield  {journal}
  {\bibinfo  {journal} {Phys. Rev. Research}\ }\textbf {\bibinfo {volume}
  {3}},\ \bibinfo {pages} {013045} (\bibinfo {year} {2021})}\BibitemShut
  {NoStop}%
\bibitem [{\citenamefont {Galperin}\ \emph {et~al.}(2006)\citenamefont
  {Galperin}, \citenamefont {Altshuler}, \citenamefont {Bergli},\ and\
  \citenamefont {Shantsev}}]{Galperin_Echo_TLFs}%
  \BibitemOpen
  \bibfield  {author} {\bibinfo {author} {\bibfnamefont {Y.~M.}\ \bibnamefont
  {Galperin}}, \bibinfo {author} {\bibfnamefont {B.~L.}\ \bibnamefont
  {Altshuler}}, \bibinfo {author} {\bibfnamefont {J.}~\bibnamefont {Bergli}},\
  and\ \bibinfo {author} {\bibfnamefont {D.~V.}\ \bibnamefont {Shantsev}},\
  }\bibfield  {title} {\bibinfo {title} {\textit{Non-Gaussian Low-Frequency
  Noise as a Source of Qubit Decoherence}},\ }\href
  {https://doi.org/10.1103/PhysRevLett.96.097009} {\bibfield  {journal}
  {\bibinfo  {journal} {Phys. Rev. Lett.}\ }\textbf {\bibinfo {volume} {96}},\
  \bibinfo {pages} {097009} (\bibinfo {year} {2006})}\BibitemShut {NoStop}%
\bibitem [{\citenamefont {McCourt}\ \emph {et~al.}()\citenamefont {McCourt},
  \citenamefont {Neill}, \citenamefont {Lee}, \citenamefont {Quintana},
  \citenamefont {Chen}, \citenamefont {Kellyand}, \citenamefont {Smelyanskiy},
  \citenamefont {Dykman}, \citenamefont {Korotkov}, \citenamefont {Chuang},\
  and\ \citenamefont {Petukhov}}]{Petukhov_nonGaussianDD}%
  \BibitemOpen
  \bibfield  {author} {\bibinfo {author} {\bibfnamefont {T.}~\bibnamefont
  {McCourt}}, \bibinfo {author} {\bibfnamefont {C.}~\bibnamefont {Neill}},
  \bibinfo {author} {\bibfnamefont {K.}~\bibnamefont {Lee}}, \bibinfo {author}
  {\bibfnamefont {C.}~\bibnamefont {Quintana}}, \bibinfo {author}
  {\bibfnamefont {Y.}~\bibnamefont {Chen}}, \bibinfo {author} {\bibfnamefont
  {J.}~\bibnamefont {Kellyand}}, \bibinfo {author} {\bibfnamefont {V.~N.}\
  \bibnamefont {Smelyanskiy}}, \bibinfo {author} {\bibfnamefont {M.~I.}\
  \bibnamefont {Dykman}}, \bibinfo {author} {\bibfnamefont {A.}~\bibnamefont
  {Korotkov}}, \bibinfo {author} {\bibfnamefont {I.~L.}\ \bibnamefont
  {Chuang}},\ and\ \bibinfo {author} {\bibfnamefont {A.~G.}\ \bibnamefont
  {Petukhov}},\ }\bibfield  {title} {\bibinfo {title} {\textit{Learning Noise
  via Dynamical Decoupling of Entangled Qubits } (2022)},\ }\Eprint
  {https://arxiv.org/abs/arXiv:2201.11173} {arXiv:2201.11173} \BibitemShut
  {NoStop}%
\bibitem [{\citenamefont {Itakura}\ and\ \citenamefont
  {Tokura}(2003)}]{Tokura_dephasing}%
  \BibitemOpen
  \bibfield  {author} {\bibinfo {author} {\bibfnamefont {T.}~\bibnamefont
  {Itakura}}\ and\ \bibinfo {author} {\bibfnamefont {Y.}~\bibnamefont
  {Tokura}},\ }\bibfield  {title} {\bibinfo {title} {\textit{Dephasing due to
  Background Charge Fluctuations}},\ }\href
  {https://doi.org/10.1103/PhysRevB.67.195320} {\bibfield  {journal} {\bibinfo
  {journal} {Phys. Rev. B}\ }\textbf {\bibinfo {volume} {67}},\ \bibinfo
  {pages} {195320} (\bibinfo {year} {2003})}\BibitemShut {NoStop}%
\end{thebibliography}%
\end{document}


\title{Supplemental Material for ``High-Order Qubit Dephasing at Sweet Spots by\\ Non-Gaussian Fluctuators: Symmetry Breaking and Floquet Protection''}
\author{Ziwen Huang}
\address{Superconducting Quantum Materials and Systems Center,
Fermi National Accelerator Laboratory (FNAL), Batavia, IL 60510, USA}
\author{Xinyuan You}
\address{Superconducting Quantum Materials and Systems Center,
Fermi National Accelerator Laboratory (FNAL), Batavia, IL 60510, USA}
\author{Ugur Alyanak}
\address{Superconducting Quantum Materials and Systems Center,
Fermi National Accelerator Laboratory (FNAL), Batavia, IL 60510, USA}
\author{Alexander Romanenko}
\address{Superconducting Quantum Materials and Systems Center,
Fermi National Accelerator Laboratory (FNAL), Batavia, IL 60510, USA}
\author{Anna Grassellino}
\address{Superconducting Quantum Materials and Systems Center,
Fermi National Accelerator Laboratory (FNAL), Batavia, IL 60510, USA}
\author{Shaojiang Zhu}
\address{Superconducting Quantum Materials and Systems Center,
Fermi National Accelerator Laboratory (FNAL), Batavia, IL 60510, USA}
\maketitle
\setcounter{page}{8}
\thispagestyle{headings}
\usetagform{supplementary}
\section{Model}
\subsection{Qubit}
This section provides more details of the qubit and the fluctuators studied in the main text. As a recap of the main text, the Hamiltonian of the qubit is $\hat{H}_q(\lambda) = \hat{H}_{q}(0) + \lambda \hat{x}$, where $\lambda$ is a control parameter of the qubit. The Hamiltonian $\hat{H}_{q}(0)$ satisfies a $\mathbb{Z}_2$ symmetry, i.e., $\hat{R}\hat{H}_{q}(0)\hat{R}^\dagger = \hat{H}_{q}(0)$, where $\hat{R}$ is defined by the transformation $\hat{R}\hat{x}\hat{R}^\dagger = -\hat{x}$. Using this property, one can prove
\begin{align}
    (\partial\omega^{\mathrm{bare}}_{ge}/\partial \lambda)|_{\lambda = 0} =\langle e\vert \hat{x}\vert e \rangle|_
    {\lambda = 0}- \langle g\vert \hat{x}\vert g \rangle|_{\lambda = 0} = 0.
\end{align}
The last equation holds because the $\mathbb{Z}_2$ symmetry renders $\langle j\vert \hat{x}\vert j\rangle _{\lambda = 0} = -\langle j\vert \hat{R}^\dagger\hat{x}\hat{R}\vert j\rangle _{\lambda = 0} = -\langle j\vert \hat{x}\vert j\rangle _{\lambda = 0} = 0$ for any non-degenerate eigenstates $\vert j\rangle$ with eigenenergy  $\omega^{\mathrm{bare}}_j$ of bare Hamiltonian $\hat{H}_{q}(\lambda)$. (We use $j$ to index the eigenstates.)

Important for the discussion below, here we evaluate perturbatively the matrix elements of $\hat{x}$ for $\lambda\approx 0$. We find
\begin{align}
    \vert j\rangle_{\lambda} = \vert j\rangle_{\lambda=0} + \lambda\sum_{j'\neq j}\left(\frac{x_{j'j}}{\omega^{\mathrm{bare}}_j-\omega^{\mathrm{bare}}_{j'}}\vert j'\rangle\right)\Bigg|_{\lambda=0} + O(\lambda^2),
\end{align}
where we define $x_{j'j}\equiv \langle j'\vert \hat{x}\vert j\rangle$. This equation approximates $x_{gg}|_{\lambda} \approx \sum_{j\neq g}2\lambda [|x_{gj}|^2/(\omega^{\mathrm{bare}}_g - \omega^{\mathrm{bare}}_j)]|_{\lambda=0}$ as well as $x_{ee}|_{\lambda}$ by switching $e$ and $g$.

\subsection{Two-level fluctuators}
This subsection focuses on the correlation functions of a single TLF. As a recap of the main text, the TLF $\delta\xi_T(t)$ can only take two values, $\pm |\xi_{T}| - \overline{\xi}_{T}$, and the probabilities of finding the TLF in these states are $P_{\pm}$. (We omit the subscript $\mu$ that indexes the TLFs, since only one TLF is under investigation.) The coefficient $\overline{\xi}_T$ is set by $\overline{\xi}_T = |\xi_T|(P_+-P_-)$ to ensure $\overline{\delta\xi(t)} =0$. The rate of the downward (upward) flipping is denoted by $\kappa_{-(+)}$. The state probabilities and flipping rates satisfy  $\kappa_{-}P_+=\kappa_{+}P_-$. For convenience, we denote the sum rate by $\kappa \equiv \kappa_{+}+\kappa_{-}$. 

To derive the correlation functions, we find it convenient to first study the equal-amplitude but nonzero-average TLF: $\xi_T(t)\equiv \delta\xi_T(t)+\overline{\xi}_T$. If we can find the correlation functions of $\xi_T(t)$, those of $\delta\xi_T(t)$ can be easily derived. An important fact for $\xi_T(t)$ is that, if we find the TLF $\xi_T(t)$ in $+|\xi_T|$ state at time $t$, then at time $t+\Delta t$ ($\Delta t>0$), the probability of finding it still in the same state is $f_1(\Delta t)\equiv P_{-}e^{-\kappa\Delta t}+P_+$. Using this relation, we can derive the multi-time correlation functions as  products of such probabilities based on the formula for conditional probability:
\begin{align}
    &\overline{\Big[\xi_{T}(t)\!+\!|\xi_T|\Big]\Big[\xi_{T}(t_1)\!+\!|\xi_T|\Big]\cdots\Big[\xi_{T}(t_n)\!+\!|\xi_T|\Big]\Big[\xi_{T}(0)\!+\!|\xi_T|\Big]} \nonumber\\
    &=|2\xi_T|^{n+1}P_+ f_1(t-t_1)f_1(t_1-t_2)\cdots f_1(t_n-0).\label{eq:TLFcorrelation}
\end{align}
Note that the times shown above are ordered by $t\geq t_1 \geq\cdots\geq t_n\geq 0$. Then, the correlation functions of $\xi_T(t)$ and $\delta\xi_T(t)$ can be obtained to arbitrary orders. Specifically, we show the results up to the fourth order:
\begin{align}
&\overline{\delta\xi_T(t)} = 0,\label{eq:correlations}\\
    &\overline{\delta\xi_{T}(t)\delta\xi_{T}(0)} =|2\xi_T|^2P_{-}P_+ e^{-\kappa t},\nonumber\\
    &\overline{\delta\xi_{T}(t)\delta\xi_{T}(t_1)\delta\xi_{T}(0)} = -|2\xi_T|^3(P_+\!-\!P_{-})P_{-}P_+e^{-\kappa t},\nonumber\\
    &\overline{\delta\xi_{T}(t)\delta\xi_{T}(t_1)\delta\xi_{T}(t_2)\delta\xi_{T}(0)}=|2\xi_T|^4(P_{-}P_+)^2e^{-\kappa(t-t_1)\!-\!\kappa t_2}\nonumber\\
    &\,\,\qquad\qquad\qquad\qquad\qquad\quad~~+|2\xi_T|^4P_{-}P_+(P_+\!-\!P_{-})^2 e^{-\kappa t}.\nonumber
\end{align}
Obviously, the three-time correlation function of the TLF is nonzero in general, while such function of a Gaussian fluctuator is strictly zero.

Although a single TLF shows clear non-Gaussian features, a sufficiently large number of independent TLFs can be approximated as Gaussian noise according to the central limit theorem. Therefore,  the Gaussian noise in our numerical simulation is emulated by 2001 independent weak TLFs. Using this Gaussian-noise model, we find great agreement between the  numerical simulation and the analytical prediction.

\section{Keldysh Formalism}
\subsection{General Keldysh discussion}
The Keldysh technique is a powerful tool that evaluates the formal expression of ${\hat{\rho}}_I(t) = \overline{\hat{U}^\dagger_I(t)\hat{\rho}_I(0)\hat{U}_I(t)}$ perturbatively and diagrammatically. This perturbative treatment starts with the expansion of $\tilde{U}_I(t) \equiv \mathcal{T}\!\exp[-i\int_0^tdt'\tilde{H}_I(t')]$ in powers of the perturbation Hamiltonian  $\tilde{H}_I(t) = \hat{U}_0^\dagger(t)\hat{H}_I(t)\hat{U}_0(t)$, where the bare propagator is $\hat{U}_0(t) \equiv \exp[-i\hat{H}_q(\lambda)t]$ and the noise coupling term is $\hat{H}_I(t) = \delta\xi(t) \hat{x}$. [This specific choice of $\hat{U}_0(t)$ is suitable for describing the free-induced (Ramsey) dephasing of the qubit.] The expansion is $\tilde{U}_I(t) = \sum_\nu \tilde{U}_I^\nu(t)$, where the $\nu$th term $\tilde{U}_I^\nu(t)$ is given by
\begin{align}
    \tilde{U}^\nu_I(t)\!=&(-i)^\nu\!\int_0^t\! dt_1 \tilde{H}_I(t_1)\int_0^{t_1} \tilde{H}_I(t_2) dt_2\cdots\nonumber\\
    &\times\int_0^{t_{\nu-1}} \tilde{H}_I(t_\nu)dt_\nu.
    \label{eq:prop_nu}
\end{align}
It contains a product of $\nu$ times of $\tilde{H}_I$. 

Using Eq.~\eqref{eq:prop_nu}, we expand the density matrix elements as
\begin{align}
{\rho}_{I,jk}(t) =\!\sum_{j'k'\nu}\!{\rho}_{I,j'k'}(0)\,\Pi^{\nu}_{jk\leftarrow j'k'}(t),
\label{eq:rho_exp}
\end{align}
where the $\nu$th order Keldysh projector is defined by $\Pi^{\nu}_{jk\leftarrow j'k'}(t) \equiv \sum_{\nu'+\nu'' = \nu}\overline{\langle j\vert \tilde{U}^{\nu'}_I(t) \vert j'\rangle \langle k'\vert \, \tilde{U}^{\nu''\dagger}_I(t)\vert k\rangle}|$. With no noise coupling, we have $\Pi^{(0)}_{jk\leftarrow j'k'}(t)=\delta_{j,j'}\delta_{k,k'}$, which is also the zeroth-order term in the expansion \eqref{eq:rho_exp}. Note that not all the projectors are important. Because we are interested in the dephasing of the qubit, which is related to the decay of the off-diagonal matrix element, we only focus on the projectors $\Pi^{\nu}_{eg\leftarrow jk}(t)$.  To measure the evolution of  $\rho_{I,eg}(t)$, the qubit is usually prepared in an equal superposition state $\vert e\rangle$ and $\vert g\rangle$, which sets  $\rho_{I,eg}(0) = \rho_{I,ge}(0) = \rho_{I,gg}(0) =\rho_{I,ee}(0) = 1/2$ and other initial matrix elements zero. This specific protocol further restricts our attention to projectors with $j=e,g$ and $k=e,g$.

To extract the noise-mediated qubit frequency and the dephasing rate, the quantity we focus on is the exponent of the evolution  $-i\omega'_q t-\Phi(t)$ in $\rho_{eg}(t)\approx\rho_{eg}(0)\exp[-i\omega'_q t-\Phi(t)]$. In the literature on Keldysh formalism \cite{Clerk_non_Gaussian,Shnirman_Keldysh_dephasing_1,Shnirman_Keldysh_dephasing}, this exponent is related to the self energy. For the specific measurement protocol we describe above, the “self energy” in this paper is defined as
\begin{align}
    \Sigma_{eg}(t) \equiv& \,\mathrm{ln}\left[\sum_{i,j=e,g} \Pi_{eg\leftarrow ij}(t)\right]\label{eq:self-energy}\\
    =&\!\!\!\sum_{i,j=e,g\,\,\,\!}\!\!\sum_{\nu>0}\Pi^{(\nu)}_{eg\leftarrow ij}(t) \!-\! \frac{1}{2} \left[\!\sum_{i,j=e,g\,\,\,\!}\!\!\sum_{\nu>0}\Pi^{(\nu)}_{eg\leftarrow ij}(t)\right]^2\!+\cdots\nonumber
\end{align}
To obtain useful expressions for the dephasing profile and qubit frequency, in the next subsection we introduce a truncation and approximation protocol to perturbatively evaluate formal expansion above.

\subsection{Truncation and approximation}
In the main text, we mainly focus on the parameter region $\lambda \sim |\delta\xi(t)|\equiv \Big[\overline{\delta\xi^2(t)}\Big]^{\frac{1}{2}}$, where we find the mismatch of the extrema of the qubit frequency and dephasing rate. In this parameter regime, the quantities $\delta\xi(t)$ and $\lambda$ should be treated on equal footing in the perturbation theory. To track the order of the magnitude of the terms we derive in the following, we add a prefactor $\epsilon$ before the two small parameters $\lambda$ and $\delta\xi(t)$, i.e., $\lambda\rightarrow \epsilon\lambda$ and $\delta\xi(t)\rightarrow\epsilon\delta\xi(t)$, and finally set $\epsilon=1$. 

Under this ordering scheme, the real and imaginary parts of $\Sigma_{eg}(t)$ are found to be of different orders, i.e.,
\begin{align}
    \mathrm{Re}\{\Sigma_{eg}(t)\}\sim O(\epsilon^4),\quad\mathrm{Im}\{\Sigma_{eg}(t)\}\sim O(\epsilon^2).\nonumber
\end{align}
Beyond the leading orders, we can certainly use \eqref{eq:self-energy} to include higher-order contributions, but we find it difficult to cast the high-order terms into simple expressions with clear physical meaning. Therefore, we will approximate the real and imaginary parts of $\Sigma_{eg}(t)$ only to their respective leading order in $\epsilon$, which we find sufficient to capture the quantitative behaviors of the qubit frequency and dephasing rate around the sweet spot. Because of this approximation scheme, we only need to keep the Keldysh expansion $\Pi^{\nu}_{eg\leftarrow ij}(t)$ up to $\nu = 4$, and only require imaginary parts up to  $\nu = 2$. This approximation scheme is nicknamed as “$\epsilon${R4I2}”, convenient for referencing below.  Using the assumption of the low-frequency noise, we can further neglect the contributions from $\Pi_{eg\leftarrow ij}(t)$ for $i\neq e$ or $j\neq g$ in this approximation scheme  -- their contributions are oscillatory and do not grow with time. This further focus our attention on $\Pi_{eg\leftarrow eg}(t)$.

\subsection{Two-point Keldysh diagrams}
\begin{figure}
    \centering
    \includegraphics[width = 16.0cm]{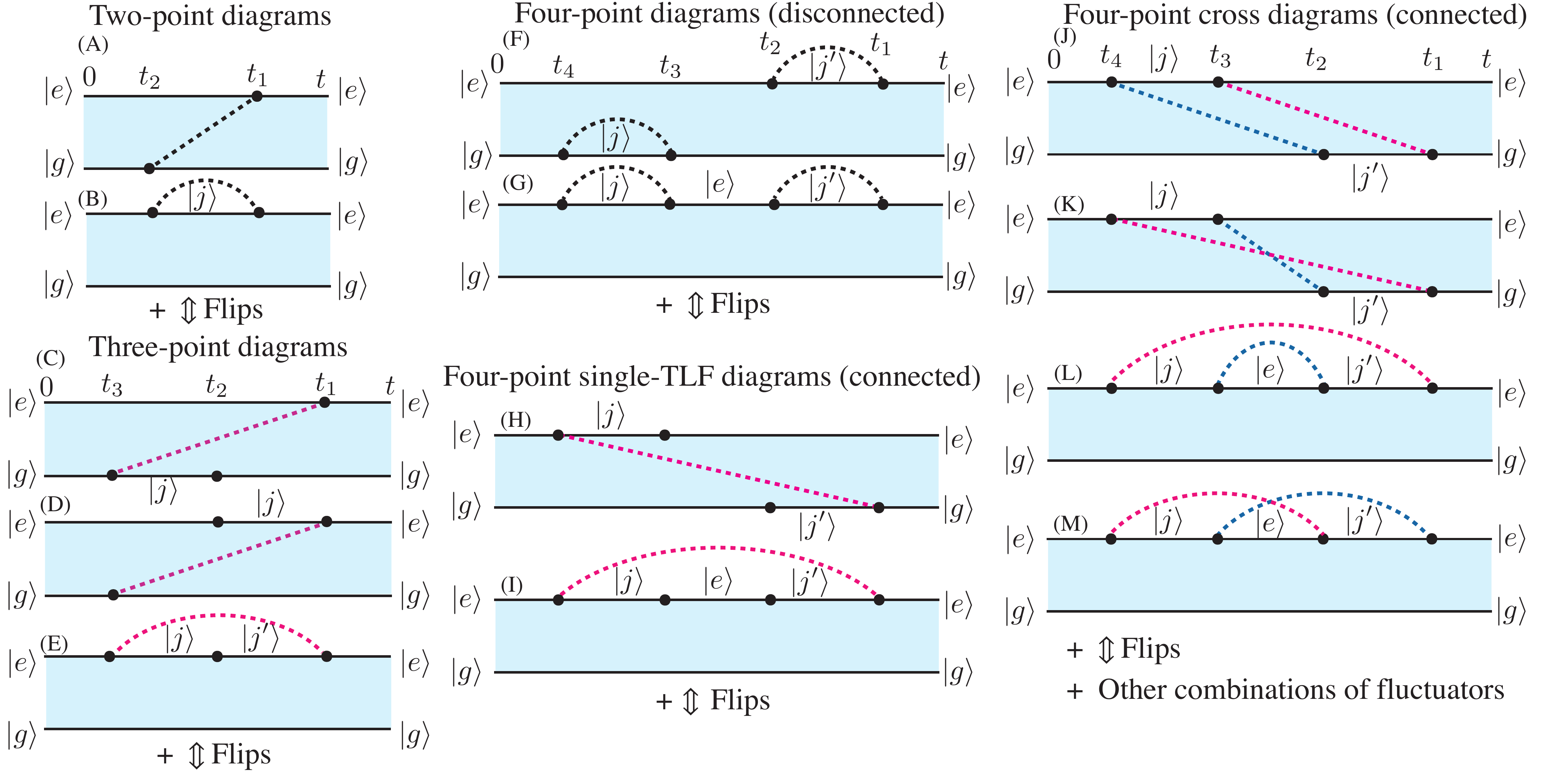}
    \caption{Keldysh diagrams for $\Pi^{\nu=2,3,4}_{eg\leftarrow eg}(t)$ that are relevant under the $\epsilon$R4I2 approximation scheme. The black dashed lines represent the correlation functions of the full fluctuation $\delta\xi(t)$, while the blue and pink lines represent those of the Gaussian fluctuator, $\delta\xi_G(t)$, and a single TLF, $\delta\xi_{T\!\mu}(t)$.}
    \label{fig:Keldysh}
\end{figure}

In this and the following subsections, we expand the projectors according to Eq.~\eqref{eq:prop_nu} for the concrete noise model considered in the main text. Each subsection addresses one specific type of diagrams. We start with the two-point ones due to the vanishing of $\overline{\delta\xi_T(t)}$. There are in total four such diagrams. Two of them are shown in FIG.~S\ref{fig:Keldysh} (A) and (B), while the remaining can be obtained by flipping the dots between the upper and lower branches. 

To help readers better follow our derivation, we next show calculation details of (A) for example. The two-point diagram, FIG.~S\ref{fig:Keldysh} (A), corresponds to the integral
\begin{align}
    \text{(A)}\! =&(-i)(i)\epsilon^2\!\int_{0}^t\!\!\!dt_1\!\int_{0}^{t_1}\!\!\!dt_2{\langle e\vert \tilde{x}(t_2)\vert e\rangle}{\langle g\vert \tilde{x}(t_1)\vert g\rangle}\overline{\delta\xi(t_1)\delta\xi(t_2)}\nonumber\\
    =&\epsilon^2\!\!\int_{0}^t\!\!\!dt_1\!\int_{0}^{t_1}\!\!\!dt_2x_{ee}x_{gg}\overline{\delta\xi(t_1)\delta\xi(t_2)}\nonumber\\
    =&\epsilon^2\!\int_{0}^t\!\!\!dt_1\!\int_{0}^{t_1}\!\!\!dt_2x_{ee}x_{gg}\int_{-\infty}^{\infty}\frac{d\omega}{2\pi} S(\omega)e^{-i\omega(t_1-t_2)}\nonumber\\
    =&\epsilon^2\!\,x_{ee}x_{gg}\!\int_{-\infty}^{\infty}\!\!\!\frac{d\omega}{2\pi} S(\omega)[K_R(\omega,t)\!+\!iK_I(\omega,t)].
    \label{eq:(A)}
\end{align}
Above, $\delta\xi(t) =  \delta\xi_G(t) +\sum_\mu \delta\xi_{T\!\mu}(t)$ is the full fluctuator, whose two-point spectrum is $S(\omega)\equiv S_G(\omega)+\sum_{\mu}S_{\mu}(\omega)$. The filter functions $K^R(\omega,t)$ and $K^I(\omega,t)$ are real and imaginary parts of the integral $\int_0^tdt_1\int_0^{t_1}dt_2e^{-i\omega(t_1-t_2)}$, which are derived as
$K^R(\omega,t) \equiv t^2 \mathrm{sinc}^2\left(\omega t/2\right)/2$ and $K^I(\omega,t) \equiv -(t/\omega) [1- \mathrm{sinc}(\omega t)]$. The rotated operator $\tilde{x}(t)$ is given by 
\begin{align}
\tilde{x}(t) \equiv \hat{U}_0^\dagger(t)\hat{x}\hat{U}_0(t) = \sum x_{jj'} \vert j\rangle\langle j'\vert  e^{i(\omega_j-\omega_{j'})t}.
\label{eq:rot_x}
\end{align}

\onecolumngrid
\newpage

\twocolumngrid

\begin{figure}[t!]
    \centering
    \includegraphics[width =7.1 cm ]{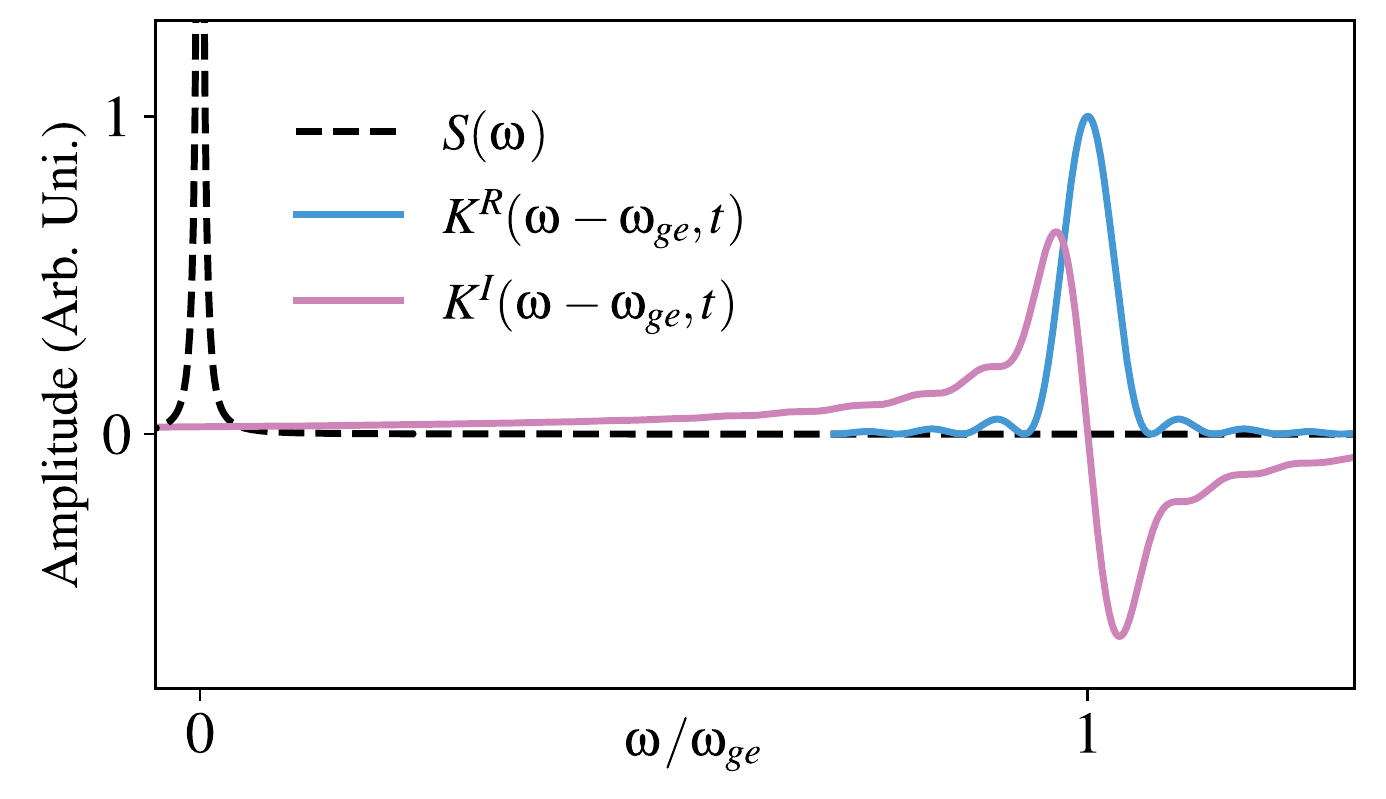}
    \caption{Filter functions $K^R(\omega - \omega_{ge},t)$ and $K^I(\omega - \omega_{ge},t)$, and the noise spectrum $S(\omega)$. We choose $t=1\,\mu$s for this plot, and the $\omega_{ge}/2\pi\approx 14$ MHz. The peak widths of the filter functions are given by $\sim 2\pi/t$. As time grows, the filter functions will turn narrower.}
    \label{fig:FilterFunctions}
\end{figure}

Following the same procedure,  (B) gives us
\begin{align}
    \text{(B)} = &\,(-i)^2\epsilon^2|x_{ej}|^2\!\int_{-\infty}^{\infty}\!\frac{d\omega}{2\pi}S(\omega)\nonumber\\
    &\times [K_R(\omega\!+\!\omega_j\!-\!\omega_e,t)\!+\!iK_I(\omega\!+\!\omega_j\!-\!\omega_e,t)].
    \label{eq:(B)}
\end{align}
(In this and following expressions, we will drop the superscript “bare” to save space). Note that at this order, it is unnecessary to differentiate between the Gaussian and non-Gaussian noise. To further simplify the expressions, we next approximate the exact results in Eqs.~\eqref{eq:(A)} and \eqref{eq:(B)}. These approximations are based on the structures of both the filter functions $K^{R,I}(\omega, t)$ and the noise spectrum $S(\omega)$. First, the spectrum $S(\omega)$ assumed in this study is symmetric (due to the classical nature of the noise) and is concentrated around $\omega=0$, as shown in FIG.~S\ref{fig:FilterFunctions} (black-dashed curves). Such structured spectrum motivates us to assume that at any non-zero qubit transition frequency $\omega=\omega_{j}-\omega_{j'}$ ($j\neq j'$), the noise power $S(\omega)$ is negligible. Second, we are interested in the timescale that is much longer than the qubit oscillation period, i.e., $t\gg 2\pi/(\omega_j-\omega_{j'})$ ($j\neq j'$). In this way, the widths of the filter functions are much smaller than the qubit frequencies, and any filter function centered at a nonzero qubit transition frequency negligibly sample the low-frequency part of the noise spectrum (see FIG.~S\ref{fig:FilterFunctions}).

With these assumptions, we first simplify the real parts of these diagrams. The most important approximations are made for (B) with $j\neq e$. The real-valued contribution from high frequencies is negligible due to the assumption of low-frequency noise. (For the Floquet case discussed later, such contribution is potentially more important due to the smaller quasi-energy difference.) After neglecting it, the magnitude of the real part of such diagram is approximately $\sim\epsilon^2|x_{ej}|^2\int d\omega S(\omega)/(2\pi\omega_{je}^2)$, which does not grow with time and is negligible if the noise amplitude is sufficiently weak. After dropping such contributions, the sum of the remaining real-valued terms from (A) and (B) and their flipped diagrams is approximated by
\begin{align}
    -\Phi^{(2)}(t)\approx&\,-\epsilon^2(x_{ee}-x_{gg})^2\int_{-\infty}^{\infty} \frac{d\omega}{2\pi}K^R(\omega, t) S(\omega)\nonumber\\
        =&-\epsilon^2 D_{1,\lambda}^2\int_{-\infty}^{\infty} \frac{d\omega}{2\pi}K^R(\omega, t) S(\omega)\nonumber\\
    \approx & -\epsilon^4D_{2,0}^2\lambda^2\int^{\infty}_{-\infty}\frac{d\omega}{2\pi}K^R(\omega, t)S(\omega),\label{eq:Phi2}
\end{align}
where we have used 
\begin{align}
    D_{1,\lambda}\equiv \frac{\partial\omega^{\mathrm{bare}}_{ge}}{\partial\lambda} = x_{ee}-x_{gg}\label{eq:1stdrv}
\end{align}
and $D_{1,\lambda}\approx D_{2,\lambda=0}\lambda$. Eq.~\eqref{eq:Phi2} is reminiscent of those derived using the Bloch-Redfield theory \cite{Koch_transmon_theory,Ithier_decoherence_analysis}, which describes how the dephasing of the qubit is related to the first-order derivative. Note that in Eq.~\eqref{eq:Phi2} we add the prefactor $\epsilon^2$ to track the order of the small parameter $\lambda$. This term is of the order $O(\epsilon^4)$, and grows with $t$ according to the expression of $K^{R}(\omega,t)$. For example, if $S(\omega)$ can be treated flat within the width of the filter function, $\Phi^{(2)}(t)$ grows linearly with $t$, while for $1/f$ noise, such growth is approximately $\sim t^2$ \cite{Ithier_decoherence_analysis}.

We next turn to the imaginary parts of (A) and (B). Contrary to $K^R(\omega,t)$, the function $K^I(\omega, t)$ is an odd function in the frequency domain. The symmetric spectrum of the classical noise then guarantees the vanishing of the integral $\int_{-\infty}^{\infty}(d\omega/2\pi)S(\omega)K^I(\omega, t)$. Therefore, the imaginary terms we collect are from the diagrams with $j\!\neq\!e$ in (B). Using the assumption of the low-frequency noise again, we only focus on the integration over $\omega\approx 0$, which allows us to approximate $K^I(\omega+\omega_j-\omega_e)\approx -t/(\omega_{j}-\omega_e)$. This step simplifies the expression of the imaginary part by
\begin{align}
    -i\delta\omega^{(2)}t\approx \! &-i\epsilon^2 \int_{-\infty}^{\infty}\frac{d\omega}{2\pi}\Bigg[\sum_{j\neq g}|x_{ej}|^2K^I(\omega\!+\!\omega_j\!-\!\omega_e) S(\omega) \nonumber\\ &\qquad\qquad+           \sum_{j'\neq e}|x_{gj'}|^2K^I(\omega\!+\!\omega_{g}\!-\!\omega_{j'})S(\omega)\Bigg]\nonumber\\
    \approx& \!-\!it\epsilon^2\Bigg[\sum_{j\neq e}\frac{|x_{ej}|^2}{\omega_e\!-\!\omega_j} \! - \!\!  \sum_{j'\neq g}\frac{|x_{gj'}|^2}{\omega_g\!-\!\omega_j}  \Bigg]\int_{-\infty}^{\infty}\frac{d\omega}{2\pi}S(\omega)\nonumber\\
    \approx&-\!\frac{it}{2}\epsilon^2 D_{2,\lambda=0}\!\int_{-\infty}^{\infty}\!\frac{d\omega}{2\pi}  S(\omega)\label{eq:LambShif}.
\end{align}
Above, we have used the relation
\begin{align}
    D_{2,\lambda}\equiv\frac{\partial^2\omega_{ge}^{\mathrm{bare}}}{\partial \lambda^2} =2\left[ \sum_{j\neq e}\frac{|x_{ej}|^2}{\omega_e\!-\!\omega_j}  -   \sum_{j'\neq g}\frac{|x_{gj'}|^2}{\omega_g\!-\!\omega_{j'}}\right].\label{eq:2nddrv}
\end{align}
Eq.~\eqref{eq:LambShif}  is related to the Lamb shift caused by the fluctuator $\delta\xi(t)$. To summarize, the two-point diagrams contribute $\Pi_{eg\leftarrow eg}^{(2)}(t) = -\Phi^{(2)}(t) - i\delta\omega^{(2)}t$ to the full projector, where $\Phi^{(2)}(t)$ is of the order $\epsilon^4$ and $\delta\omega^{(2)}$ is of the order $\epsilon^2$. 

\subsection{Three-point Keldysh diagrams}
Different from in Refs.~\cite{Muller_Keldysh,Shnirman_Keldysh_dephasing,Shnirman_Keldysh_dephasing_1}, it is necessary to calculate diagrams with odd interaction points due to the non-zero odd-time correlation functions. This subsection focuses on these diagrams. According to the $\epsilon$R4I2 scheme described in Sec.II.B and the expansion Eq.~\eqref{eq:self-energy}, we only need to focus on real parts of these diagrams.

The relevant diagrams are FIG.~S\ref{fig:Keldysh} (C)-(E). Again, we show the calculation details of one of them. The diagram (C) corresponds to the integral
\begin{align}
\text{(C)} =&\, (-i)(i)^2\epsilon^3\int_0^t\!dt_1\int_0^{t_1}\!dt_2\int_0^{t_2}\!dt_3 x_{ee}|x_{gj}|^2\nonumber\\
&\times\overline{\delta\xi(t_1)\delta\xi(t_2)\delta\xi(t_3)}e^{i(\!\omega_j\!-\!\omega_g\!)t_2+i(\!\omega_g\!-\!\omega_j\!)t_3}\label{eq:3pE}.
\end{align}
In general, we need the information of the bispectrum associated with the correlation $\overline{\delta\xi(t_1)\delta\xi(t_2)\delta\xi(t_3)}$ to proceed with the calculation \cite{Clerk_non_Gaussian}. Fortunately, the three-time correlation functions of the TLFs, shown in Eq.~\eqref{eq:correlations}, have a simple structure. As one can check using Eq.~\eqref{eq:TLFcorrelation}, the three-time correlation function of one TLF can be expressed as
\begin{align}
    \overline{\delta\xi_T(t_1)\delta\xi_T(t_2)\delta\xi_T(t_3)}=&-\!2\overline{\xi}_TS_T(t_1-t_3)\label{eq:3pto2p}\\
    =&-\!2\overline{\xi}_T\!\int_{-\infty}^{\infty}\!\frac{d\omega}{2\pi}S_T(\omega)e^{-i\omega(t_1-t_3)}.\nonumber
\end{align}
Since the Gaussian noise does not contribute odd-time correlation functions, the full correlation $\overline{\delta\xi(t_1)\delta\xi(t_2)\delta\xi(t_3)}$ is just the sum of all TLF contributions. In the following, we will first focus on one TLF's contribution. 

For a single TLF, we insert Eq.~\eqref{eq:3pto2p} into Eq.~\eqref{eq:3pE}. The result is more concisely expressed by a Laplace transform:
\begin{align}
    &\mathcal{L}\{\mathrm{(C)}\}(s) = -i\epsilon^3 x_{ee}|x_{gj}|^2\int_{-\infty}^{\infty}\!\frac{d\omega}{2\pi}2\overline{\xi}_TS_T(\omega)\nonumber\\
    &\times\frac{1}{s^2}\frac{1}{s+i\omega} \frac{1}{s-i(\omega_j\!-\!\omega_g\!-\!\omega)}.
\end{align}
Note that we are not interested in diagrams with $j=g$ because their leading order is $O(\epsilon^6)$. Then for $j\neq g$, the last line of the expression above is approximated by
\begin{align}
    &\mathcal{L}^{-1}\left\{\frac{1}{s^2}\frac{1}{s+i\omega}\frac{1}{s-i(\omega_j-\omega_g-\omega)}\right\}(t)\\
    = &\,  \frac{iK^R(\omega, t)}{\omega_j-\omega_g} - \frac{iK^R(\omega-\omega_j+\omega_e, t)}{\omega_j-\omega_g}. \nonumber
\end{align}
The second term contributes negligibly due to the structures of $K^R(\omega-\omega_j+\omega_g, t)$ and $S_T(\omega)$ as presented in FIG.~S\ref{fig:FilterFunctions}, therefore we can focus only on the first one. 

We can calculate (D) and (E) following similar procedures. Again, we will neglect diagrams with $j=g$ in (D) and $j= j'=e$ in (E) according to the $\epsilon^4$ approximation. On the other hand, (E) with $j\neq e$ and $j'\neq e$ does not contribute significantly to the real part, so we will also omit the evaluation of this case. To summarize, we find 
\begin{align}
    &\mathrm{Re}\{\mathrm{(C)}|_{j\neq g}\} \!\approx\! -\frac{\epsilon^3x_{ee}|x_{gj}|^2}{(\omega_g-\omega_j)}\int\!\frac{d\omega}{2\pi}2\overline{\xi}_T K^R(\omega, t) S_T(\omega),\nonumber\\
    &\mathrm{Re}\{\mathrm{(D)}|_{j\neq e}\} \approx -\frac{\epsilon^3x_{gg}|x_{ej}|^2}{(\omega_e-\omega_j)}\int\!\frac{d\omega}{2\pi}2\overline{\xi}_T K^R(\omega, t) S_T(\omega),\!\nonumber\\
    &\mathrm{Re}\{\mathrm{(E)}|_{j=e,j'\neq e}\}\! \approx\! \frac{\epsilon^3 x_{ee}|x_{ej'}|^2}{(\omega_e-\omega_{j'})}\int\!\frac{d\omega}{2\pi}2\overline{\xi}_TK^R(\omega, t)S_T(\omega).\nonumber\\
    &\mathrm{Re}\{\mathrm{(E)}|_{j\neq e,j'= e}\}\! \approx\! \frac{\epsilon^3 x_{ee}|x_{ej}|^2}{(\omega_e-\omega_{j})}\int\!\frac{d\omega}{2\pi}2\overline{\xi}_TK^R(\omega, t)S_T(\omega).
\end{align}

Finally,  using Eqs.~\eqref{eq:1stdrv} and \eqref{eq:2nddrv}, we summarize (C)-(E) and their flipped diagrams as
\begin{align}
-\Phi^{(3)}\!(t)\!\approx & \epsilon^3D_{1,\lambda}D_{2,\lambda}\!\sum_{\mu}\!\int\! \!\frac{d\omega}{2\pi}2\overline{\xi}_{T\!\mu}K^R(\omega, t)S_{T\!\mu}(\omega),\nonumber\\
\approx& \epsilon^4D^2_{2,\lambda = 0}\! \sum_{\mu}\!(2\lambda\overline{\xi}_{T\!\mu})\!\int\! \!\frac{d\omega}{2\pi}K^R(\omega, t)S_{T\!\mu}(\omega).
\end{align}
The three-point diagrams contribute $\mathrm{Re}\{\Pi^{(3)}_{eg\leftarrow eg}(t)\} \approx -\Phi^{(3)}(t)$ to the projector up to $\epsilon^4$.

\subsection{Four-point Keldysh diagrams}
The calculation of the four-point diagrams is more complicated than the previous cases, because here both Gaussian and non-Gaussian noise contribute. Some diagrams only involve one of them, while the others involve both. The four-time correlation functions of these two types of fluctuators can both be derived from the two-time ones, but through different relations. Specifically, they are respectively expanded by
\begin{align}
    \overline{\delta\xi_G\!(t)\delta\xi_G\!(t_1)\delta\xi_G\!(t_2)\delta\xi_G\!(0)} 
    =&\, \overline{\delta\xi_G\!(t)\delta\xi_G\!(t_1)}\,\overline{\delta\xi_G\!(t_2)\delta\xi_G\!(0)}\nonumber\\
    +&\overline{\delta\xi_G\!(t)\delta\xi_G\!(t_2)}\,\overline{\delta\xi_G\!(t_1)\delta\xi_G\!(0)}\nonumber\\
    +&\overline{\delta\xi_G\!(t)\delta\xi_G\!(0)}\,\overline{\delta\xi_G\!(t_1)\delta\xi_G\!(t_2)},\nonumber\\
    \overline{\delta\xi_T(t)\delta\xi_T(t_1)\delta\xi_T(t_2)\delta\xi_T(0)} =& \overline{\delta\xi_T(t)\delta\xi_T(t_1)}\,\overline{\delta\xi_T(t_2)\delta\xi_T(0)}\nonumber\\
    +&\Big(2\overline{\xi}_T\Big)^2\overline{\delta\xi_T(t)\delta\xi_T(0)}.
\end{align}
In the following, we will carefully address such distinction, and outline the results of various types of combinations.

As discussed in Ref.~\cite{Muller_Keldysh}, certain diagrams should not appear in the self energy due to the cancellation by the products of aforementioned diagrams, according to Eq.~\eqref{eq:self-energy}. We first focus on these redundant ones, which are called the disconnected diagrams, e.g., (F) and (G). Without showing more details, we list the results of these integrals below
\begin{align}
    \mathrm{Re}\{\text{(F)}_{j\neq g,j'\neq e}\}\! \approx&  \frac{\epsilon^4t^2|x_{ej'}|^2|x_{gj}|^2}{2(\omega_e\!-\!\omega_{j'})(\omega_g\!-\!\omega_{j})} \nonumber\\
    &\times\iint\!\!\frac{d\omega}{2\pi}\frac{d\omega'}{2\pi}S(\omega)S({\omega'}),\nonumber\\
    \mathrm{Re}\{\text{(G)}_{j\neq e, j'\neq e}\}\! \approx&  \frac{-\epsilon^4 t^2|x_{ej'}|^2|x_{ej}|^2}{2(\omega_e\!-\!\omega_{j'})(\omega_e\!-\!\omega_{j})}\nonumber\\&\times\iint \!\!\frac{d\omega}{2\pi}\frac{d\omega'}{2\pi}S(\omega)S({\omega'}).\!\!
\end{align}
These diagrams will be cancelled by the products of two-point ones. For example, one can check $\mathrm{Re}\{\mathrm{(G)}\} \approx [i\,\mathrm{Im}\{\mathrm{(B)}\}]^2/2$. Therefore, according to Eq.~\eqref{eq:self-energy}, $\mathrm{Re}\{\mathrm{(G)}\}$ should not contribute to the self energy. The same conclusion holds for (F) as well as their flipped diagrams.

On the contrary, FIG.~S\ref{fig:Keldysh} (H)-(M) are connected and contribute to the self energy under our approximation scheme. Among them, (H) and (I) correspond to the contributions by a single TLF. They are evaluated as
\begin{align}
    \mathrm{Re}\{\mathrm{(H)}_{j\neq e, j'\neq g}\}\! \approx& \frac{\epsilon^4|x_{ej}|^2|x_{gj'}|^2}{(\omega_{e}\!-\!\omega_j)(\omega_g\!-\!\omega_{j'})}\nonumber\\
    &\times\int\!\!\frac{d\omega}{2\pi}(2\overline{\xi}_T)^2 K^R(\omega, t) S_T(\omega)\nonumber\\
    \mathrm{Re}\{\mathrm{(I)}_{j\neq e, j'\neq e}\}\! \approx&
    \frac{-\epsilon^4|x_{ej}|^2|x_{ej'}|^2}{(\omega_{e}\!-\!\omega_{j'})(\omega_e\!-\!\omega_{j})}\nonumber\\
    &\times\int\!\!\frac{d\omega}{2\pi}(2\overline{\xi}_T)^2 K^R(\omega, t) S_T(\omega).
\end{align}

Differently, FIG.~S\ref{fig:Keldysh} (J)-(M) can describe both the mutual contribution by a Gaussian fluctuator-TLF pair or a TLF-TLF pair and the contribution by the Gaussian noise itself. In the Gaussian fluctuator-TLF case, the four diagrams are evaluated as
\begin{align}
    \mathrm{Re}\{\mathrm{(J)}_{j\neq e,j'\neq g}\}  \approx& \frac{\epsilon^4|x_{ej}|^2|x_{gj'}|^2}{(\omega_{e}\!-\!\omega_j)(\omega_g\!-\!\omega_{j'})}\nonumber\\
    \times&\iint\!\!\frac{d\omega}{2\pi}\frac{d\omega'}{2\pi} K^R(\omega+\omega', t) S_{T\!\mu}(\omega)S_G(\omega')\nonumber\\
    \mathrm{Re}\{\mathrm{(K)}_{j\neq e,j'\neq g}\}\! \approx&\,\mathrm{Re}\{\mathrm{(J)}_{j\neq e,j'\neq g}\}\nonumber\\
    \mathrm{Re}\{\mathrm{(L)}_{j\neq e,j'\neq g}\}\! \approx& \frac{-\epsilon^4|x_{ej}|^2|x_{ej'}|^2}{(\omega_{e}\!-\!\omega_{j'})(\omega_e\!-\!\omega_{j})}\nonumber\\
    \times&\iint\!\!\frac{d\omega}{2\pi}\frac{d\omega'}{2\pi} K^R(\omega+\omega', t) S_{T\!\mu}(\omega)S_G(\omega'),\nonumber\\
    \mathrm{Re}\{\mathrm{(M)}_{j\neq e,j'\neq g}\}\!\approx&\,\mathrm{Re}\{\mathrm{(L)}_{j\neq e,j'\neq g}\}.
\end{align}
These expressions correspond to the mutual decoherence terms contributed by the Gaussian fluctuator $\delta\xi_G(t)$ and the $\mu$th TLF $\delta\xi_{T\!\mu}(t)$. The TLF-TLF contribution can be obtained by replacing $S_G(\omega')$ by the noise spectrum of another TLF $S_{T\!\mu'}(\omega')$, while the pure Gaussian contribution is obtained by replacing $S_{T\!\mu}(\omega)$ by $S_{G}(\omega)$.

Finally, using Eq.~\eqref{eq:2nddrv} again, we can summarize all real-valued contributions from four-point diagrams up to $\epsilon^4$ as
\begin{align}
    -\Phi^{(4)}\!&(t)\!\approx\!    -\epsilon^4D^2_{2,\lambda = 0} \sum_{\mu}\overline{\xi}_{T\!\mu}^2\!\!\int \!\frac{d\omega}{2\pi}K^R(\omega, t)S_{\mu}(\omega)\\
   \! -\epsilon^4&\frac{D^2_{2,\lambda = 0}}{2} \iint \!\frac{d\omega}{2\pi}\frac{d\omega'}{2\pi} K^R(\omega+\omega', t)S_{G}(\omega)S_G(\omega')\nonumber\\
\!-\epsilon^4&\frac{D^2_{2,\lambda = 0}}{2}\! \sum_{\mu\neq\mu'}\!\iint \!\frac{d\omega}{2\pi}\frac{d\omega'}{2\pi}K^R(\omega+\omega', t)S_{T\!\mu}(\omega)S_{T\!\mu'}(\omega')\nonumber\\
\!-\epsilon^4&D^2_{2,\lambda = 0}\! \sum_{\mu}\!\iint \!\frac{d\omega}{2\pi}\frac{d\omega'}{2\pi}K^R(\omega+\omega', t)S_{G}(\omega)S_{T\!\mu}(\omega')\nonumber.
\end{align}
Again, its order is $\epsilon^4$.

\subsection{Summary}
After adding the terms from Sec.~II.C-E, we finally find the leading-order expressions of $\mathrm{Re}\{\Sigma_{eg}(t)\} = -\Phi(t)$ and $\mathrm{Im}\{\Sigma_{eg}(t)\} = -\delta\omega_{LS}$ as
\begin{align}
\delta\omega_{LS} \approx & \frac{D_{2,\lambda=0}}{2}\!\!\int\!\!\frac{d\omega}{2\pi}S(\omega).\label{eq:lamb}\\
    \Phi(t) \approx&
     {D^2_{2,\lambda = 0}}\lambda^2 \!\!\int\! \!\frac{d\omega}{2\pi} K^R(\omega, t)S_{G}(\omega)\label{eq:dephasing}\\
   + &\frac{D^2_{2,\lambda = 0}}{2} \iint \!\frac{d\omega}{2\pi}\frac{d\omega'}{2\pi} K^R(\omega+\omega', t)S_{G}(\omega)S_G(\omega')\nonumber\\
   +&D^2_{2,\lambda = 0} \sum_{\mu}(\lambda-\overline{\xi}_{T\!\mu})^2\!\!\int \!\frac{d\omega}{2\pi}K^R(\omega, t)S_{\mu}(\omega)\nonumber\\
+&\frac{D^2_{2,\lambda = 0}}{2}\! \sum_{\mu\neq\mu'}\!\iint \!\frac{d\omega}{2\pi}\frac{d\omega'}{2\pi}K^R(\omega+\omega', t)S_{T\!\mu}(\omega)S_{T\!\mu'}(\omega')\nonumber\\
+&\,D^2_{2,\lambda = 0}\! \sum_{\mu}\!\iint \!\frac{d\omega}{2\pi}\frac{d\omega'}{2\pi}K^R(\omega+\omega', t)S_{G}(\omega)S_{T\!\mu}(\omega').\nonumber
\end{align}
\subsection{Generalization to the Floquet dephasing}
The expressions derived above can be generalized to describe the qubit dephasing under a periodic drive. In the following, we will outline such generalization and predict the Floquet dephasing profile using the Keldysh method.

Compared to the diagrams for static (Ramsey) dephasing, those in this driven case have two major differences. First, the qubit is prepared in a superposition of the two computational Floquet states, rather than its bare eigenstates. Second, due to the inclusion of the periodic drive, the bare propagator $\hat{U}_0(t)$ is
\begin{align}
    \hat{U}_0(t)=\sum_{k}e^{-i\varepsilon_k t}\vert w_k(t)\rangle \langle w_k(0)\vert,
\end{align}
where $\vert w_k(t)\rangle$ is the $k$th independent Floquet state, and $\varepsilon_k$ is its quasi-energy. As a result, the rotated operator in Eq.~\eqref{eq:rot_x} is replaced by
\begin{align}
    \tilde{x}(t) =&\, \sum_{k,k'}\langle w_{k}(t)\vert \hat{x}\vert w_{k'}(t)\rangle e^{i(\varepsilon_k-\varepsilon_{k'}) t} \vert w_{k}(0)\rangle\langle w_{k'}(0)\vert \nonumber\\
    =&\!\! \sum_{k,k',n\in \mathbb{Z}} x_{kk',n} e^{i(\varepsilon_k-\varepsilon_{k'}-n\omega_d) t} \vert w_{k'}(0)\rangle\langle w_k(0)\vert,
    \label{eq:rotxfloquet}
\end{align}
where $\omega_d=2\pi/ T_d$ denotes drive frequency, and $x_{kk',n}$ is the $k$th Fourier coefficient of the time-dependent matrix element defined by
\begin{align}
    x_{kk',n} \equiv \frac{1}{T_d}\int_0^{T_d}\!\!dt\langle w_{k}(t)\vert \hat{x}\vert w_{k'}(t)\rangle e^{in\omega_d t}.
    \label{eq:fourier_coef}
\end{align}
Note that the expansion Eq.~\eqref{eq:rotxfloquet} is analogous to Eq.~\eqref{eq:rot_x}, except that there is an additional index $n$ to enumerate. The analogy allows us to use the Keldysh diagrams in FIG.~S\ref{fig:Keldysh} to perturbatively calculate the Floquet dephasing profile. Useful for this calculation,  here the matrix elements are related to the derivatives by
\begin{align}
    D^F_{1,\lambda}\equiv \frac{\partial\varepsilon_{01}}{\partial \lambda} =& \, x_{11,0}-x_{00,0},\nonumber\\
    D^F_{2,\lambda}\equiv \,\frac{\partial\varepsilon^2_{01}}{\partial \lambda^2} 
    =&\,2\Bigg[ \sum_{k\neq 1,n\in\mathbb{Z}} \frac{|x_{1k,n}|^2}{\varepsilon_1-\varepsilon_k-n\omega_d}\nonumber\\
    &-\!\sum_{k'\neq 0,n'\in\mathbb{Z}} \frac{|x_{0k',n'}|^2}{\varepsilon_0-\varepsilon_{k'}-n'\omega_d}\Bigg].
    \label{eq:floquet_drvs}
\end{align}

Before proceeding with the approximations, we notice one important difference in the Floquet calculation -- the vanishing of $D^F_{1,\lambda=0}$ is not guaranteed for a general periodic drive. Such vanishing is the basis of the protection scheme introduced in the main text, which we must ensure. Fortunately, we prove that, this condition is satisfied if we choose $H_d(t) = f(t)\hat{x}$ and set $f(t+T/2) = -f(t)$. [The protocol used in the main text for the triple protection clearly satisfies this condition.] This vanishing is again related to the $\mathbb{Z}_2$ symmetry of the qubit Hamiltonian. In detail, the drive we choose  ensures that noise-free Hamiltonian at $\lambda = 0$, $\hat{H}_{q,d}(0, t) = \hat{H}_q(0) + \hat{H}_d(t)$, satisfies
\begin{align}
    \hat{R}\hat{H}_{q,d}(0,t)\hat{R}^\dagger = \hat{H}_{q,d}\left(0,t+\frac{T}{2}\right).
\end{align}
Again, if there is no degeneracy in the quasi-energy spectrum, the Floquet states must preserve a certain parity, i.e., $\hat{R}\vert w_k(t)\rangle =\pm \vert w_{k}\!\left(t+T/2\right)\rangle$. Then using the definition \eqref{eq:fourier_coef} and the relation $\hat{R}\hat{x}\hat{R}^\dagger = -\hat{x}$, one can find $x_{kk,0}|_{\lambda = 0}=0$, which further leads to $D^F_{1,\lambda =0}=0$ according to Eq.~\eqref{eq:floquet_drvs}.

After ensuring $D^F_{1,\lambda=0}=0$, we can follow a similar perturbative calculation in the previous sections to derive the Floquet dephasing profile. Conveniently, if the quasi-energy differences are sufficiently large, the profile can be obtained by replacing $D_{2,\lambda =0}$ by $D^{F}_{2,\lambda =0}$ in Eq.~\eqref{eq:dephasing}. In the main text, the Floquet mitigation protocol improves the dephasing time of a static qubit by 10 times, which is limited by the non-negligible Floquet depolarization rather than pure dephasing \cite{Dynamical_sweet_spot}. Such limitation goes beyond the approximation made above by neglecting the contribution from noise at qubit frequency, because this frequency (quasi-energy difference) is reduced at the triple protection point. In this way, the qubit is more susceptible to depolarization. However, such limitation can be potentially lifted by engineering larger $\varepsilon_{01}$ at the triple protection point through optimizing the qubit parameters and driving protocols.

\onecolumngrid
\newpage

\section{Adiabatic approximation and echo dephasing profile}
In this section, we derive the qubit dephasing profile using a different method, which is based on the adiabatic approximation \cite{Ithier_decoherence_analysis}. The purpose of this derivation is twofold. First, we want to check whether the intuitive picture, which assumes that the noise only shifts the qubit frequency, reproduces the rigorously derived results in Eq.~\eqref{eq:lamb} and \eqref{eq:dephasing}. Second, we can derive an expression of the dephasing profile by the Echo protocol around the qubit sweet spot, which is difficult to  obtain using the Keldysh expansion. 

Under the adiabatic approximation, the qubit dephasing profile is approximated by $\exp[-\mathcal{K}(t)]$, where the kernel is expressed as
\begin{align}
    \mathcal{K}(t) = \ln\left[\overline{\exp\left(-i\int_0^t\chi(t')\sum_{\nu}\frac{D_{\nu,\lambda}}{\nu!}\delta\xi^{\nu}(t')\right) } \right].\label{eq:K(t)}
\end{align}
Above, $\chi(t')$ describes the external control pulses specific to different measurement protocols. For a Ramsey measurement, we take the control function  to be $\chi_R(t') =  1$, while in the Echo case, we set it by $\chi_E(t) = 1$ for $0<t'<t/2$ and $\chi_E(t') = -1$ for $t/2<t'<t$. Similar to the previous section, we will expand Eq.~\eqref{eq:K(t)} to $\delta\xi^4(t)$. The terms that contain $\nu$ times of $\delta\xi(t)$ are collected into  $\mathcal{K}^{(\nu)}(t)$, which is given by
\begin{align}
    \mathcal{K}^{(2)}(t) =&\, -\frac{D^2_{1,\lambda}}{2}\overline{\iint_{0}^tdt_1dt_2\chi(t_1)\chi(t_2)\delta\xi(t_1)\delta\xi(t_2)}\\
    &-i\frac{D_{2,\lambda}}{2}\overline{\int_0^tdt_1\chi(t_1)\delta\xi^2(t_1)},\\
    \mathcal{K}^{(3)}(t) =& -i\frac{D^3_{1,\lambda}}{6}\overline{\iiint_{0}^tdt_1dt_2dt_3\chi(t_1)\chi(t_2)\chi(t_3)\delta\xi(t_1)\delta\xi(t_2)\delta\xi(t_3)}\nonumber\\
    &-\frac{D_{1,\lambda}D_{2,\lambda}}{2}\overline{\iint_0^tdt_1dt_2\chi(t_1)\chi(t_2)\delta\xi(t_1)\delta\xi^2(t_2)}\\
    &-i\frac{D_{3,\lambda}}{6}\overline{\int_0^tdt_1\chi(t_1)\delta\xi^3(t_1)},\nonumber\\
    \mathcal{K}^{(4)}(t) =&\, \frac{D^4_{1,\lambda}}{24}\overline{\iiiint_{0}^tdt_1dt_2dt_3dt_4\chi(t_1)\chi(t_2)\chi(t_3)\chi(t_4)\delta\xi(t_1)\delta\xi(t_2)\delta\xi(t_3)\delta\xi(t_4)}\nonumber\\
    &-\frac{D_{1,\lambda}D_{3,\lambda}}{6}\overline{\iiiint_0^tdt_1dt_2\chi(t_1)\chi(t_2)\delta\xi(t_1)\delta\xi^3(t_2)}\nonumber\\
    &-\frac{D^2_{2,\lambda}}{8}\overline{\iint_0^tdt_1dt_2\chi(t_1)\chi(t_2)\delta\xi^2(t_1)\delta\xi^2(t_2)}-\frac{1}{2}[\mathcal{K}^{(2)}(t)]^2\\
    &-i\frac{D_{4,\lambda}}{24}\overline{\int_0^tdt_1\chi(t_1)\delta\xi^4(t_1)}\nonumber.
\end{align}
We provide several key points in proceeding with the derivation. First, some correlation functions are Fourier transformed by 
\begin{align}
    &\overline{\delta\xi^2(t_1)} =\int\!\!\frac{d\omega}{2\pi} S(\omega),\quad \overline{\delta\xi(t_1)\xi(t_2)} = \int\!\!\frac{d\omega}{2\pi} S(\omega)e^{-i\omega(t_1-t_2)},\quad
    \overline{\delta\xi(t_1)\delta\xi^2(t_2)} =\sum_{\mu} (-2\overline{\xi}_T)\!\!\int\!\!\frac{d\omega}{2\pi} S_{T\!\mu}(\omega) e^{-i\omega(t_1-t_2)},\nonumber\\
    &\overline{\delta\xi^2(t_1)\delta\xi^2(t_2)} = \left[\int\!\! \frac{d\omega}{2\pi}S(\omega)\right]^2 + 2\!\iint\!\!\frac{d\omega}{2\pi}\frac{d\omega'}{2\pi}S_G(\omega)S_G(\omega')e^{-i(\omega+\omega')(t_1-t_2)} +\sum_{\mu}(2\overline{\xi}_T)^2\!\!\int\!\!\frac{d\omega}{2\pi}S_{T\!\mu}(\omega)e^{-i\omega(t_1-t_2)}\nonumber\\
    &\qquad\qquad\qquad+ 4\!\sum_{\mu}\!\iint\!\!\frac{d\omega}{2\pi}\frac{d\omega'}{2\pi}S_G(\omega)S_{T\!\mu}(\omega')e^{-i(\omega+\omega')(t_1-t_2)} + 2\!\!\sum_{\mu\neq \mu'} \iint\!\!\frac{d\omega}{2\pi}\frac{d\omega'}{2\pi}S_{T\!\mu}(\omega)S_{T\!\mu'}(\omega')e^{-i(\omega+\omega')(t_1-t_2)}.
\end{align}
Second, for the Ramsey case, the integral
$I(\omega, t) = \int_0^tdt_1\int_0^tdt_2 \chi(t_1)\chi(t_2)e^{-i\omega(t_1-t_2)}$ gives us the filter function $2K^R(\omega,t)$ by setting $\chi(t)=\chi_R(t)$. Using these expressions, if we still truncate the expressions according to the $\epsilon$R4I2 scheme, we can recover Eq.~\eqref{eq:lamb} and \eqref{eq:dephasing}. [Note that only the terms that survive this approximation scheme are numbered.]

\newpage
\twocolumngrid
 For the Echo protocol, the filter function is replaced by $I(\omega, t) = 2K^E(\omega,t)$ considering the choice $\chi(t)=\chi_E(t)$. [As a reminder, we define $K^R(\omega, t) \equiv t^2\mathrm{sinc}^2(\omega t/2)/2$, and $K^E(\omega, t) \equiv t^2\mathrm{sinc}^2(\omega t/4)\mathrm{sin}^2(\omega t/4)/2$.] Conveniently, the imaginary parts up to $\epsilon^4$ in this Echo case vanish due to the introduction of $\chi_E(t)$. Therefore, from this method, the Echo dephasing profile is obtained by replacing $K^R(\omega, t)$ by $K^E(\omega, t)$ in Eq.~\eqref{eq:dephasing}. We note that, although the basis of this derivation is less rigorous than the Keldysh formalism, the expression we obtain here well fits the Echo dephasing rates for $\lambda\approx 0$.

\section{Dephasing rate in the TLF beating regime}
In this section, we study the dephasing rates of the qubit for $\lambda \gg |\delta\xi(t)|$, where the TLF has entered the beating regime \cite{Altshuler_telegraph_TLF}. (We will define this regime later.) In this regime, we find that the dephasing rates extracted from the calculated dephasing profile \eqref{eq:dephasing} do not agree with the numerical simulation. To analytically capture the dephasing rate in this regime, we need to study the qubit evolution non-perturbatively. 

The dephasing rate in this regime can be obtained using an intuitive picture. In this picture, the first-order fluctuation of the qubit frequency overwhelms that of the second order, i.e., $|D_{1,\lambda}||\delta\xi(t)|\gg |D_{2,\lambda}||\delta\xi(t)|^2/2$, therefore the qubit frequency is approximated by $\omega_{ge}(t) \approx \omega^{\mathrm{bare}}_{ge}|_{\lambda} + D_{1,\lambda}\delta\xi(t)$ under the adiabatic approximation. To simplify the problem, here we neglect the much weaker dephasing contribution from the Gaussian noise and only focus on the $N_T$ strong TLFs. We denote the probability of finding $\mu$th TLF at $\pm|\xi_{T\!\mu}|-\overline{\xi}_{T\!\mu}$ by $P_{\mu,\pm}$ ($\mu=1,2,\cdots,N_T$), and the flipping rates at these configurations by $\kappa_{\mu,\mp}=P_{\mu,\mp}\kappa_{\mu}$, where $\kappa_{\mu}$ is the sum flipping rate of the $\mu$th TLF. The configuration of the strong TLFs is denoted as a vector $\vec{\eta} = [\eta_1,\eta_2,\cdots, \eta_{N_T}]$ ($\eta_\mu=\pm$), whose corresponding probability  is $P_{\vec{\eta}} = \Pi_{\mu}P_{\eta_\mu}$. The inverse of the characteristic time for one flip to take place from this configuration is  $\overline{\kappa}_{\vec{\eta}} = \sum_{\mu}\kappa_{\mu,-\eta_{\mu}}$. 

The dephasing rate is estimated by the following intuition. If we only focus on one TLF, it takes the time $\sim 1/\kappa_{\mu}$ for this TLF to flip back and forth, which means the phase shift introduced between the two adjacent flips is $\sim2\Delta\omega_{T\!\mu}/\kappa$. Here, the frequency difference is approximated by $2\Delta\omega_{T\!\mu}\approx 2|D_{1,\lambda}||\overline{\xi}_{T\!\mu}|$. If this phase shift is much larger than $2\pi$, the phase coherence should be completely destroyed after this process. Therefore, in this regime, the qubit dephasing time is determined by the flipping rate of the TLF, which also means that the qubit dephasing rate will not further grow with increasing $\Delta\omega_T$ \cite{Altshuler_telegraph_TLF}. For a specific TLF configuration $\vec{\eta}$, if this intuition still holds, the phase coherence time is given by $(\overline{\kappa}_{\vec{\eta}})^{-1}$. Then, if we consider all possible configurations, the average flipping rate of the TLFs sets the qubit dephasing rate by $\overline{\kappa} = \sum_{\vec{\eta}}\overline{\kappa}_{\vec{\eta}}P_{\vec{\eta}}$. In the simplest case, if we assume identical probability $P_{\mu,\pm} = 0.5$ and $\kappa_{\mu} = \kappa$, the dephasing rate  is reduced to $\overline{\kappa} = N_T\kappa/2$.

A more rigorously justification is made via the master equation method. Except for the correlation functions, we can also conveniently describe the flipping dynamics of TLFs by the jump operators in the master equation, if we promote the TLFs to quantum degrees of freedom. One can check that, this method yields identical correlation functions as in Eq.~\eqref{eq:TLFcorrelation} using the quantum regression theorem \cite{Koch_TLS_FD_thoerem}. For simplicity, we truncate the qubit degree of freedom to only two levels for a minimal model. The full qubit-TLFs Hamiltonian is given by
\begin{align}
    \hat{H}_{q\mathrm{TLF}} = \frac{\omega_q}{2}\hat{\sigma}_z +  \sum_\mu\frac{\Delta\omega_{T\!\mu}}{2}(\hat{\tau}_{z\mu} - \langle\hat{\tau}_{z\mu}\rangle)\hat{\sigma}_z,
\end{align}
where  $\hat{\sigma}_z$ and $\hat{\tau}_{z\mu}$ are the Pauli $z$ operators for the qubit and the $\mu$th TLF, respectively, and $\langle \hat{\tau}_{z\mu}\rangle = (P_{+,\mu}-P_{-,\mu})$ is the expectation of $\hat{\tau}_{z\mu}$ given the probability distribution introduced previously. Since the TLFs we address are classical, in the Hamiltonian above we have not included the bare TLF Hamiltonians, and the interaction between the qubit and TLFs is longitudinal such that the qubit and TLFs do not exchange excitations. Note that the later assumption is only reasonable for the regime $|D_{1,\lambda}||\delta\xi(t)|\gg |D_{2,\lambda}||\delta\xi(t)|^2/2$ interested in this section. Close to the sweet spot, the transverse coupling cannot be neglected. Including the flipping dynamics of the TLFs, the master equation is given by
\begin{align}
    \frac{d\hat{\rho}_{q\mathrm{TLF}}(t)}{dt} =& -i[\hat{H}_{q\mathrm{TLF}}, \hat{\rho}_{q\mathrm{TLF}}(t)] \nonumber\\
    &+ \sum_{\mu} \Big[\kappa_{\mu,+} \mathbb{D}[\hat{\tau}_{\mu,+}]\hat{\rho}_{q\mathrm{TLF}}(t) \nonumber\\
    &\quad \quad +\kappa_{\mu,-} \mathbb{D}[\hat{\tau}_{\mu,-}]\,\hat{\rho}_{q\mathrm{TLF}}(t)\Big],
    \label{eq:density_evolution}
\end{align}
where the jump operator is defined by $\mathbb{D}[\hat{L}]\hat{\rho} \equiv \hat{L} \hat{\rho}\hat{L}^\dagger - (\hat{L}^\dagger \hat{L}\hat{\rho} + \hat{\rho}\hat{L}^\dagger\hat{L})/2$ and $\hat{\tau}_{\mu,+(-)}$ is the raising (lowering) operator for the $\mu$th TLF.

The quantity that we are interested in is the partial density matrix of the qubit $\hat{\rho}_q(t) \equiv \mathrm{Tr}_{\mathrm{TLF}}\{\hat{\rho}_{q\mathrm{TLF}}(t)\}$ rather than the full matrix $\hat{\rho}_{q\mathrm{TLF}}(t)$. Before we proceed with the derivation of the evolution of $\hat{\rho}_q(t)$, it is important to first specify the initial states of the qubit-TLFs system. Since the TLFs serve as classical noise in this model, there should be no entanglement between the qubit and the TLFs at $t=0$, and the TLFs are in its equilibrium state. Therefore, the full system is initiated as
\begin{align}
    \hat{\rho}_{q\mathrm{TLF}}(0) =\hat{\rho}_q(0)\otimes  \sum_{\vec{\eta}} P_{\vec{\eta}}\vert \vec{\eta}\rangle\langle \vec{\eta}\vert,
    \label{eq:initial}
\end{align}
where $\vert \vec{\eta}\rangle$ is the state of the TLFs in configuration $\vec{\eta}$. Using Eq.~\eqref{eq:density_evolution}, one can prove that the density matrix can keep the form
\begin{align}
        \hat{\rho}_{q\mathrm{TLF}}(t) = \sum_{\vec{\eta}} \hat{\rho}_{q,\vec{\eta}}(t)\otimes P_{\vec{\eta}}\vert \vec{\eta}\rangle\langle \vec{\eta}\vert
        \label{eq:rhoqtlfexpansion}
\end{align}
for any $t>0$, where the partial density matrix satisfies $\mathrm{Tr}_q\{\hat{\rho}_{q,\vec{\eta}}(t)\} = 1$. The proof is based on the assumption of the longitudinal coupling, which does not affect the population in the TLF and qubit states, and the relation satisfied by the upward and downward flipping rates, $\kappa_{\mu,+}P_{\mu,-} = \kappa_{\mu,+}P_{\mu,+}$. This form allows us to express the qubit density matrix by
\begin{align}
    \hat{\rho}_{q}(t) = \sum_{\vec{\eta}} P_{\vec{\eta}}\,\hat{\rho}_{q,\vec{\eta}}(t).
\end{align}

Eq.~\eqref{eq:density_evolution} can be further simplified by rotation of
$\hat{\rho}_{q\mathrm{TLF}}(t)$ according to
\begin{align}
\tilde{\rho}_{q\mathrm{TLF}}(t) = \hat{U}^\dagger_{q\mathrm{TLF}}(t)\hat{\rho}_{q\mathrm{TLF}}(t)\hat{U}_{q\mathrm{TLF}}(t),\label{eq:density_rotation}
\end{align}
where the propagator is given by $\hat{U}_{q\mathrm{TLF}}(t) = \exp[-i\hat{H}_{q\mathrm{TLF}}\,t]$. Using Eq.~\eqref{eq:rhoqtlfexpansion}, the rotated density matrix is expanded by
\begin{align}
    \tilde{\rho}_{q\mathrm{TLF}}(t) = \sum_{\vec{\eta}} \hat{U}_{q,\vec{\eta}}^\dagger(t)\hat{\rho}_{q,\vec{\eta}}(t)\hat{U}_{q,\vec{\eta}}(t) \otimes P_{\vec{\eta}}\vert \vec{\eta}\rangle\langle\vec{\eta}\vert,
\end{align}
where we define  the partial $\vec{\eta}$-dependent unitary $\hat{U}_{q,\vec{\eta}}(t) \equiv \exp[-i(\omega_q + \Delta\omega_{\vec{\eta}}) \hat{\sigma}_z t/2]$ and $\Delta\omega_{\vec{\eta}}\equiv \sum_{\mu}\Delta\omega_{T\!\mu}\langle \vec{\eta}\vert \hat{\tau}_{z\mu}-\langle\hat{\tau}_{z\mu}\rangle\vert \vec{\eta}\rangle$. The evolution of the rotated partial density matrix $\tilde{\rho}_{q}(t) \equiv \mathrm{Tr}_{\mathrm{TLF}}\{\tilde{\rho}_{q\mathrm{TLF}}(t)\}$ is governed by
\begin{align}
    \frac{d\tilde{\rho}_{q\mathrm{TLF}}(t)}{dt} =
    &\sum_{\mu} \Big[\kappa_{\mu,+} \mathbb{D}[\tilde{\tau}_{\mu,+}(t)]\tilde{\rho}_{q\mathrm{TLF}}(t) \nonumber\\
    &\quad \quad +\kappa_{\mu,-} \mathbb{D}[\tilde{\tau}_{\mu,-}(t)]\,\tilde{\rho}_{q\mathrm{TLF}}(t)\Big],
    \label{eq:master_rot}
\end{align}
where the jump operators are also rotated by the unitary $\hat{U}_{q\mathrm{TLF}}(t)$. For example, the raising operator of the $\mu$th TLF is rotated by
\begin{align}
    \tilde{\tau}_{\mu,+}(t) = &\, \hat{U}^\dagger_{q\mathrm{TLF}}(t)\hat{\tau}_{\mu,+}\hat{U}_{q\mathrm{TLF}}(t)\label{eq:rot_tau}\\
    =&\,\vert +_{\mu}\rangle\langle -_{\mu}\vert \otimes \hat{\mathbbm{1}}^{\prime\mu}_{\mathrm{TLF}}\otimes \vert e\rangle\langle e\vert \exp(i\Delta\omega_{T\!\mu}t)\nonumber\\
    &+ \vert +_{\mu}\rangle\langle -_{\mu}\vert \otimes \hat{\mathbbm{1}}^{\prime\mu}_{\mathrm{TLF}}\otimes \vert g\rangle\langle g\vert \exp(-i\Delta\omega_{T\!\mu}t),\nonumber
\end{align}
where $\hat{\mathbbm{1}}^{\prime\mu}_{\mathrm{TLF}}$ is the identity operator of the partial Hilbert space for the degrees of freedom of the TLFs with the $\mu$th TLF excluded. 

\textit{Asymptotic solution.--} The expansion \eqref{eq:rot_tau} allows us to simplify the master equation \eqref{eq:master_rot} by neglecting  fast-rotating terms. Using Eq.~\eqref{eq:rot_tau}, we can expand the jump term $\mathbb{D}[\tilde{\tau}_{\mu,+}(t)]\tilde{\rho}_{q\mathrm{TLF}}(t)$  into terms with different oscillating frequencies. If the strong TLFs are in their beating regime $\Delta\omega_{T\!\mu}\gg\sum_{\mu} \kappa_{\mu}$, the oscillatory terms can all be regarded as fast rotating and can therefore be neglected (rotating-wave approximation). This leads to the much simplified jump term
\begin{align}
    &\mathbb{D}[\tilde{\tau}_{\mu,+}(t)]\tilde{\rho}_{q\mathrm{TLF}}(t)\nonumber\\
    \approx&\, \mathbb{D}[\hat{\tau}_{e,\mu,+}]\tilde{\rho}_{q\mathrm{TLF}}(t) + \mathbb{D}[\hat{\tau}_{g,\mu,+}]\tilde{\rho}_{q\mathrm{TLF}}(t),
\end{align} 
where we define $\hat{\tau}_{e(g),\mu,+}\equiv \vert +_{\mu}\rangle\langle -_{\mu}\vert \otimes \mathbbm{1}^{\prime\mu}_{\mathrm{TLF}}\otimes \vert e(g)\rangle\langle e(g)\vert$. The partial trace of the approximated jump term gives us
\begin{align}
    &\mathrm{Tr}_{\mathrm{TLF}}\{\mathbb{D}[\tilde{\tau}_{\mu,+}(t)]\tilde{\rho}_{q\mathrm{TLF}}(t)\}\nonumber\\
    \approx&\,\left(\mathbb{D}[\vert e\rangle\langle e\vert]+\mathbb{D}[\vert g\rangle\langle g\vert]\right)\sum_{\vec{\eta}|_{\eta_{\mu} = -}}P_{\vec{\eta}}\tilde{\rho}_{q,\vec{\eta}}(t),\label{eq:etadephasing}
\end{align}
where the $\vec{\eta}$-dependent partial density matrix is $\tilde{\rho}_{q,\vec{\eta}}(t) = \hat{U}_{q,\vec{\eta}}^\dagger(t)\hat{\rho}_{q,\vec{\eta}}(t)\hat{U}_{q,\vec{\eta}}(t)$. The operation $\mathbb{D}[\vert e(g)\rangle\langle e(g)\vert]$  in Eq.~\eqref{eq:etadephasing}  does not affect the qubit population [diagonal matrix elements of $\tilde{\rho}_{q}(t)$], but exponentially reduces the magnitude of the off-diagonal matrix elements. As a result, we find
\begin{align}
    \tilde{\rho}_{q,eg}(t) \approx \tilde{\rho}_{q,eg}(0)\sum_{\vec{\eta}} P_{\vec{\eta}}\exp(-\kappa_{\vec{\eta}} t).
    \label{eq:rhoeg}
\end{align}
The lab-frame density matrix elements can be obtained by the reverse transformation of Eq.~\eqref{eq:density_rotation}, which only multiplies a $\vec{\eta}$-dependent oscillatory factor to the exponential decay function on the right-hand side of Eq.~\eqref{eq:rhoeg}. In the short-time limit ($\kappa_{\vec{\eta}}t\ll 1$), the effective dephasing rate is then approximated by $\overline{\kappa} = \sum_{\vec{\eta}}\kappa_{\vec{\eta}}P_{\vec{\eta}}$, which confirms the expression obtained previously.

The results given above are derived for the Ramsey measurement. However, we check that including the Echo pulse in the qubit Hamiltonian does not affect the conclusion of the dephasing rate in the beating limit \cite{Altshuler_telegraph_TLF,Galperin_Echo_TLFs}. In this parameter regime, it is expected that one random flip of the strong TLFs already occurs during the qubit coherence time, which impedes the refocusing of the phase of the qubit.

\textit{Exact solution.--} The asymptotic solution obtained above gives the saturation dephasing rates and clearly explains the beating behavior in the qubit oscillation. Meanwhile, an exact analytical solution for the density matrix of this longitudinally coupled qubit-TLFs model is also possible \cite{Altshuler_telegraph_TLF,Petukhov_nonGaussianDD}.

The derivation of the solution presented below follows the aforementioned references, although the framework used to describe this problem is not exactly the same. Also, we do not take the simplification of even probability distribution \cite{Tokura_dephasing}. 

For the ODE \eqref{eq:master_rot}, we propose a solution of $\tilde{\rho}_{eg,\mathrm{TLF}}(t)\equiv \langle e\vert\tilde{\rho}_{q\mathrm{TLF}}(t) \vert g\rangle$ as
\begin{align}
    \tilde{\rho}_{eg,\mathrm{TLF}}(t)\! =\!\rho_{q,eg}(0)\Pi_{\mu}\!\otimes\! \Big[\!h_{\mu+}\!(t)\vert +_{\mu}\rangle\langle +_{\mu}\vert\! +\! h_{\mu-}\!(t)\vert  -_{\mu}\rangle\langle -_{\mu}\vert\!\Big],
\end{align}
whose initial condition is set by \eqref{eq:initial} as $h_{\mu\pm}(0) = P_{\mu,\pm}$. For this solution to satisfy Eq.~\eqref{eq:master_rot}, the coefficients should satisfy the following differential equation:
\begin{align}
    \frac{dh_{\mu\pm}(t)}{dt} = -\kappa_{\mu,\mp} h_{\mu\pm}(t) + \kappa_{\mu,\pm} h_{\mu\mp}(t)e^{\pm 2i\Delta\omega_{T\!\mu}t}.
    \label{eq:hmupm}
\end{align}
Note that the rotating wave approximation we have made above is equivalent to neglecting the second term on the right-hand side. This approximation directly leads to the asymptotic solution in Eq.~\eqref{eq:rhoeg}. For the exact solution, we do not omit any term. Then to remove the time dependence in Eq.~\eqref{eq:hmupm}, we find it convenient to instead investigate the evolution of $h'_{\mu,\pm}(t) = h_{\mu,\pm}(t)e^{\pm i\Delta\omega_{T\!\mu}t}$, which satisfy
\begin{align}
    \frac{d}{dt}\!\begin{bmatrix}
h'_{\mu,+}(t)\\
h'_{\mu,-}(t)
\end{bmatrix} = 
\begin{bmatrix}
-i\Delta\omega_{T\!\mu}-\kappa_{\mu,-}&\kappa_{\mu,+}\\
\kappa_{\mu,-}&i\Delta\omega_{T\!\mu}-\kappa_{\mu,+}
\end{bmatrix}\!\begin{bmatrix}
h'_{\mu,+}(t)\\
h'_{\mu,-}(t)
\end{bmatrix}. 
\end{align}
This differential equation set can be solved conveniently after diagonalizing the 2$\times$2 matrix, if the determinant $S\equiv (\kappa_{\mu}^2-4\Delta\omega^2_{T\!\mu}-4i\Delta\kappa_{\mu}\Delta\omega_{T\!\mu})^{\frac{1}{2}}$ is not zero ($\Delta\kappa_{\mu}$ is the difference in decay rates $\Delta\kappa_{\mu} \equiv \kappa_{\mu,+}-\kappa_{\mu,-}$). We first focus on the scenario with nonzero $S$, where the two eigenvalues are
\begin{align}
    \Omega_{\mu,\pm} = \frac{1}{2}(-\kappa_\mu \pm S ),
\end{align}
and the corresponding eigenvectors (not normalized) are
\begin{align}
    \mathbf{v}_{\pm} = \Bigg[ \frac{\Delta\kappa -2i\Delta\omega_{T\!\mu} \pm S }{\kappa-\Delta\kappa}, 1\Bigg]^{\mathrm{T}}\!,
\end{align}
The solutions for $h'_{\mu,\pm}(t)$ are then obtained by matching the initial condition. Toward this goal, it is convenient to first find the inverse of $\mathbf{V} = [\mathbf{v}_+, \mathbf{v}_-]$, which is
\begin{align}
    \mathbf{V}^{-1} = \frac{1}{2S}\begin{bmatrix}
    \kappa-\Delta\kappa&-\Delta\kappa+2i\Delta\omega_{T\!\mu}+S\\
\Delta\kappa-\kappa& \Delta\kappa - 2i\Delta\omega_{T\!\mu}+S
\end{bmatrix}.
\end{align}
The initial population of $\mathbf{v}_{\pm}$ is denoted by $\mathbf{p} = \mathbf{V}^{-1}\mathbf{P}$, where $\mathbf{P} = [P_{\mu,+}, P_{\mu,-}]^{\mathrm{T}}$. Using these notations, we express the solutions as
\begin{align}
    h'_{\mu+}(t) =&\, (\mathbf{V}^{-1}\mathbf{P})_+(\mathbf{V})_{++}e^{\Omega_{\mu,+}t} + (\mathbf{V}^{-1}\mathbf{P})_-(\mathbf{V})_{+-}e^{\Omega_{\mu,-}t},\nonumber\\
    h'_{\mu-}(t) =&\, (\mathbf{V}^{-1}\mathbf{P})_-(\mathbf{V})_{--}e^{\Omega_{\mu,-}t} + (\mathbf{V}^{-1}\mathbf{P})_+(\mathbf{V})_{-+}e^{\Omega_{\mu,+}t}.
\end{align}

The exact solutions presented above are complicated. If we only consider TLFs with even probability distribution ($\Delta\kappa_{\mu}=0$, $P_{\mu,\pm}=1/2$), these solutions are simplified as
\begin{align}
    h'_{\mu+}(t) =& \frac{1}{4S}\Big[( S-2i\Delta\omega_{T\!\mu}+\kappa_{\mu})e^{\frac{1}{2}(-\kappa_{\mu}+S)t}\nonumber\\
    &\quad+(S+2i\Delta\omega_{T\!\mu}-\kappa_{\mu})e^{\frac{1}{2}(-\kappa_{\mu}-S)t}\Big],\nonumber\\
    h'_{\mu-}(t) =& \frac{1}{4S}\Big[( S-2i\Delta\omega_{T\!\mu}-\kappa_{\mu})e^{\frac{1}{2}(-\kappa_{\mu}-S)t}\nonumber\\
    &\quad+(S+2i\Delta\omega_{T\!\mu}+\kappa_{\mu})e^{\frac{1}{2}(-\kappa_{\mu}+S)t}\Big],
    \label{eq:TLFdephasingbalance}
\end{align}
which reproduce the results in Refs.~\cite{Altshuler_telegraph_TLF,Petukhov_nonGaussianDD} after tracing the TLF degrees of the freedom.

The special case with $S=0$, which is equivalent to $\Delta\kappa_{\mu} = 0$ and $\kappa_{\mu} = 2|\Delta\omega_{T\!\mu}|$, gives a critical solution
\begin{align}
    h'_{\mu\pm}(t) = \Big[\frac{1}{4}(1\mp i)\kappa_{\mu}t+\frac{1}{2}\Big]e^{-\kappa_{\mu}t/2}.
\end{align}
This solution can also be obtained by taking the limit $S\!\rightarrow\! 0$ in Eq.~\eqref{eq:TLFdephasingbalance}.

\bibliography{mybib}